\documentclass[twocolumn,iop]{emulateapj}
\usepackage{apjfonts}

\usepackage{amsmath,graphicx,longtable,hyperref}
\usepackage{natbib,threeparttable}
\hypersetup{
    bookmarks=true,         
    unicode=true,          
    pdftoolbar=true,        
    pdfmenubar=true,        
    pdffitwindow=false,     
    pdftitle={},    
    pdfauthor={Yue Shen},     
    pdfsubject={},   
    pdfcreator={Yue Shen},   
    pdfproducer={},  
    pdfkeywords={} {} {}, 
    pdfnewwindow=true,      
    colorlinks=false,       
    linkcolor=red,          
    citecolor=green,        
    filecolor=magenta,      
    urlcolor=cyan           
}

\newcommand{\fracd}[2]{\frac{\displaystyle{#1}}{\displaystyle{#2}}}
\newcommand{\etal}{et al.}
\newcommand{\hbeta}{H{$\beta$}}

\newcommand{\CIV}{C{\sevenrm IV}}

\newcommand{\CIII}{C{\sevenrm III]}}

\def\FeII{Fe\,{\sc ii}}
\def\MgII{Mg\,{\sc ii}}

\def \OIII {[O\,{\sc iii}]}

   \font\sevenrm=cmr7 scaled 1000

\def\kms{{\rm km\,s^{-1}}}

\begin{document}

\title{The Sloan Digital Sky Survey Reverberation Mapping Project: Technical Overview}


\author{Yue Shen$^{1,2,3}$, W.~N. Brandt$^{4,5}$, Kyle S.~Dawson$^{6}$, Patrick B.~Hall$^{7}$, Ian D.~McGreer$^{8}$, Scott F.~Anderson$^{9}$, Yuguang Chen$^{10}$, Kelly D.~Denney$^{11}$, Sarah Eftekharzadeh$^{12}$, Xiaohui Fan$^{8}$, Yang Gao$^{13,14}$, Paul J. Green$^{15}$, Jenny E.~Greene$^{16}$, Luis C.~Ho$^{3,10}$, Keith Horne$^{17}$, Linhua Jiang$^{18,2}$, Brandon C.~Kelly$^{19}$, Karen Kinemuchi$^{20}$, Christopher S.~Kochanek$^{11,21}$, Isabelle P\^aris$^{22}$, Christina M.~Peters$^{23}$, Bradley M.~Peterson$^{11,21}$, Patrick Petitjean$^{24}$, Kara Ponder$^{25}$, Gordon T.~Richards$^{23}$, Donald P.~Schneider$^{4,5}$, Anil Seth$^{6}$, Robyn N.~Smith$^{23}$, Michael A.~Strauss$^{16}$, Charling Tao$^{26,27}$, Jonathan R.~Trump$^{4,5,2}$, W.~M. Wood-Vasey$^{25}$, Ying Zu$^{11}$, Daniel J. Eisenstein$^{15}$, Kaike Pan$^{20}$, Dmitry Bizyaev$^{20}$, Viktor Malanushenko$^{20}$, Elena Malanushenko$^{20}$, Daniel Oravetz$^{20}$ }



\altaffiltext{1}{Carnegie Observatories, 813 Santa Barbara Street, Pasadena,
CA 91101, USA}
\altaffiltext{2}{Hubble Fellow}
\altaffiltext{3}{Kavli Institute for Astronomy and Astrophysics, Peking University, Beijing 100871, China}
\altaffiltext{4}{Department of Astronomy \& Astrophysics, The Pennsylvania
State University, University Park, PA, 16802, USA}
\altaffiltext{5}{Institute for Gravitation and the Cosmos, The Pennsylvania
State University, University Park, PA 16802, USA}
\altaffiltext{6}{Department of Physics \& Astronomy, University of Utah, 115 South 1400 East, Salt Lake City, UT 84112, USA}
\altaffiltext{7}{Department of Physics and Astronomy, York University, Toronto, ON M3J 1P3, Canada}
\altaffiltext{8}{Steward Observatory, The University of Arizona, 933 North Cherry Avenue, Tucson, AZ 85721-0065, USA}
\altaffiltext{9}{Astronomy Department, University of Washington, Box 351580, Seattle WA 98195, USA}
\altaffiltext{10}{Department of Astronomy, School of Physics, Peking University, Beijing 100871, China}
\altaffiltext{11}{Department of Astronomy, The Ohio State University, 140 West 18th Avenue,
Columbus, OH 43210, USA}
\altaffiltext{12}{Department of Physics and Astronomy, University of Wyoming, 1000 E. University Ave., Laramie, WY, 82071, USA}
\altaffiltext{13}{Department of Engineering Physics and Center for Astrophysics, Tsinghua University, Beijing 100084, China}
\altaffiltext{14}{Key Laboratory of Particle \& Radiation Imaging (Tsinghua University), Ministry of Education, Beijing 100084, China}
\altaffiltext{15}{Harvard-Smithsonian Center for Astrophysics, 60 Garden Street, Cambridge, MA 02138, USA}
\altaffiltext{16}{Department of Astrophysical Sciences, Princeton University, Princeton, NJ 08544, USA}
\altaffiltext{17}{SUPA Physics/Astronomy, Univ. of St. Andrews, St. Andrews KY16 9SS, Scotland, UK}
\altaffiltext{18}{School of Earth and Space Exploration, Arizona State University, Tempe, AZ 85287-1504, USA}
\altaffiltext{19}{Department of Physics, Broida Hall, University of California, Santa Barbara, CA 93107, USA}
\altaffiltext{20}{Apache Point Observatory and New Mexico State, University, P.O. Box 59, Sunspot, NM 88349-0059, USA}
\altaffiltext{21}{Center for Cosmology and AstroParticle Physics, The Ohio State University, 191 West Woodruff Avenue, Columbus, OH 43210, USA}
\altaffiltext{22}{INAF - Osservatorio Astronomico di Trieste, Via G. B. Tiepolo 11, I-34131 Trieste, IT}
\altaffiltext{23}{Department of Physics, Drexel University, 3141 Chestnut Street, Philadelphia, PA 19104, USA}
\altaffiltext{24}{Institut d'Astrophysique de Paris, Universit\'e Paris 6 and CNRS, 98bis Boulevard Arago, 75014 Paris, France}
\altaffiltext{25}{Pittsburgh Particle Physics, Astrophysics, and Cosmology Center (PITT PACC), Physics and Astronomy Department, University of Pittsburgh, Pittsburgh PA, 15260, USA}
\altaffiltext{26}{Centre de Physique des Particules de Marseille, Aix-Marseille Universit\'e , CNRS/IN2P3, 163, avenue de Luminy - Case 902 - 13288 Marseille Cedex 09, France}
\altaffiltext{27}{Tsinghua Center for Astrophysics, Tsinghua University, Beijing 100084, China}

\shorttitle{SDSS-RM: TECHNICAL OVERVIEW}


\shortauthors{SHEN ET~AL.}

\begin{abstract}

The Sloan Digital Sky Survey Reverberation Mapping project (SDSS-RM) is a dedicated multi-object RM experiment that has spectroscopically monitored a sample of 849 broad-line quasars in a single 7\,deg$^2$ field with the SDSS-III Baryon Oscillation Spectroscopic Survey (BOSS) spectrograph. The RM quasar sample is flux-limited to $i_{\rm psf}=21.7$ mag, and covers a redshift range of $0.1<z <4.5$ without any other cuts on quasar properties. Optical spectroscopy was performed during 2014 Jan--Jul dark/grey time, with an average cadence of $\sim 4$ days, totaling more than $30$ epochs. Supporting photometric monitoring in the $g$ and $i$ bands was conducted at multiple facilities including the Canada-France-Hawaii Telescope (CFHT) and the Steward Observatory Bok telescope in 2014, with a cadence of $\sim 2$ days and covering all lunar phases. The RM field (RA, DEC=14:14:49.00, +53:05:00.0) lies within the CFHT-LS W3 field, and coincides with the Pan-STARRS 1 (PS1) Medium Deep Field MD07, with three prior years of multi-band PS1 light curves. The SDSS-RM 6-month baseline program aims to detect time lags between the quasar continuum and broad line region (BLR) variability on timescales of up to several months (in the observed frame) for $\sim 10\%$ of the sample, and to anchor the time baseline for continued monitoring in the future to detect lags on longer timescales and at higher redshift. SDSS-RM is the first major program to systematically explore the potential of RM for broad-line quasars at $z>0.3$, and will investigate the prospects of RM with all major broad lines covered in optical spectroscopy. SDSS-RM will provide guidance on future multi-object RM campaigns on larger scales, and is aiming to deliver more than tens of BLR lag detections for a homogeneous sample of quasars. We describe the motivation, design and implementation of this program, and outline the science impact expected from the resulting data for RM and general quasar science. 

\end{abstract}


\section{Introduction}

Reverberation mapping (RM) is a technique for studying the structure and
kinematics of the broad-line regions (BLRs) of active galactic nuclei (AGNs) and quasars 
\citep[e.g.,][]{Blandford_McKee_1982,Peterson_1993,Peterson_2013}. RM is a
particularly important tool as the BLRs generally project to angular sizes of only tens of
microarcseconds or less, much too small to be resolved directly by any current
or near-future technology.
The broad emission lines in AGN spectra are often observed to have flux
variations correlated with that of the AGN continuum, but with a time delay interpreted as the
mean light-travel time across the BLR: the line-emitting gas 
appears to ``reverberate'' in response to the continuum flux variation. 

\begin{figure*}
\centering
	 \includegraphics[width=0.45\textwidth]{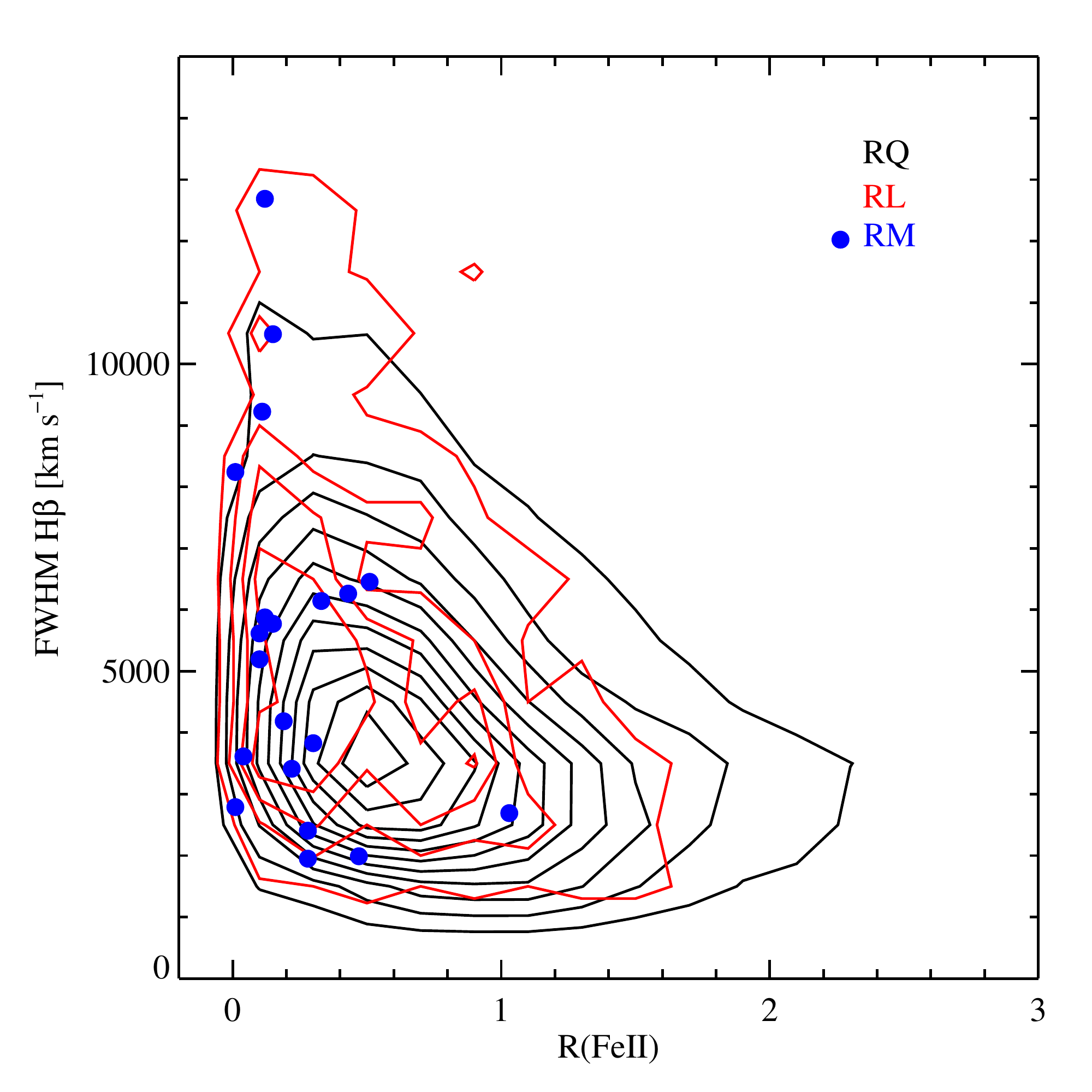}
	 \includegraphics[width=0.45\textwidth]{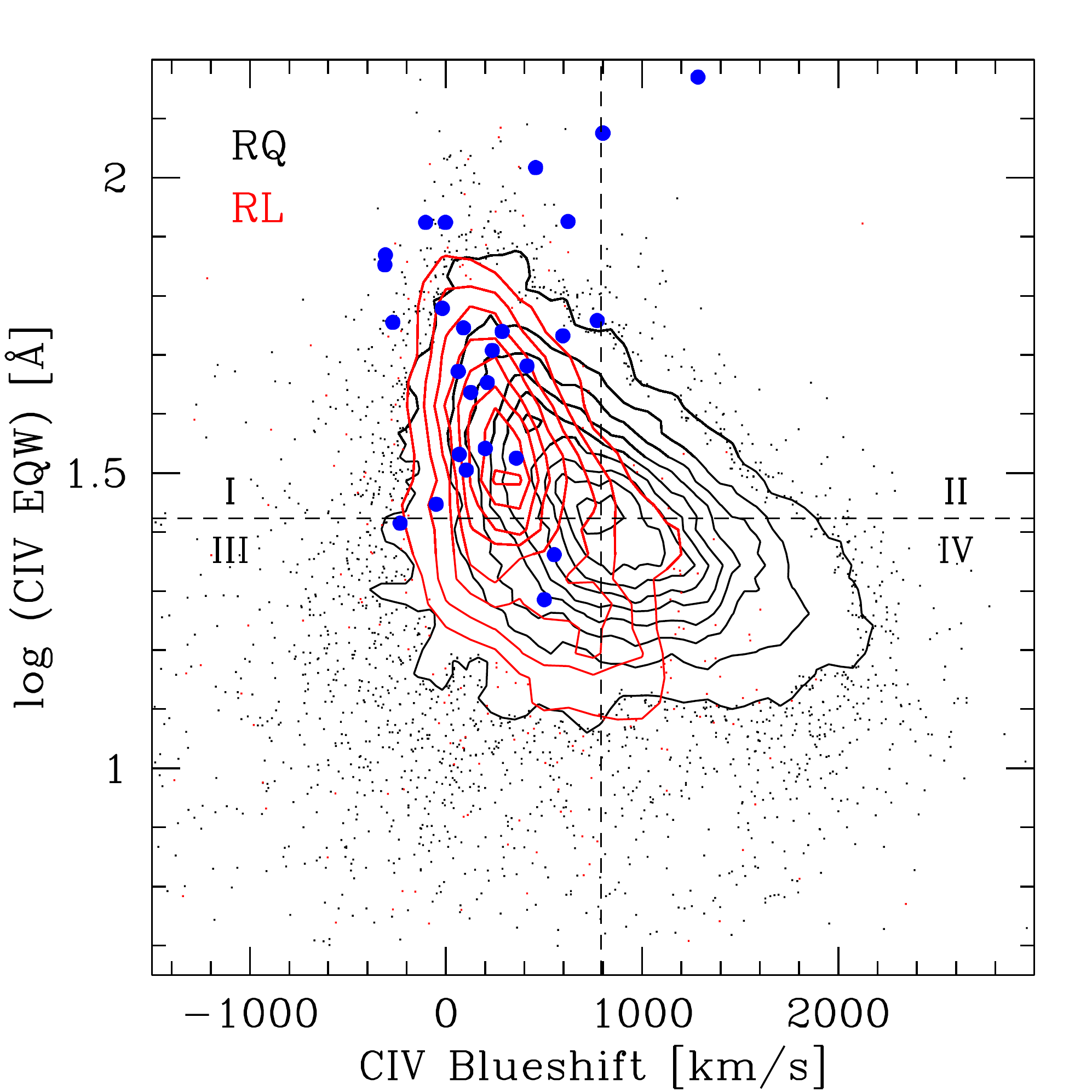}
\caption{{\em Left}: The distribution of \hbeta\ FWHM against the \FeII\ strength (measured as $R({\rm \FeII})\equiv {\rm EW_{\FeII 4434-4684}/EW_{H\beta}}$) for $z<0.8$ quasars. The distribution of quasars in this figure is related to the so-called Eigenvector 1 (EV1, e.g., Boroson \& Green 1992). The current AGN with RM measurements are shown as blue circles (only a representative subset of them with publicly available emission line measurements are shown). The current RM AGN sample lacks objects in the lower-right of the sequence, and hence the extrapolation of the current RM results to objects in that regime is uncertain. {\em Right}: There is a similar bias of the current RM sample, in terms of \CIV\ properties \citep[e.g.,][]{Richards_etal_2011}. The current RM sample lacks objects that have large \CIV\ blueshift or lower-than-average EWs, which are both typical of high-redshift quasars. \citet{Richards_etal_2011} divided this 2D plane into four quadrants (I-IV), to classify quasars based on \CIV\ properties. In both panels the contours are from SDSS quasars in Shen et~al.\ (2011), and the measurements for the subset of the current RM sample are from \citet{Sulentic07}. The black and red contours are for radio-quiet and radio-loud objects, respectively. }
\label{fig:EV1}
\end{figure*}

Because, by definition, the broad emission
lines are resolved in line-of-sight (or Doppler) velocity, it is in principle possible to
obtain additional information about the BLR structure and kinematics
by determining the time delay as a function of Doppler velocity.
The observational requirements for obtaining a ``velocity--delay map'' (i.e., the
projection of the BLR into the two observables, line-of-sight velocity and time delay) are quite
demanding in terms of data quality 
and time resolution and duration \citep{Horne_etal_2004}. Consequently it is only
recently that reliable velocity--delay maps have begun to appear in the literature
\citep[e.g.,][]{Bentz_etal_2010b, Grier_etal_2013}.
However, the somewhat simpler task of measuring only the mean time delay 
for emission lines remains important because it is possible to
measure to modest precision ($\sim 0.3\,{\rm dex}$) the mass of the
central black hole by combining the mean time delay
with the emission-line velocity width. This measurement assumes, of course, that the 
BLR is virialized in the gravitational potential of the black hole. The observed relationship between time delay and velocity width for
multiple emission lines in the same source is apparently consistent with this
hypothesis\footnote{The possible role of radiation pressure, also an
inverse square force, remains a source
of systematic uncertainty \citep{Marconi_etal_2008,Netzer_Marziani_2010}.}
\citep{Peterson_Wandel_1999, Peterson_Wandel_2000,
Kollatschny_2003, Peterson_etal_2004, Bentz_etal_2010a}. This virial hypothesis can be further tested with
velocity-resolved RM results \citep[e.g.,][]{Denney_etal_2009a}, or dynamical modeling of the structure and kinematics of
the BLR with velocity-resolved RM data \citep[e.g.,][]{Pancoast_etal_2012,Pancoast_etal_2013,Li_etal_2013}.

While dependent on a number of assumptions that require
further verification, RM is currently the primary 
method of BH mass measurement beyond the local ($z<0.1$) Universe, and anchors
other indirect methods of quasar BH mass estimation  that can be used, again
with assumptions that require further scrutiny, 
to arbitrarily high redshift \citep[for a recent review, see, e.g.,][]{Shen_2013}. Of particular interest are the so-called
single-epoch virial BH mass estimators utilizing the AGN luminosity and broad line width measured from single-epoch spectroscopy, which rely on the tight empirical relation between the BLR size and the AGN luminosity (the $R-L$ relation) discovered in RM studies \citep[e.g.,][]{Kaspi_etal_2000,Bentz_etal_2009a}.

RM experiments are time-consuming exercises that require a large
amount of telescope time and coordinated effort. Accurate determination of
an emission-line lag typically requires 30--50 well-spaced epochs of observation,
and some non-negligible fraction of RM experiments fail simply because
the AGN variations were too small or too unfavorable (e.g., a monotonic 
increase in flux throughout the campaign) to measure an emission-line lag.
Moreover, over much of its history, RM has been regarded as a technique
that is still in an experimental and developmental state.
Taken together, these circumstances necessitated undertaking
RM programs on telescopes that are relatively modest in size
(1--2\,m for spectroscopy, even smaller for continuum monitoring) where
it is realistically possible to assign large blocks of observing time to single projects.

Given the necessity of proving the value of the technique,
RM monitoring programs have been  ``success-oriented,'' 
targeting the apparently brightest, and in many cases, the most variable, AGNs in the sky. The subset of AGNs for which
RM measurements have been thus obtained is not, and was never intended to be,
representative of the quasar population. The current sample of AGN with reliable RM measurements is
highly biased toward local, low-luminosity AGNs. In addition, the current RM AGN sample does not probe the AGN parameter space uniformly. 

This point is illustrated in Fig.\ \ref{fig:EV1}. The left panel of Fig.\ \ref{fig:EV1} shows the distribution of \hbeta\ FWHM against the \FeII\ strength, for $z<0.8$ Sloan Digital Sky Survey (SDSS, York et~al.\ 2000) quasars. The distribution of quasars in this plot is related to the so-called Eigenvector 1 \citep[EV1, e.g.,][]{Boroson_Green_1992, Sulentic_etal_2000}, a physical sequence that is related to the accretion process of quasars \citep[e.g.,][]{Shen_Ho_2014}. The current RM AGNs are under-represented in the lower-right end of the sequence, and hence the extrapolation of the current RM results to objects in other regimes (in particular, narrow-line Seyfert 1, NLS1, e.g., Osterbrock \& Pogge 1985) is uncertain. The right panel of Fig.\ \ref{fig:EV1} presents a similar bias of the current RM sample, in terms of \CIV\ properties (e.g., Richards et~al.\ 2011). The current RM sample lacks objects that have large \CIV\ blueshifts or lower-than-average EWs, which are typical of high-redshift, high-luminosity quasars; this bias in the current RM sample is mostly a luminosity effect, and may secondarily depend on the sample radio properties. Thus expanding the current RM AGN sample to cover the AGN parameter space more uniformly is critical to test if we can apply RM results to the general broad-line quasar population \citep[][]{Shen_etal_2008,Richards_etal_2011,Denney_2012}. It is encouraging to note, however, the estimation of BLR sizes using gravitational microlensing in lensed, $z>1$ quasars is roughly consistent with the expectation from the current RM results \citep[e.g.,][]{Guerras_etal_2013}.


Since the first major reverberation campaign some 25 years ago
\citep{Clavel_etal_1991,Peterson_etal_1991,Dietrich_etal_1993},
RM measurements have been obtained for $\sim50$ AGNs, most often
only for the \hbeta\ emission line and almost exclusively for
AGNs at $z < 0.3$. Beyond $z\sim 1$, \hbeta\ shifts out of the optical band, and the major broad lines of RM interest are \MgII\ and \CIV\ (and \CIII). However, despite limited attempts \citep[e.g.,][]{Kaspi_etal_2007,Trevese_etal_2007}, there has not been any systematic RM investigation of these UV broad lines at high redshift. In fact, there is only one reliable detection of a \MgII\ lag in NGC 4151 \citep{Metzroth_etal_2006}, and two marginal \MgII\ lag detections in NGC 5548 \citep{Clavel_etal_1991} and in NGC 3783 \citep{Reichert_etal_1994}, despite the fact that \MgII\ shares many similarities with the Balmer lines \citep[e.g.,][]{Shen_etal_2008,Shen_Liu_2012}. It is therefore important to evaluate the potential of \MgII\ and \CIV\ RM for $z>1$ quasars with dedicated programs. 

RM campaigns have been carried out by
many different groups \citep[e.g.,][]{Peterson_etal_1998,Kaspi_etal_2000,Peterson_etal_2002,
Peterson_etal_2004,Kaspi_etal_2007,Bentz_etal_2009b,Denney_etal_2009a,Rafter_etal_2011,Rafter_etal_2013,
Barth_etal_2011a,Barth_etal_2011b,Barth_etal_2013,Du_etal_2014,Wang_etal_2014}. In all of these cases, the observations have been executed in 
a serial mode, observing one object at a time, although some efficiency
and risk mitigation is obtained by the use of queue observing.
However, strategies that are appropriate for brighter AGNs and smaller telescopes are simply
not extendable to higher redshift and fainter objects. 
Aside from being apparently fainter at higher redshift, higher-luminosity quasars
have lower amplitudes of variability \citep[e.g.,][]{Vandenberk_2004,MacLeod_etal_2010}, longer time delays for a given emission line,
and response times that are lengthened by cosmological time dilation. It would be extremely 
inefficient to pursue a similar strategy to obtain a substantial sample
of $z>1$ AGNs with \MgII\ or \CIV\ RM data.

Given the importance of RM in
understanding the BLR structure and measuring AGN BH masses,
it is crucial to consider alternative, more efficient, RM observing strategies
that will better sample from low to high redshift the broad characteristics of
the quasar population.
One possibility is ``photometric RM'', where the lag between an emission-line-free
bandpass and another bandpass containing an emission line is measured \citep[e.g.,][]{Haas_etal_2011,Chelouche_Daniel_2012,Chelouche_etal_2014,Zu_etal_2014}. However,
given the typical amplitude of emission-line variability on reverberation time scales
and the small fraction of the flux that
an emission line contributes to the total flux in the bandpass, this technique has challenges 
as it can be expected to succeed only in cases with some favorable combination of
large equivalent-width emission lines, 
narrow photometric bandpasses, unusually large variations, and extraordinarily precise
photometry \citep[e.g.,][]{Zu_etal_2014}.
Another possibility is multi-object spectroscopy. Given the low sky surface density of quasars, this obviously requires larger telescopes (with a suitable field-of-view, FoV) than have been used in the past for RM as it is necessary to obtain high-quality spectra of faint quasars in order to realize the multiplexing
advantage. 

In this work, we consider the practical difficulties in implementing a RM
program of multi-object spectroscopy. Such a program must
develop a strategy that is 
optimized to observe as many quasars as possible simultaneously to
an appropriate depth, with the optimal cadence and time baseline. In
addition, a significant amount of telescope time must be allocated to
this program, and the required observing strategy is best achieved
with service-mode observation. Currently, SDSS \citep[][]{SDSS} provides an ideal facility
for such a program, given its unique characteristics, including (a) a large
field-of-view (7 deg$^2$), (b) sufficiently numerous fibers to simultaneously observe nearly a
thousand quasars, (c) routinely executed queue observations by the
survey team, (d) a data reduction pipeline that is automatic and uniform, and 
(e) well-characterized calibration.

This paper presents a technical overview of the SDSS-RM project, the
first major multi-object RM program attempted to date, executed with the SDSS-III Baryon Oscillation Spectroscopic Survey (BOSS) spectrograph. This program was approved as one of the extra dark
time projects during the last observing season (2014A) of the SDSS-III
survey \citep[][]{Eisenstein_etal_2011}, and received significant time allocations from the Canada-France-Hawaii Telescope (CFHT) and the
Steward Observatory Bok 2.3-m telescope to obtain accompanying
photometric light curves. The baseline program had a length of 6 months (2014A semester) over
7 dark/grey runs, with 3-6 spectroscopic epochs per run (a cadence of a few days). Each spectroscopic epoch performed simultaneous spectroscopy for a flux-limited sample of $\sim 850$ quasars in a single, 7 deg$^2$ field. The photometric monitoring had a cadence of $\sim 2$ days over the same period, covering both dark and bright times. Given the large spectral coverage of BOSS spectroscopy and the time baseline, this program aims to detect $\sim 100$ lags on timescales up to a few months, and to systematically investigate the prospects of RM with UV broad lines at high redshift, for a uniformly selected quasar sample over a wide redshift range. It will also perform the first large, systematic investigation of \hbeta\ RM at $z>0.3$. With substantial science merit on its own, this program will also serve as a pathfinder for future multi-object RM projects on larger scales. 

This paper is structured in two main parts. In the first part (\S\ref{sec:design}), we present a general description of the optimal design of a multi-object RM program using simulations, with the goal to provide some guidance for similar programs in the future; these investigations also drove the design of the SDSS-RM program and provided forecasts on the yields. Readers not interested in the technical details can skip this section. The second part of the paper (\S\ref{sec:implem}) describes the specifics of the SDSS-RM program: its implementation, observations, data processing and organization. We discuss the science impact of our program
in \S\ref{sec:disc} and summarize in \S\ref{sec:sum}. By default the term ``quasar'' refers to
unobscured, Type 1 broad-line quasars. All coordinates are in the J2000 system, and the quasar magnitudes are the best-fit Point Spread Function (PSF) magnitudes from SDSS. We adopt a flat $\Lambda$CDM cosmology with $\Omega_0=0.3$ and $H_0=70\,{\rm km\,s^{-1}Mpc^{-1}}$.



\section{Designing a multi-object RM program}\label{sec:design}

To design the optimal program and estimate the expected yields, we performed simulations of mock detections of time lags for a sample of quasars covering a wide redshift and magnitude range under various observing schemes. We start by simulating a grid of quasars and their variability properties, then generate mock light curves (LCs), and compute cross-correlation functions to estimate the detection probability on the grid. We then impose the expected flux-limited quasar sample and flux measurement errors to predict the detection yields under different cadences and lengths of the program. Throughout the simulation there are simplified assumptions -- we give justifications for these assumptions whenever possible, but some assumptions can only be tested with actual observations. These simulations serve as a guide for the SDSS-RM program, and are also of general interest for future multi-object RM programs.  

\subsection{SDSS and BOSS characteristics}

The SDSS I/II/III used a dedicated 2.5-m wide-field telescope
\citep{Gunn_etal_2006} with a drift-scan camera with 30 $2048 \times 2048$
CCDs \citep{Gunn_etal_1998} to image the sky in five broad bands
\citep[$u\,g\,r\,i\,z$;][]{Fukugita_etal_1996}. The imaging data were taken on
dark photometric nights of good seeing \citep{Hogg_etal_2001}, calibrated
photometrically \citep{Smith_etal_2002, Ivezic_etal_2004, Tucker_etal_2006, Padmanabhan_etal_2008}
and astrometrically \citep{Pier_etal_2003}, and object parameters were
measured \citep{Lupton_etal_2001}. 

The BOSS survey is one major component of the SDSS III survey
\citep{Eisenstein_etal_2011}, which obtained spectra for massive galaxy
and quasar targets selected using photometry from SDSS I/II and new imaging
data in the South Galactic Cap (SGC) in SDSS III. BOSS targets are observed with
an upgraded version of the pair of multi-object fiber spectrographs for SDSS I/II
\citep{Smee_etal_2012}, with a circular FoV of 3$^\circ$ in diameter. The BOSS spectra are processed and classified by an
automatic pipeline described in \citet{Bolton_etal_2012}, and the final
public data release of BOSS spectra is Data Release 12
\citep[DR12,][]{Ahn_etal_2015}. The wavelength coverage of BOSS spectroscopy is 3650--10,400 \AA, with a spectral resolution of $R\sim 2000$. The typical S/N per pixel in $g$ band in a 2-hr exposure is $\sim 4.5$ at $g_{\rm psf}=21.2$.

\begin{figure}
\centering
    \includegraphics[width=0.45\textwidth]{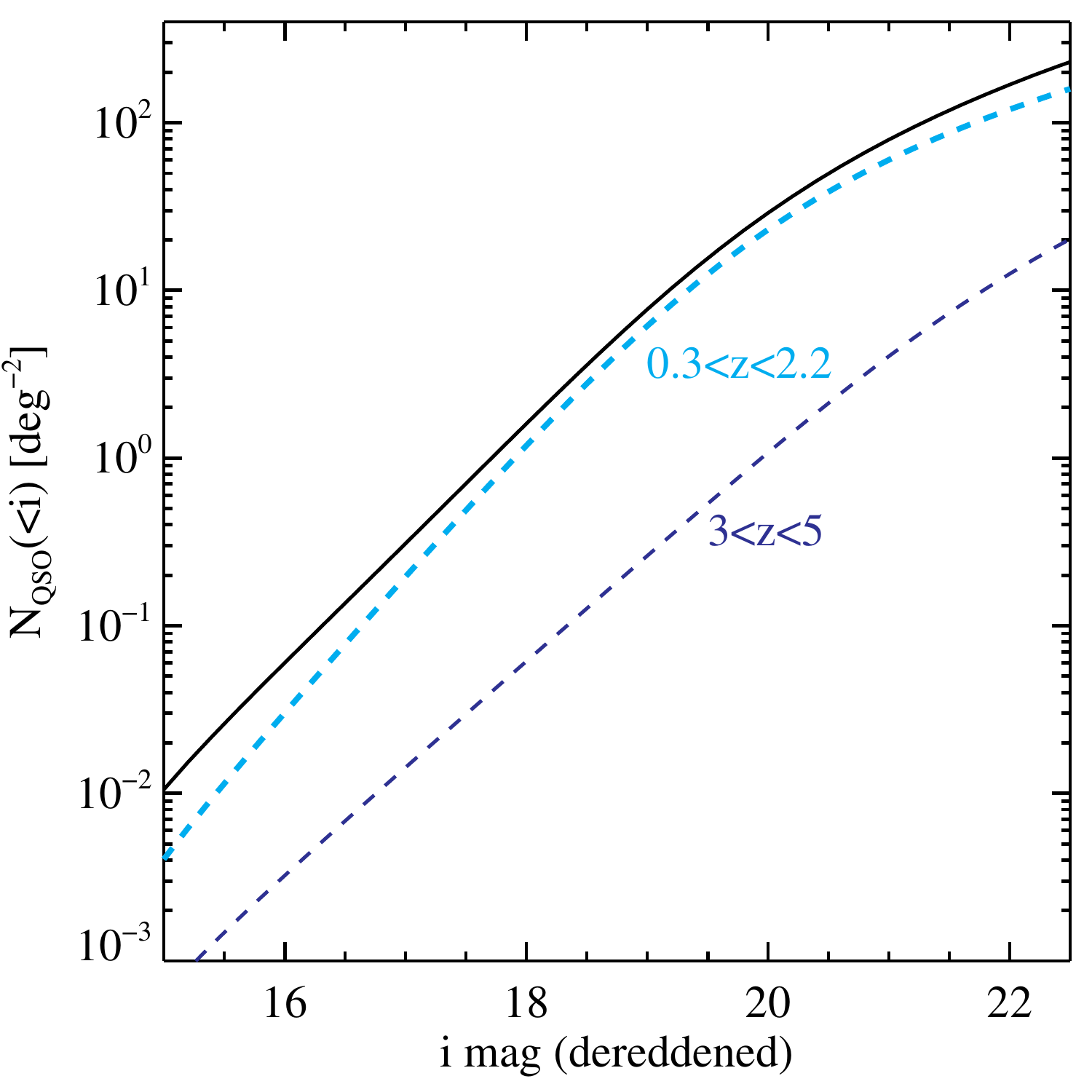}
    \caption{Cumulative quasar number density, calculated using the Hopkins et al. (2007) luminosity function, converted to $i$-band. The black line is for quasars at all redshifts. These estimates agree well with the observed number counts in Richards et al. (2006) at $i\gtrsim 17$, the range relevant to our project. There are $\sim 180$ quasars per square degree down to dereddened $i=22$.} \label{fig:qso_den}
\end{figure}

Fig. \ref{fig:qso_den} shows the expected surface density of quasars as a function of limiting $i$-band magnitude, using the quasar luminosity function (LF) compiled in \citet{Hopkins_etal_2007}. There are $\sim 180$ quasars per deg$^2$ at $i<22$. The BOSS spectrograph has 1000 fibers in total, therefore we limit our simulations to $i<22$, where the number of observable quasars significantly exceeds the number of available fibers. Quasars fainter than $i=22$ are also less favorable, as their spectra with a 2.5m telescope will be of insufficient quality for RM purposes. 

\begin{figure}
\centering
    \includegraphics[width=0.48\textwidth]{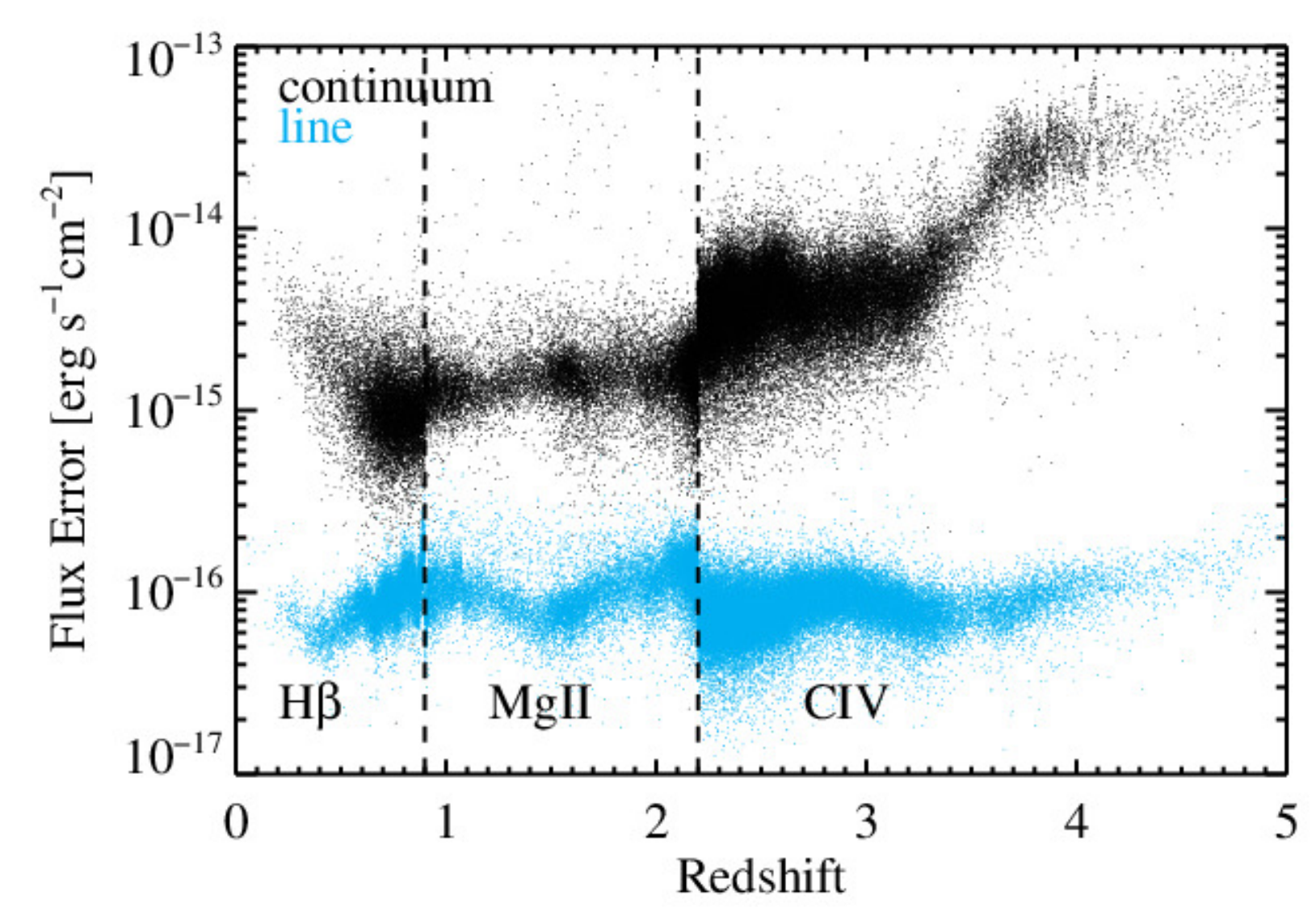}
    \includegraphics[width=0.48\textwidth]{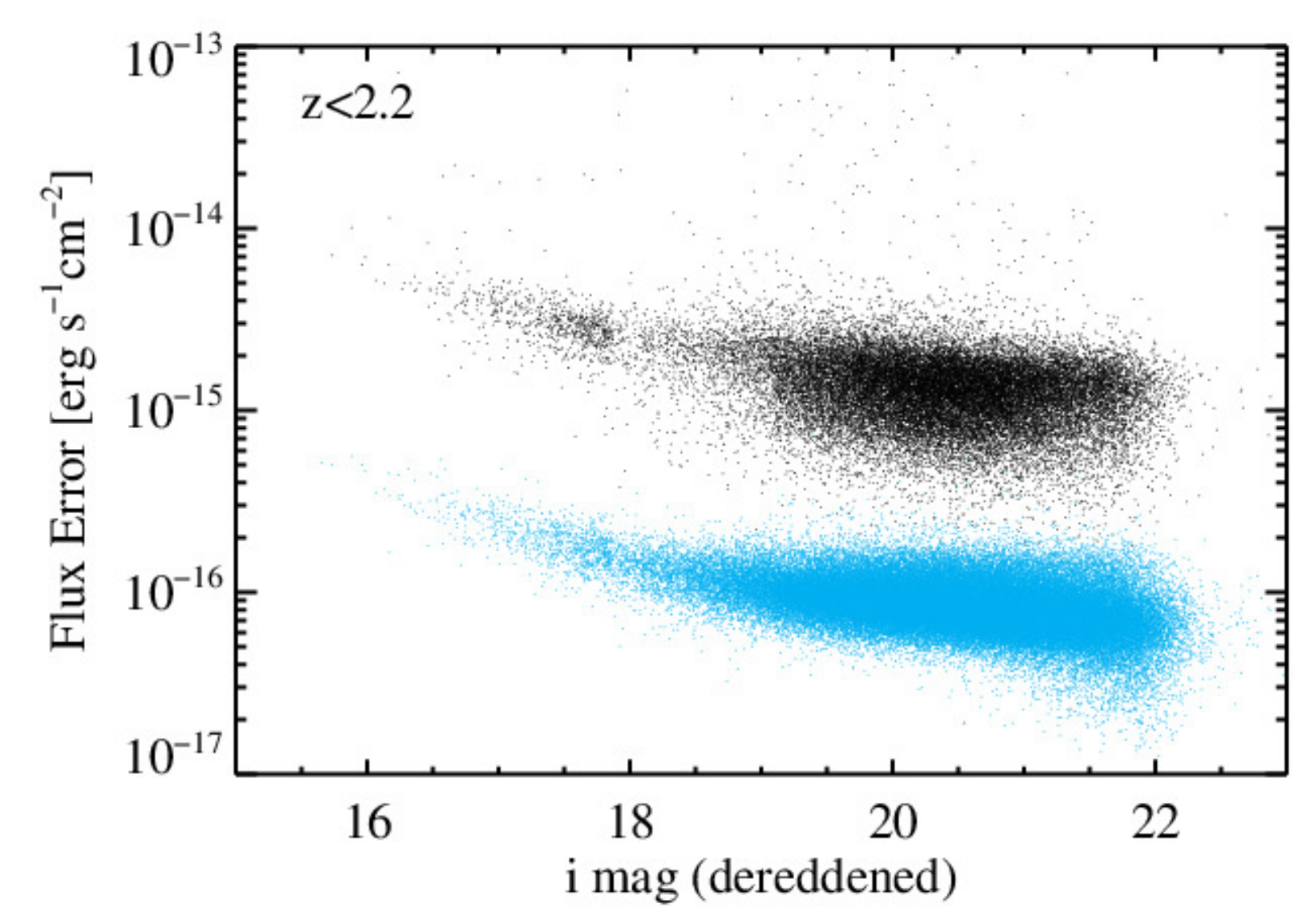}
    \caption[caption_2]{Estimates of the continuum and line flux measurement errors using the DR9 BOSS quasar catalog \citep{Paris_12}. The top panel shows the trends with redshift, where the two vertical lines show the transition from one broad line to another. The sudden jumps across the transition are due to the switch from one line (and its corresponding continuum) to another. The bottom panel shows the dependence of flux errors on magnitude for $z<2.2$ objects. The $z>2.2$ objects have systematically larger continuum flux errors (at restframe $1350$\,\AA), but are also less important in the SDSS-RM program given their long time lags. Given these results, fiducial 1$\sigma$ flux errors of
    $10^{-15}$ (continuum) and $10^{-16}\,{\rm erg\,s^{-1}cm^{-2}}$ (line) are used in the simulations unless otherwise specified. }
    \label{fig:dr9_err}
\end{figure}

\subsection{Simulated quasar properties}

We generate a uniform grid of quasars in the redshift and (unreddened) $i$-band magnitude ($i-z$) plane with $0<z<5$ and $15<i<22$, and assign absolute $i$-band magnitudes using $K$-corrections in \citet{Richards_etal_2006}. We then
assume a fixed power-law slope $\alpha_\nu=-0.5$ \citep[e.g.,][]{Vandenberk_2001} to generate monochromatic continuum luminosities $L_{5100}$, $L_{3000}$, and $L_{1350}$, the corresponding continuum luminosities ($\lambda L_{\lambda}$, in units of ${\rm erg\,s^{-1}}$) at rest-frame wavelength 5100\,\AA, 3000\,\AA\ and 1350\,\AA, for \hbeta, \MgII, and \CIV\ RM, respectively. Throughout this simulation we consider \hbeta\ RM at $z<1$ (since the BOSS spectra extend to $\sim$1 $\mu$m, \S\ref{sec:specphoto}), \MgII\ RM at $1<z<2.2$, and \CIV\ RM at $z>2.2$. This is only for demonstration purposes; in practice \MgII\ and \CIV\ RM can be performed at lower redshifts whenever these lines are covered in the spectrum. 

For each quasar, we generate line luminosities and line widths for the three lines using the restframe equivalent width (EW) and FWHM distributions (mean and scatter) found for SDSS Data Release 7 (DR7) quasars \citep{Shen_etal_2011}. These distributions are approximately log-normal. SDSS DR7 quasars (in the uniformly-selected sample) are flux-limited to $i=19.1$ and $i=20.2$ at $z\lesssim 2.9$ and $z\gtrsim 2.9$ \citep[][]{Richards_etal_2002,Schneider_etal_2010,Shen_etal_2011}. We assume that the extrapolations of these EW and FWHM distributions to fainter magnitudes ($i<22$) are reasonable. We do not model the Baldwin effect \citep{Baldwin_1977} for \MgII\ and \CIV\ (for \hbeta\ there is no obvious Baldwin effect), but this is acceptable because the scatter in the \MgII\ and \CIV\ EWs is quite substantial and dominates the dynamic range in line EWs. There is no obvious dependence of FWHM on quasar luminosity for all three lines within the dynamic range of SDSS \citep[][]{Shen_etal_2011}.

Given the continuum luminosities and line widths, we assign a ``virial BH mass'' to each quasar using the fiducial relations in Shen et al. (2011, eqns.\ 2,5,6,8). Continuum and line fluxes (in units of ${\rm erg\,s^{-1}cm^{-2}}$) are calculated using luminosities and the corresponding redshift, under the default cosmology. We verified that the distributions of properties of the simulated quasars (when
restricted to the DR7 flux limit) are similar to what was observed in DR7.

We assign the BLR size using the size-luminosity relation for \hbeta\ in \citet{Bentz_etal_2009a} with a nominal Gaussian scatter of $40\%$. We assume \MgII\ and \CIV\ have the same BLR size as \hbeta, which might be an overestimate for \CIV\footnote{For the low-redshift RM AGN sample, \CIV\ often shows lags shorter than \hbeta\ by about a factor of two \citep[e.g.,][]{Peterson_Wandel_1999,Peterson_Wandel_2000}. However, the situation is unclear for high-luminosity and high-redshift quasars where the \CIV\ likely has more complicated origins \citep[e.g.,][]{Richards_etal_2011}, as hinted by some recent RM results for \CIV\ at $z>1$ and for the most luminous quasars by Kaspi et al.\ (2007, 2014). We conservatively assume the same lags for \CIV\ and for \hbeta\ for our intermediate redshift/luminosity quasars, and note that the science yield will be enhanced if the actual \CIV\ lags are shorter than our assumed values. } but
is a reasonable assumption for \MgII, since \hbeta\ and \MgII\ share many similarities such as ionization potential and line width \citep[e.g.,][]{Shen_etal_2008,Shen_Liu_2012}. By default we will use the observed lag $\tau$ instead of the rest-frame lag $\tau_{\rm rest}$ (reduced by a factor of $1+z$) unless explicitly specified. 

\subsection{Quasar light curves}

We adopt the ``damped random walk'' (DRW, also known as an Ornstein-Uhlenbeck process) model to generate quasar continuum light curves \citep[][]{Kelly_etal_2009,Kelly_etal_2014}, which has been demonstrated to provide a reasonably good description of stochastic quasar continuum variability on the timescales of interest here \citep[e.g.,][]{Kozlowski_etal_2010,MacLeod_etal_2010,Zu_etal_2013}. This model is described by two parameters, $\hat{\sigma}$ and $\tau_{\rm DRW}$, which characterize the long-term variability amplitude and the characteristic damping timescale, respectively. We assign DRW parameters for each simulated quasar using the relations found in \citet[][eqn.\ 7]{MacLeod_etal_2010}, which were determined based on quasars from a repeatedly imaged region in SDSS called Stripe 82, and depend on the rest wavelength of the band, the luminosity, and the virial BH mass of the quasar.

Once we have the DRW parameters for each quasar, we generate quasar continuum light curves LC$_{\rm cont}(t)$ in the observed frame, densely sampled on a daily basis. The emission-line light curve LC$_{\rm line}(t)$ is related to the continuum light curve through a transfer function \citep[e.g.,][]{Blandford_McKee_1982} $\Psi(\tau)$:

\begin{equation}
{\rm LC_{line}}(t)=\int_{-\infty}^{\infty}\Psi(\tau){\rm LC_{cont}}(t-\tau)d\tau\ ,
\end{equation}
where we only consider the integrated line flux in the simulation (i.e., the transfer function has been integrated over velocity of the line). 

To generate the emission-line light curves, we assume a Gaussian transfer function, with a width of one tenth of the actual time lag, $\sigma_{\rm tf}=0.1\tau$. The choice of the transfer function is motivated by reverberation mapping results \citep[e.g.,][]{Grier_etal_2013}, and we have tested that changing the width of the Gaussian transfer function within a factor of 2-3 does not significantly affect the detection efficiency.

\begin{figure*}
\centering
    \includegraphics[width=\textwidth]{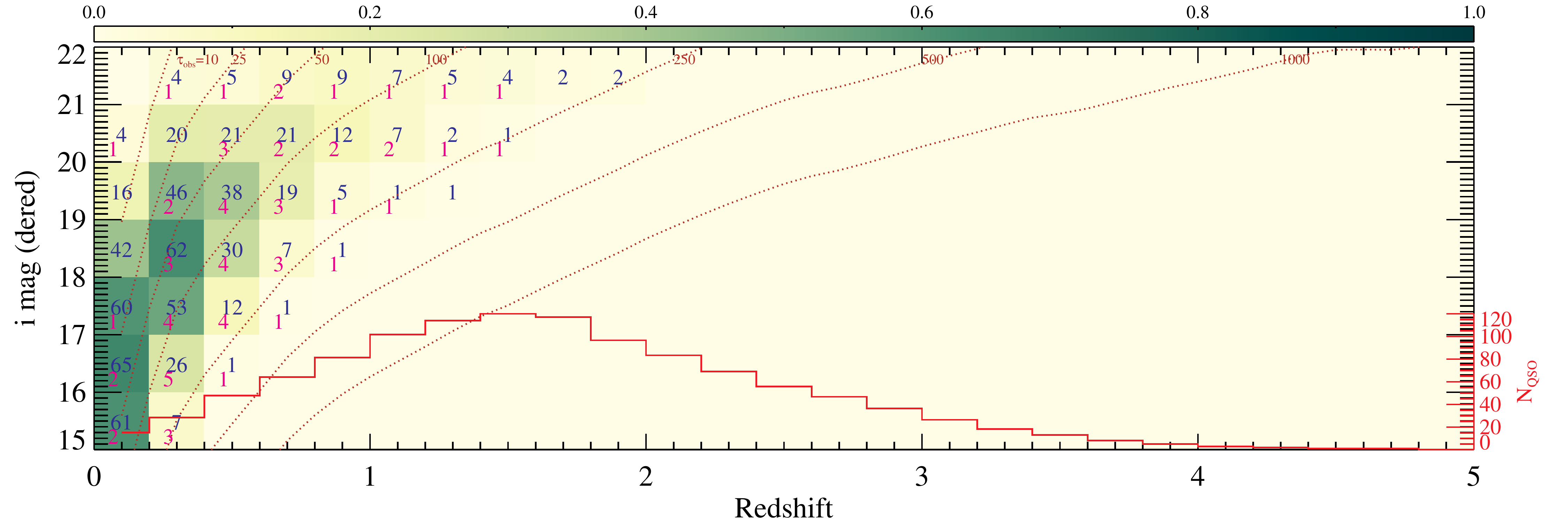}
    \caption[caption_2]{An example of the detection efficiency (fraction of objects with a robustly detected time lag) for the simulated grid of quasars described in 
    \S\ref{sec:design}, using 15 epochs obtained with a cadence of 12 days. The colormap denotes the detection efficiency in each bin. The
    blue numbers are the percentage of real detections in each bin, and the magenta numbers are the percentage
    of false detections (which can be removed by manual inspection and/or with more robust lag measuring methods). 
    The red histogram shows the redshift distribution of real quasars on a single SDSS plate (see \S\ref{sec:opt}). The dotted lines are the (approximate) loci of constant lags in the observed frame (marked by numbers in units of days), using the mean $R-L$ relation from \citet{Bentz_etal_2009a}. Due to the scatter in the $R-L$ relation, there can be detections in bins beyond the nominal $\tau_{\rm obs}=180$ days contour, because the actual lags there are shorter than the program length. }
    \label{fig:de_examp}
\end{figure*}

\begin{figure*} 
\centering
\includegraphics[width=0.95\textwidth]{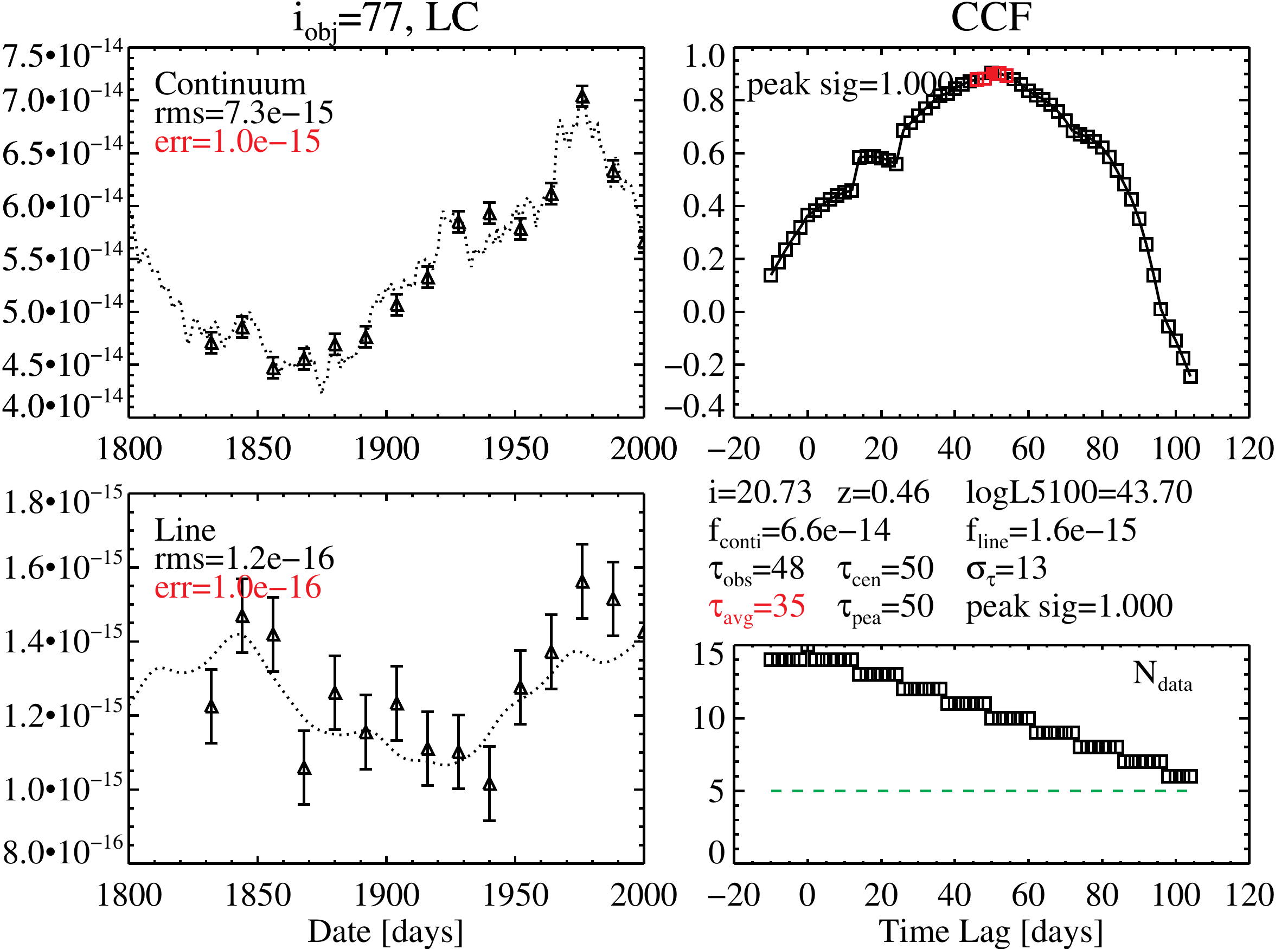} 
  \caption{An example of a real detection for a
quasar (with index $i_{\rm obj}$) in the simulation grid, using 15 epochs obtained with a cadence of 12 days. The two panels in the left column show
the simulated continuum and line light curves, where the points are the 15 epochs of
observations and the dotted lines are the underlying intrinsic (noiseless) LCs. The right column
displays the CCF (top) and the number of data points used in computing the CCF as a
function of time lag (bottom). The five red squares in the top-right panel are the points
used to compute the centroid of the CCF peak. Fluxes (and errors) are in units of 
${\rm erg\,s^{-1}cm^{-2}}$ and time is in units of days. Additional diagnostic information is provided in the middle of the right column; \textbf{row 1}: mean $i$-band magnitude, redshift, and continuum luminosity $L_{5100}$ (in units of ${\rm erg\,s^{-1}}$); \textbf{row 2}: continuum flux $f_{\rm conti}$ and line flux $f_{\rm line}$ (in this case \hbeta) in units of ${\rm erg\,s^{-1}cm^{-2}}$; \textbf{row 3}: actual lag $\tau_{\rm obs}$, centroid of the CCF $\tau_{\rm cen}$ and its uncertainty $\sigma_{\tau}$, all in units of days; \textbf{row 4}: expected lag from the mean $R-L$ relation in \citet{Bentz_etal_2009a} $\tau_{\rm avg}$, peak of the CCF $\tau_{\rm pea}$, and the statistical significance of the CCF peak. This object has a true lag of 48 days; the measured lag is $50\pm 13$ days. The statistical significance of the CCF peak is nearly 1, and the ratios
of rms flux variability amplitude during the 15 epochs to the flux error are 7.3 and 1.2
for the continuum and line fluxes, respectively. }
  \label{fig:ccf_sim}
\end{figure*}

\subsection{Host starlight}

We assume a constant host starlight luminosity of $8\times 10^{43}\,{\rm erg\,s^{-1}}$ at 5100\,\AA\ contained within the SDSS fiber. This choice is consistent with the average host contamination within the 3\arcsec diameter fiber at 5100\,\AA\ estimated for
DR7 quasars in Shen et al. (2011), but it is nevertheless a simplification as one may expect more luminous quasars to have more luminous hosts in general \citep[e.g.,][]{McLeod_Rieke_1995, Bentz_etal_2009c}. In principle, the host starlight has no
effect on the cross-correlation since it is a constant; in practice, due to
aperture effects and seeing/pointing variations, the host fraction within the fiber may vary in different epochs. This detail has negligible effect on the line flux measurement since the underlying continuum is subtracted. However, if we only have spectroscopy to measure the continuum light curve at 5100\,\AA, dedicated modeling is required to remove the variable host contamination from epoch to epoch. One possibility is to use spectral decomposition (e.g., Vanden Berk et~al.\ 2006) to separate the host and AGN continuum at 5100\,\AA, or to use a shorter wavelength to measure the AGN continuum (where the host contamination is less important). A better solution is to use photometric light curves, which by-passes this problem, and has other general advantages as we will discuss later. We assume no host starlight contamination for $L_{3000}$ and $L_{1350}$, which is a reasonable assumption for our objects.

Correcting for the host contamination at faint quasar luminosities is important in deriving an unbiased $R-L$ relation for \hbeta. One can use a combination of ground-based photometry and spectral decomposition (e.g., Vanden Berk et al.\ 2006) to estimate the host contamination in deriving the AGN $L_{5100}$. In addition, high-resolution imaging (such as that from the Hubble Space Telescope) for the objects with detected lags (hence RM BH masses) will be highly desirable to improve host correction at optical wavelengths, and to provide robust host measurements to study the correlation between RM BH mass and host properties \citep[e.g.,][]{Bentz_etal_2009a,Bentz_etal_2009c}.

\subsection{Time lag measurements}\label{sec:det_lag}

Up to this point we have not added flux errors to the simulated continuum and line light curves. Statistical errors (i.e., photon noise) are determined by the depth of the observation. To obtain a sense of the level of flux errors we expect to achieve with the BOSS spectrograph and nominal exposure times, we use the BOSS DR9 quasar sample
\citep{Paris_12}. This approach implicitly assumes that both continuum and line fluxes are measured from spectroscopy only. In practice, the continuum flux can be measured from ancillary photometric observations if available (and we will do so with the SDSS-RM sample). 

We estimate the statistical continuum and line flux uncertainties as follows: the continuum uncertainty is taken from the continuum fits performed for DR9 quasars using the methodology outlined in Shen et al. (2011) and Shen \& Liu (2012); for the line flux uncertainty, we simply use the spectral error arrays (as reported by the SDSS pipeline) in each line region \citep[e.g.,][]{Vandenberk_2001} and compute the propagated integrated flux error. 

These flux error estimates are shown in Fig.\ \ref{fig:dr9_err} for all DR9 quasars, with the same redshift divisions as above for the three lines. It appears that $10^{-15}$ and $10^{-16}\ {\rm erg\,s^{-1}cm^{-2}}$ are reasonable estimates
for the continuum and line flux errors, which were adopted as the fiducial values in the
simulation. The continuum flux error is about a factor of ten larger than the line flux error, but the continuum flux is also about a factor of 100 larger than the integrated line flux given typical EWs of the line. The S/N for the DR9 quasars is typically $>10$ and $>5$ for the continuum and line
flux measurements, respectively, suggesting that most objects near the flux limit will have insufficient spectral quality to detect intrinsic variability. The continuum errors may be larger
for \CIV\ at $z>2.2$, but a fiducial 6-month RM experiment will be insensitive to these
objects due to the limited time span of the program (see \S\ref{sec:implem}); we can improve continuum flux measurements by using simultaneous photometry. In addition, the flux 
errors are larger for brighter objects, but the bulk of our detections will be at fainter magnitudes given the many more objects there.
Thus for simplicity we adopted constant flux errors in the simulations. One could
imagine magnitude and redshift dependent flux errors, but there are not enough DR9 quasars to cover the entire simulation grid to create a detailed error map. Such sophistication is also unnecessary, as in the simulations we will vary the fiducial flux errors to test their effects on the overall lag detection efficiency. 

One caveat is that we do not account for correlated errors or systematic errors in continuum and line flux measurements from the same spectrum. In practice there should be some correlated errors for continuum and line flux measurements due to the identical scaling factor applied to the spectrum during flux calibration, or due to the spectral decomposition of continuum and emission lines; we have tested correlated continuum and line flux errors and found no significant degradation of the science return (\S\ref{sec:opt}). The systematic errors due to BOSS pipeline spectrophotometry are $\sim 6\%$ rms in broad-band $gri$ \citep{Dawson_etal_2013}, while the statistical errors for our flux measurements are typically $10\%$. Thus reduction errors will not significantly decrease our detection probability, and we have confirmed this conclusion by running parallel simulations including $6\%$ systematic flux errors. In addition, we will improve the spectrophotometry by performing a custom flux calibration (see \S\ref{sec:specphoto}). We have also acquired some real-time photometry (\S\ref{sec:imaging}) to improve spectrophotometry of the spectroscopic observations. For low-redshift objects with high S/N narrow line measurements, we will use the narrow line fluxes to perform additional spectral normalization \citep[e.g.,][also see \S\ref{sec:prepspec}]{vw_92}. The combination of these additional efforts can in principle improve the flux calibration accuracy to $< 5\%$, and perhaps even $<2\%$ for bright low-$z$ objects with strong narrow emission lines. 


\subsection{Cross-correlation of continuum and line light curves}

Adopting fiducial continuum and line flux errors of $10^{-15}$ and $10^{-16}\
{\rm erg\,s^{-1}cm^{-2}}$ respectively (Fig.\ \ref{fig:dr9_err}), we generate light curves for the simulated quasar grid, and downsample
the light curves with different cadences and total number of epochs ($N_{\rm ep}$). By default 
the continuum and line fluxes are measured simultaneously from spectroscopy at each epoch (i.e., we are
assuming a spectroscopic-only RM program). However, we also simulate a case where we have
a more densely sampled continuum light curve (e.g., from independent photometric
observations that measure the continuum), and evaluate the gains in the lag detection efficiency. 

For each pair of simulated light curves, we used the standard
interpolated-light-curve cross-correlation technique \citep[e.g.,][]{Gaskell_Peterson_1987} to compute the
correlation coefficient $r$ as a function of time lag (CCF$(\tau)$):
\begin{equation}\label{eqn:r}
r({\rm CCF})=\frac{\sum_{i=1}^N(x_i - \bar{x})(y_i - \bar{y})}{\left(\sqrt{\sum_{i=1}^N(x_i-\bar{x})^2}\right) \left(\sqrt{\sum_{i=1}^N(y_i-\bar{y})^2}\right) }\ ,
\end{equation}
where time series $x_i$ and $y_i$ are the line flux and the (shifted and interpolated) continuum flux light curves with $N$ data points each, and $\bar{x}$ and $\bar{y}$ are the mean of the data series. For programs longer than a year there will be annual gaps of about half a year due to visibility. We do not
interpolate the light curves within these gaps. 

The peak of the CCF is
located with a statistical significance depending on $r$ and the number of pairs of
data points\footnote{The number of pairs of data
points is that used for each time lag, not that for the entire time series.} used in computing $r$ (Bevington 1969):
\begin{equation}\label{eqn:peak_sig}
P_c(r;N)=2\int_{|r|}^1 p_x(x;\nu)dx\ ,
\end{equation}
where $\nu=N-2$ is the number of degrees of freedom of the data set, and 
\begin{equation}
p_x(x;\nu)=\frac{1}{\sqrt{\pi}}\frac{\Gamma[(\nu+1)/2]}{\Gamma(\nu/2)}(1-x^2)^{(\nu-2)/2}\ .
\end{equation}
$P_c(r;N)$ is the probability that a random sample of uncorrelated data points can yield a correlation coefficient as large as or larger than the observed value of $|r|$. Therefore $1- P_c(r;N)$ describes the statistical significance of the observed peak correlation, with larger values indicating a more likely correlation.  
We then compute a centroid of the CCF using the nearest five points around the peak of the CCF.
Only time lags shorter than the length of the light curves are considered.

The uncertainty of the lag measured from the CCF peak or centroid is estimated using the
``flux-redistribution and random subset sampling'' method applied in traditional RM
work \citep[e.g.,][]{Peterson_etal_2004}. Briefly, the original light curves are bootstrapped for 100 trial
samples and the fluxes are shuffled according to the nominal flux errors. For
each trial sample the CCF is re-computed and the peak/centroid is measured. The formal
$1\sigma$ uncertainty of the measured lag is the semi-quantile of the
68\% range of the peak/centroid distribution from the 100 trials.

In what follows, we adopt the centroid value as the measured time lag. 

\begin{figure} [!ht]
\centering
\includegraphics[width=0.48\textwidth]{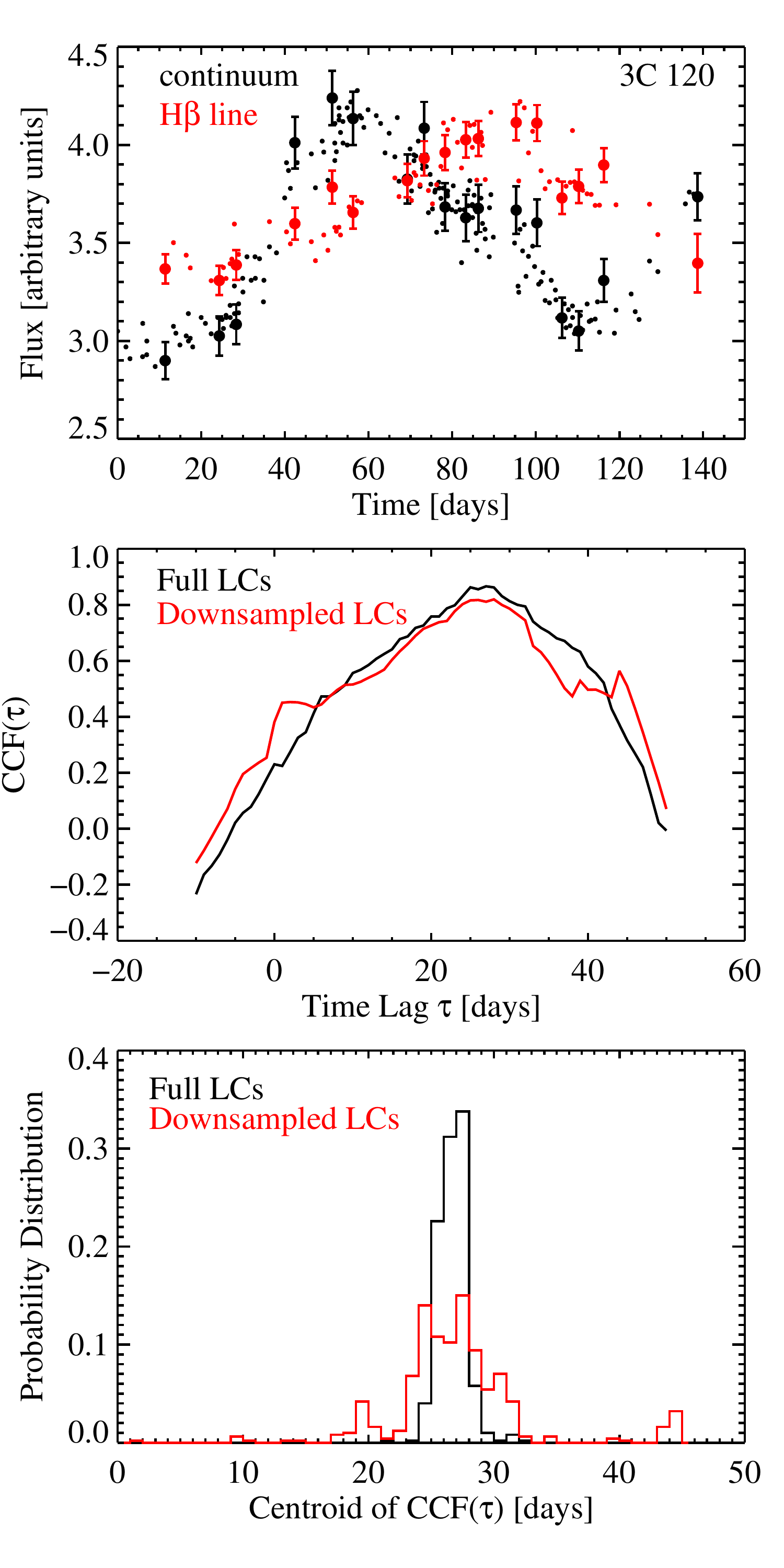}
  \caption{Testing lag detections using down-sampled light curves of 3C 120 from \citet{Grier_etal_2012}, where a lag of $\sim 30$ days was measured. {\em Top}: the down-sampled (filled circles with error bars) LCs (17 epochs), where the down-sampled continuum LC is interpolated
  onto the epochs of the down-sampled line LC. The small dots are the original LCs ($\sim 85$ epochs), and the measurement errors are suppressed for clarity. {\em Middle:} the CCFs
  computed using the original LCs (black) and the down-sampled LCs (red). In both cases a peak is unambiguously identified at similar locations. {\em Bottom:} distributions of the CCF centroid for the original LCs (black) and down-sampled LCs (red), using the flux redistribution and random subset sampling method described in the text. This example shows that with only $\sim 15$ epochs, it is still possible to make a detection, although the quality of the lag measurement is inevitably degraded with fewer epochs, as seen from the wider distribution of the CCF centroid from random trials. }
  \label{fig:3c120}
\end{figure}

\begin{figure}
\centering
    \includegraphics[width=0.45\textwidth]{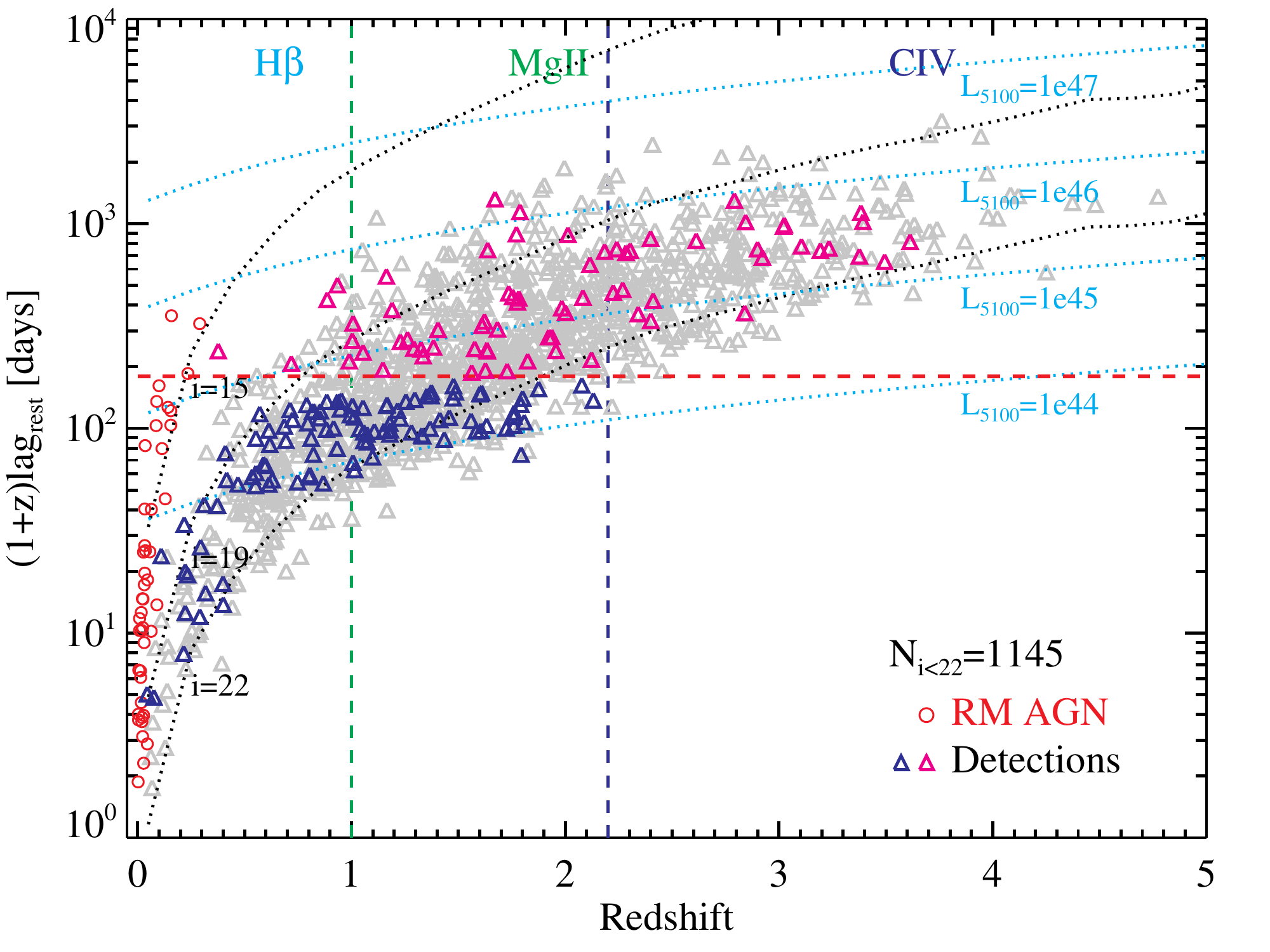}
    \caption[]{Forecast of lag detections for the SDSS-RM program. The gray points are simulated quasars in the FoV of a single SDSS plate ($\sim 7\,{\rm deg^2}$). The vertical axis is
    the expected lag in the observed frame, determined by the BLR size from the $R-L$ relation in Bentz et al. (2009) with $40\%$ scatter. The black dotted lines show the approximate correspondences to different (dereddened) $i$-band magnitudes, while the blue dotted lines show those for constant 5100\,\AA\ continuum luminosities. The red horizontal line indicates a length of 180 days in the 6-month SDSS-RM program. The current RM AGN sample \citep[from the compilation in][]{Feng_etal_2014} is shown using red circles. We consider \hbeta\ RM for $z<1$, \MgII\ RM for $1<z<2.2$, and \CIV\ RM for $z>2.2$. Given an observing length of $6$ months, the program will be mostly sensitive to lags on the order of tens of days. The blue triangles show the expected detections from the 6-month SDSS-RM program, and the magenta triangles show additional detections when the SDSS-RM program is combined with the 3-yr PS1 early photometry (see \S\ref{sec:exten}). } \label{fig:qso_plate}
\end{figure}

\subsection{Lag detection}\label{sec:detection}

To claim a ``detection'' we require the following criteria:
\begin{enumerate}
\item[1.] The statistical significance of the CCF peak must be $>0.95$,

\item[2.] The lag must be $\ge 3$ times the 1$\sigma$ error in the lag measurement.
\end{enumerate}

To decide if this is a ``real detection'', i.e., one in which it correctly recovers the actual lag, we further require the following conditions be met:
\begin{itemize}
\item[$\bullet$] The absolute difference between the measured and true lags is
    less than 3 (rest) days, or
\item[$\bullet$] The relative difference between measured and true lags is less
    than $25\%$, or
\item[$\bullet$] The absolute difference between measured and true lags is within
    2$\sigma$ of the measurement uncertainty.\footnote{The last requirement introduces a
statistical bias towards over-estimated lags, because small measured lags are
less likely to meet the 3$\sigma$ detection requirement given the same
measurement error. This effect must be considered in real data. }
\end{itemize}
Any detection that does not satisfy the above conditions is a ``false
detection''. Although in some of the cases shown below we will present the false-detection 
rate, it is not a serious concern in real data since such false detections can be reduced
with manual inspection of the LCs and CCF and/or more robust lag measurement methods. For example, for these false-detections we often see two peaks in the CCF: the false peak is at a very different location (often near the end of the time baseline) from the real lag (i.e., the false peak is an outlier from the observed $R-L$ relation), and the other, slightly weaker peak is very close to the real lag and is usually more well-defined and with more correlating data points than the false peak. In these cases we can identify the correct CCF peak by limiting the search range for the time lags (in particular for \hbeta, for which we have a well-measured $R-L$ relation to set a strong prior). In addition, alternative lag measurements to the simple CCF method, such as the \citet{Zu_etal_2011} method based on DRW models \citep[e.g.,][]{Kelly_etal_2009}, can in principle provide a more robust confidence estimation of the detected CCF peaks, and therefore reduce false detections. Finally, better S/N, longer program length, and more epochs will not only lead to more detections, but also reduce the number of false positives. In most of the cases we studied (including the SDSS-RM program), the raw false-detection rate (i.e., using the simple CCF method) is well controlled to be $\lesssim 20\%$.

\begin{figure*}
\centering
    \includegraphics[width=\textwidth]{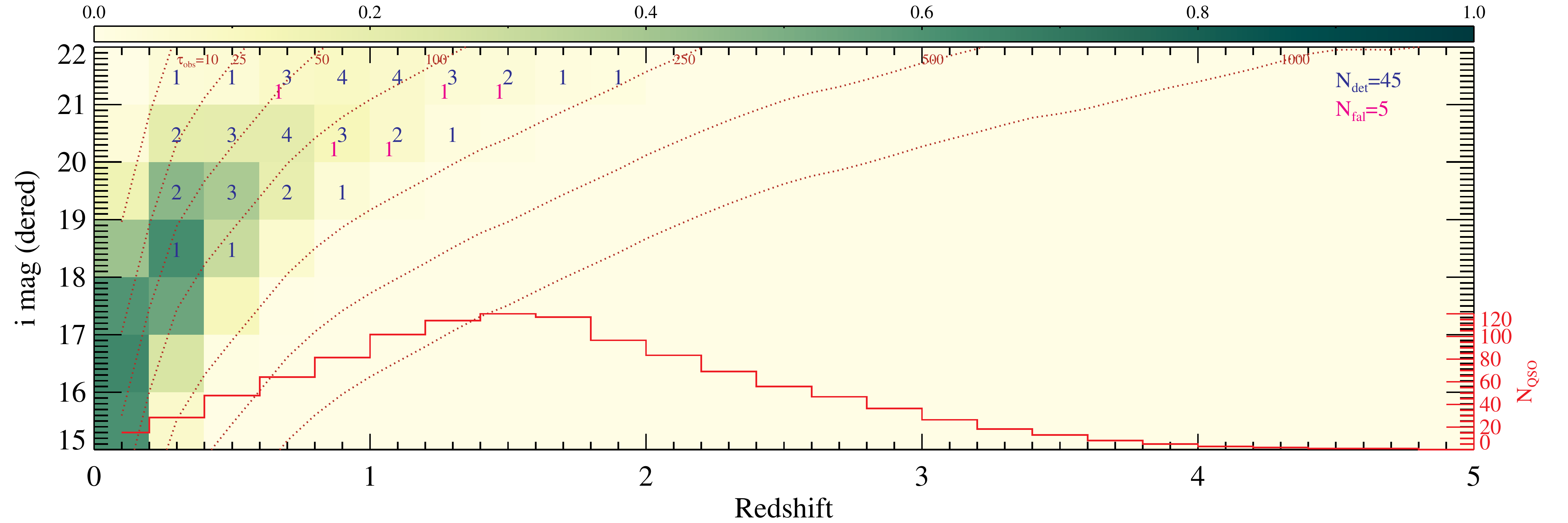}
    \caption[caption_2]{One realization of the expected number of detections for a simulated SDSS plate, 
    with a cadence of 12 days and 15 epochs, i.e., the same observing program as in Fig.\ \ref{fig:de_examp}. The colormap denotes the detection efficiency (fraction of detectable objects) in each bin. The
    cyan numbers are the number of real detections in each bin, and the magenta numbers are the number
    of false detections (which can be removed by manual inspection and more careful lag measurements). 
    The red histogram shows the redshift distribution of real quasars on a single SDSS plate. Due to the scatter in the $R-L$ relation, there can be detections in bins beyond the nominal $\tau_{\rm obs}=180$ days contour, because the actual lags there are shorter than the program length. This example represents a 6-month spectroscopic program with minimum resources and meaningful science return. }
    \label{fig:de_examp_nqso}
\end{figure*}

\begin{figure*}
\centering
    \includegraphics[width=\textwidth]{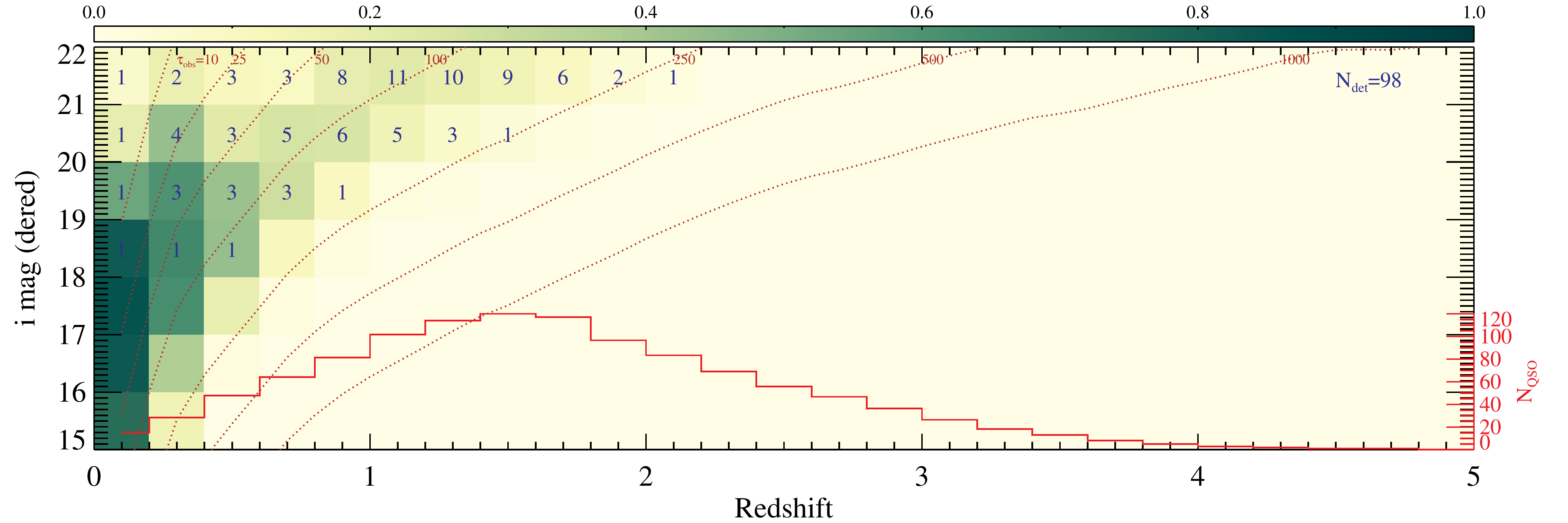}
    \caption{Another realization of the expected number of detections for a simulated SDSS plate, 
    with a cadence of $\sim 5$ days and a total of 30 epochs over 6 months. Notation is the same as Fig.\ \ref{fig:de_examp_nqso}. Due to the scatter in the $R-L$ relation, there can be detections in bins beyond the nominal $\tau_{\rm obs}=180$ days contour, because the actual lags there are shorter than the program length. The parameters considered here are representative of the executed SDSS-RM program.}
    \label{fig:de_examp_nqso_more}
\end{figure*}

Fig.\ \ref{fig:de_examp} shows an example of the detection efficiency for the simulated grid of
quasars, using 15 epochs taken with a cadence of 12 days. The detection efficiency drops rapidly
when the actual time lag is close to the time span of the monitoring program, therefore objects
with higher luminosity and/or higher redshifts are more difficult to detect. In traditional 
RM campaigns, the time span must be at least several times that of the lag to achieve a high ($\gtrsim 75\%$)
success rate. For our program, the success rate is considerably lower. Given a 6-month
time baseline (i.e., the typical visibility window within one year), such a program is most sensitive to observed lags on the order of tens of days; on the 
other hand, lags shorter than the cadence are also difficult to detect due to the lack of time resolution, but we will still be able to obtain a useful upper limit for these objects.
These points qualitatively explain the observed decrease in detection efficiency towards short and
long time lags in Fig. \ \ref{fig:de_examp}.

\begin{figure*} 
\centering
\includegraphics[width=0.45\textwidth]{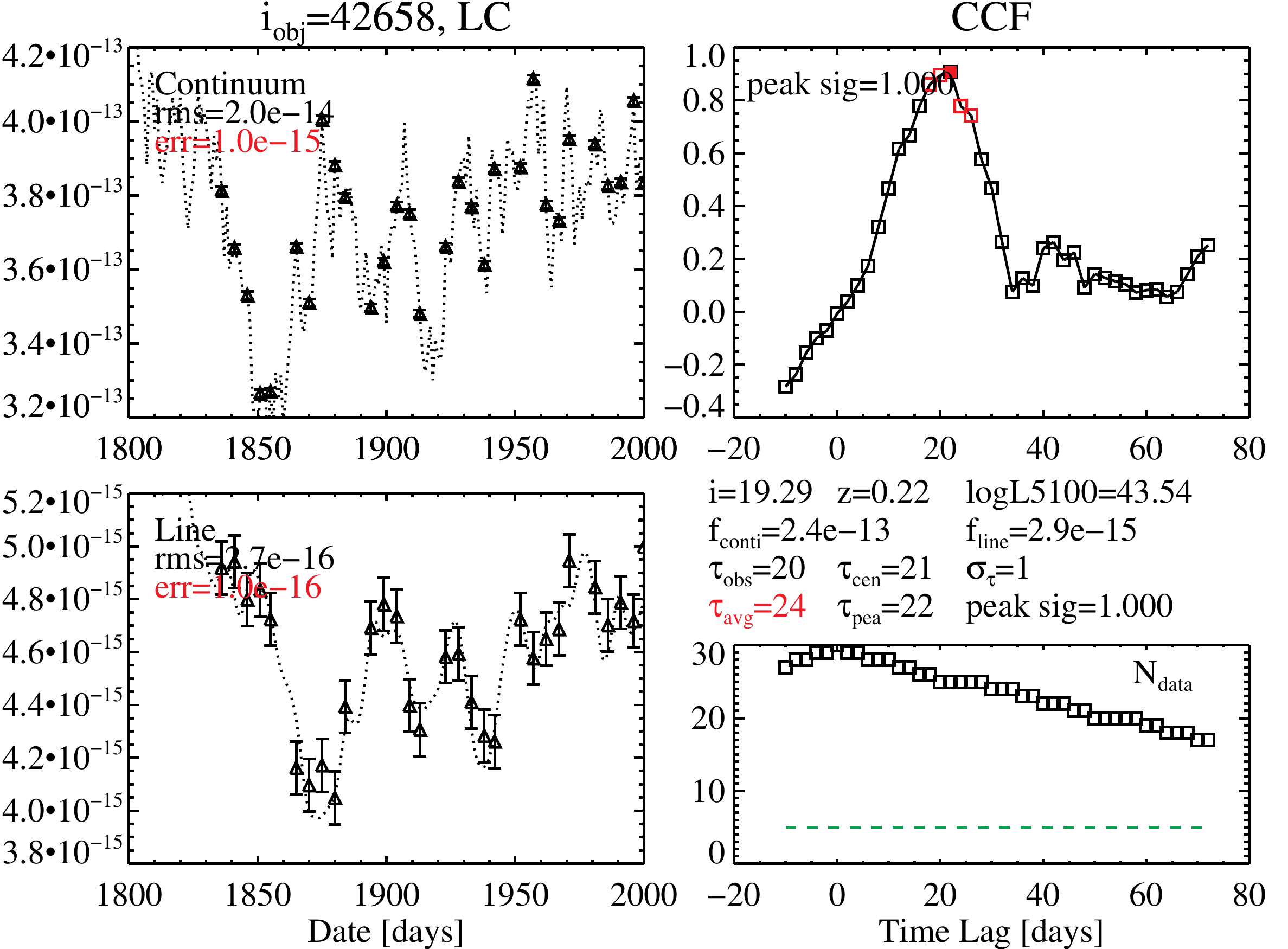} \vspace{3mm}
\includegraphics[width=0.45\textwidth]{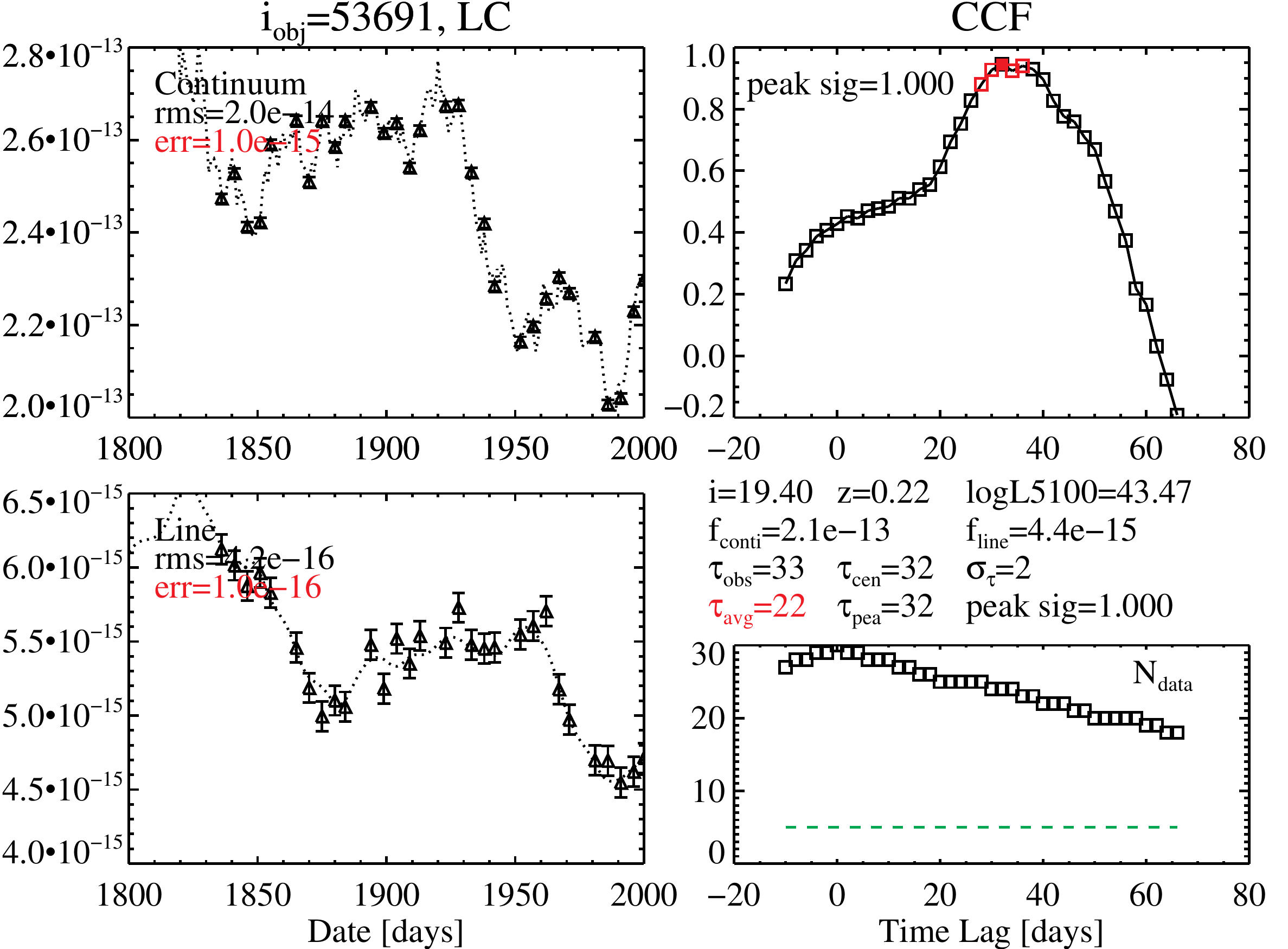} \vspace{3mm}
\includegraphics[width=0.45\textwidth]{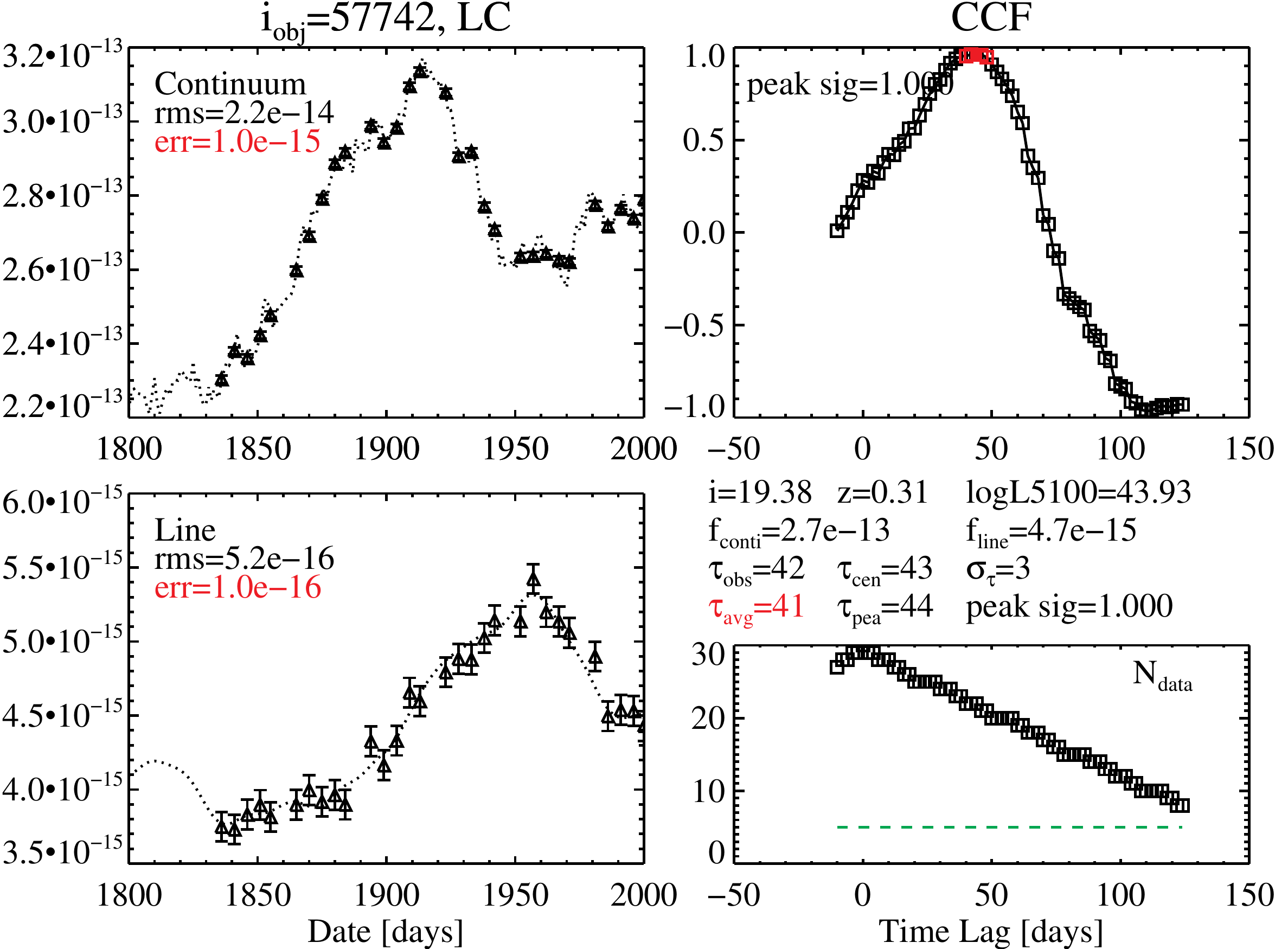} 
\includegraphics[width=0.45\textwidth]{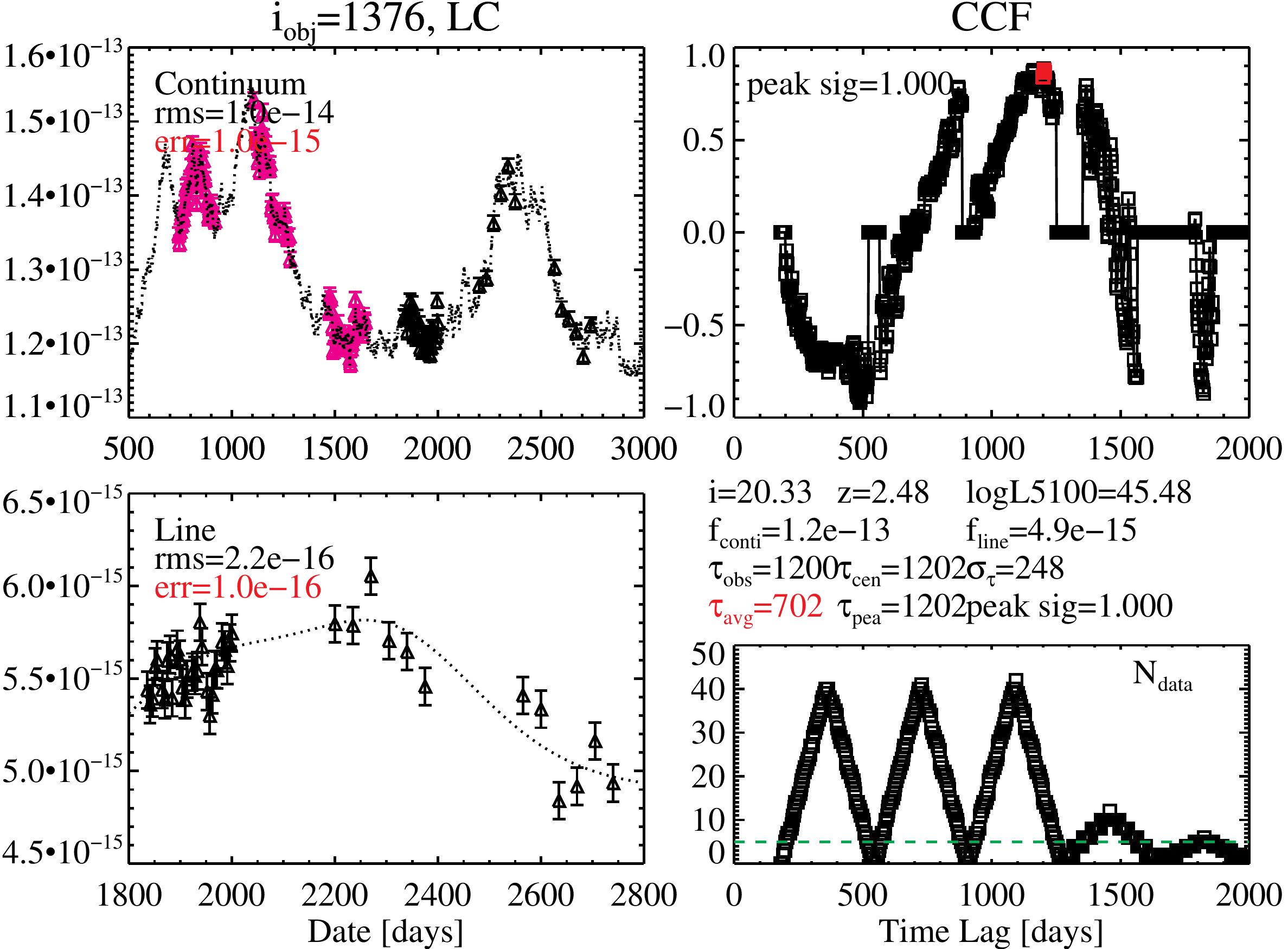} 
  \caption{Four examples of robustly detected lags with different LCs and in a different observing program from that in Fig.\ \ref{fig:ccf_sim}. Notation is the same as in Fig.\ \ref{fig:ccf_sim}. All examples are based on the simulated SDSS-RM program: 5 spectroscopic epochs each month (skipping 10 days of bright time) for 6 months (30 epochs in total), with nominal BOSS exposures. The first three examples are for the $6$-month spectroscopic program alone; the last one (lower right) is for 30 epochs (the SDSS-RM baseline program) $+$ 12 epochs in the first two years of eBOSS and with 3-yr PanSTARRS early photometry (shown in magenta points). In the last example, the number of data points used in the CCF has a periodic behavior, because we do not interpolate the light curves within annual gaps in the visibility window. In practice, we can interpolate across the seasonal gaps using the method of \citet{Zu_etal_2011}. }
  \label{fig:ccf_sim_more}
\end{figure*}

\begin{figure*}
\centering
    \includegraphics[width=0.5\textwidth]{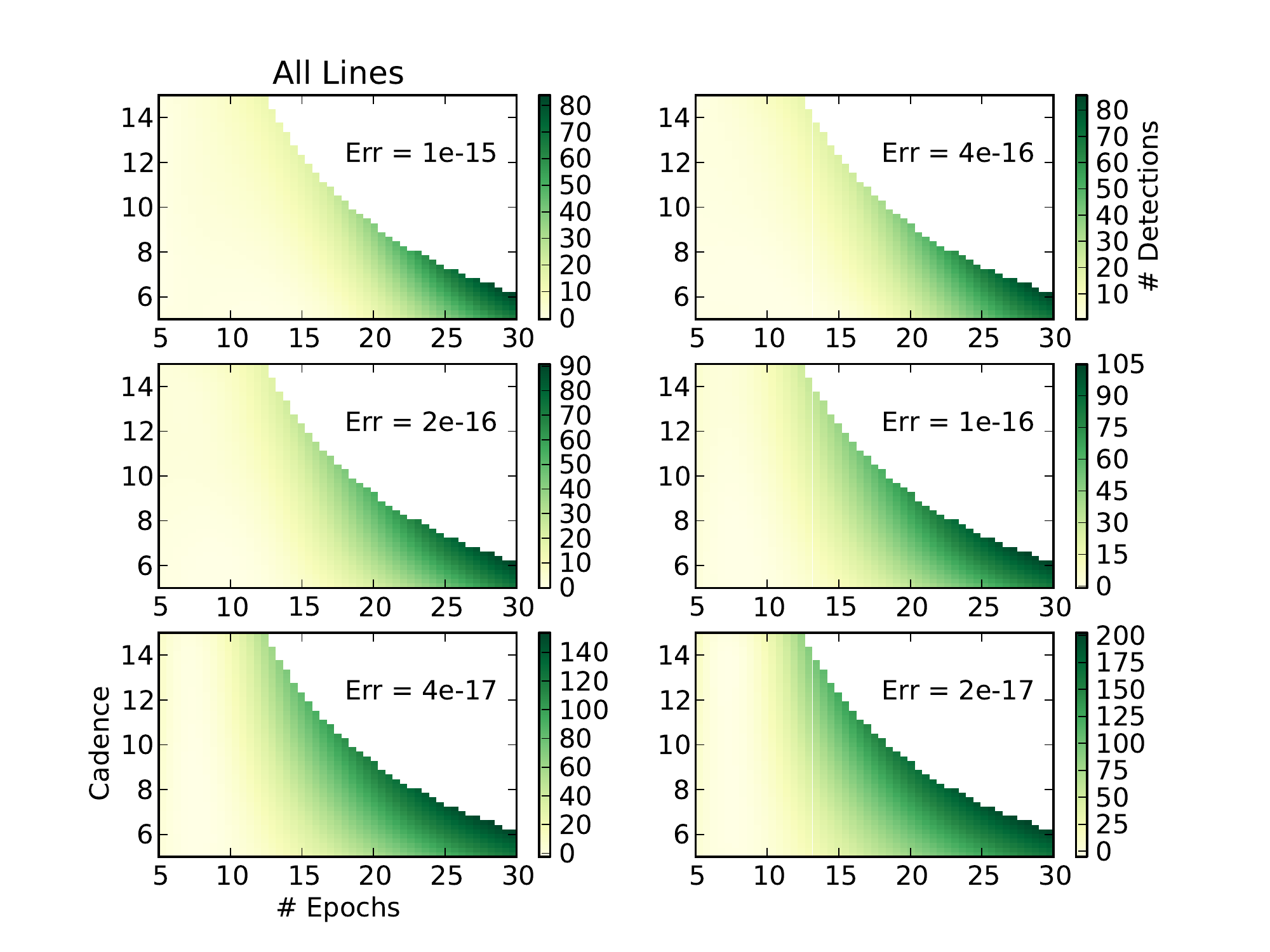}\hspace{-6mm}
    \includegraphics[width=0.5\textwidth]{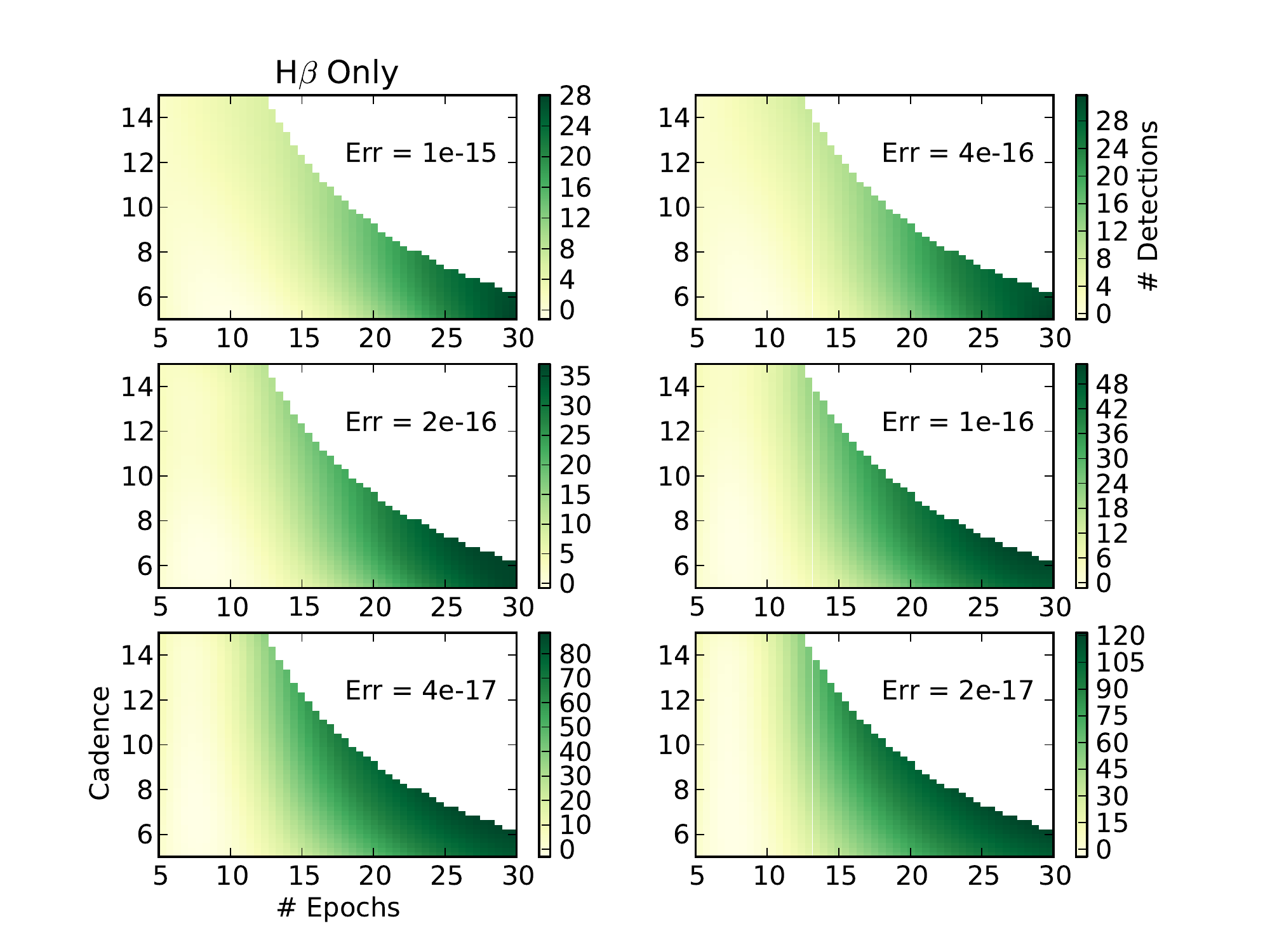}
    \caption{Total number of detections for a single plate, as functions of line flux errors (in units of ${\rm erg\,s^{-1}cm^{-2}}$), cadence $\Delta t$ (in units of days) and
    total number of epochs $N_{\rm ep}$, averaged over multiple realizations. The fiducial continuum and line flux errors are scaled by a common factor, and we simply use the line flux error as an error indicator. At fixed error, the detection efficiency increases towards both larger $N_{\rm ep}$ and larger $\Delta t$. {\em Left:} for all detections.  {\em Right:} for \hbeta\ detections only. }
    \label{fig:grid_det}
\end{figure*}

We can obtain a sense of what factors will increase the detection probability, given the actual lag $\tau$, and assuming
simultaneous continuum and line observations. The relevant factors are
(1) the errors in the flux measurements relative to the rms flux variability amplitude during the observing
period, i.e., $S_{\rm rms}/\sigma_{\rm flux}$; (2) the number of observing epochs, $N_{\rm ep}$; and (3) the time span of the 
monitoring, $t_{\rm span}$. Given a fixed cadence $\Delta t$, $t_{\rm span}=(N_{\rm ep}-1)\Delta t$. 
We can then define three figures-of-merit (FoM) for the detection:
\begin{eqnarray}
{\rm FoM1}&=&\frac{S_{\rm rms}}{\sigma_{\rm flux}},\nonumber\\ 
{\rm FoM2}&=&N_{\rm ep}\left(1-\frac{\tau}{t_{\rm span}}\right),\nonumber\\
{\rm FoM3}&=&(N_{\rm ep}-1)\frac{\tau}{t_{\rm span}}\ .
\end{eqnarray}
FoM1 describes how much of the correlation will be destroyed by random flux errors. FoM2 describes how many data points contribute to the calculation of the CCF (after
considering the lag between line and continuum LCs). FoM3 
describes how well the LC resolves the lag. Larger values of these FoMs correspond to better chance of detection. While it is 
difficult to combine these FoMs to derive a single FoM, they can at least qualitatively explain the trends we see in
Fig.\ \ref{fig:de_examp}. In the limit of $N_{\rm ep}\gg 1$, we have ${\rm FoM2}+{\rm FoM3}\approx N_{\rm ep}$,
so these two FoMs are linked. Inspecting the detections in our simulations, it appears that most of the detections are
close to having ${\rm FoM2/FoM3}\approx 2$, which indicates $t_{\rm span}\approx 3\tau$. If we substitute $t_{\rm span}=3\tau$ in FoM3 and assume $N_{\rm ep}\gg 1$, then Nyquist sampling requires ${\rm FoM3}=N_{\rm ep}/3>2$ to resolve the lag. Therefore we generally require $\gtrsim 10$ epochs to make a detection. It also appears that
the detection fraction increases faster with FoM1 than with $N_{\rm ep}$, e.g., the gain due to increasing $N_{\rm ep}$
should scale as $N_{\rm ep}^{1/2}$.

It is possible to make a detection with a FoM1 value of less than a few, albeit at low probability. Fig.\ \ref{fig:ccf_sim} shows such an example with a simulated quasar and its light curves (with the same cadence and $N_{\rm ep}=15$ as shown in Fig.\ \ref{fig:de_examp}), for which our detection pipeline successfully identified the time lag. The detection of lags with only 15 epochs is often considered by some as counterintuitive in traditional RM campaigns, which usually have several dozen epochs to secure a detection given the stochastic nature of AGN variability. To further demonstrate that a detection is possible with as few as $\sim 15$ epochs, we examine the observed
light curves for the quasar 3C 120 in \citet{Grier_etal_2012}. This object has a measured delay of $\sim 30$
days, and the light curves cover a period of a few months. We down-sample the observed light curves
to $\sim 15$ epochs, and measure the CCF. As with our simulations, we require simultaneous continuum
and line flux measurements, so we interpolated the continuum light curve at the epochs of the line
flux light curves. Fig.\ \ref{fig:3c120} shows the CCF for the original LCs and the down sampled LCs. With
17 epochs sampled on a $\sim$6-day cadence, we are still able to detect the lag, albeit with increased 
uncertainty in the lag measurement. Of course, if we are not sampling the right part of the variable
light curve (i.e., if the light curves varied monotonically throughout the monitoring), we will not be able to detect the lag. Our simulations take into account the stochasticity in 
sampling the light curves for different quasars. 

\subsection{Optimization}\label{sec:opt}

With this framework in place, we are ready to make predictions. We will consider the effectiveness of a RM program for a single SDSS III-BOSS plate. 

Fig.\ \ref{fig:qso_plate} displays all quasars down to $i=22$ for a single, simulated BOSS plate in the redshift and observed lag plane, using the LF estimates from \citet{Hopkins_etal_2007}. This figure is a useful reference for the distributions of the quasars detectable in a dedicated RM program. If the RM program is only for 6 months, we will only be sensitive to objects at $z\lesssim 2$ unless we have early photometric light curve data. However, even when the program is extended beyond one year, there will be annual gaps that affect the detection efficiency continuously across this plane. 


Fig.\ \ref{fig:de_examp_nqso} shows the expected number of detections for a 
simulated plate, for a cadence of 12 days and 15 epochs over 6 months (e.g., same as in Fig.\ \ref{fig:de_examp}),
for one realization of $\sim 1000$ simulated quasars. This example represents a {\em minimum-requirement program} with limited resources (i.e., only 15 epochs) that will provide meaningful science return (i.e., more than tens of lag detections). The expected numbers were calculated by multiplying the numbers
of expected quasars in each bin by the detection probability found for the simulated grid of quasars. In this realization we detect 45 lags and $\sim 30$ are \hbeta\ detections. 

Similarly, Fig.\ \ref{fig:de_examp_nqso_more} presents the expected lag detections for the SDSS-RM program (the detailed implementation is described in \S\ref{sec:implem}): a $\sim 5$-day cadence each month (skipping bright time) for 6 months (30 spectroscopic epochs in total). With such a program, we can detect about a hundred lags within the 6-month program length. In addition, we achieve more accurate lag measurements in the 30-epoch program than in the 15-epoch program (see Fig.\ \ref{fig:det_quality}). Several simulated examples of robustly detected cases for the SDSS-RM program are shown in Fig.\ \ref{fig:ccf_sim_more}.

\begin{figure}
\centering
    \includegraphics[width=0.48\textwidth]{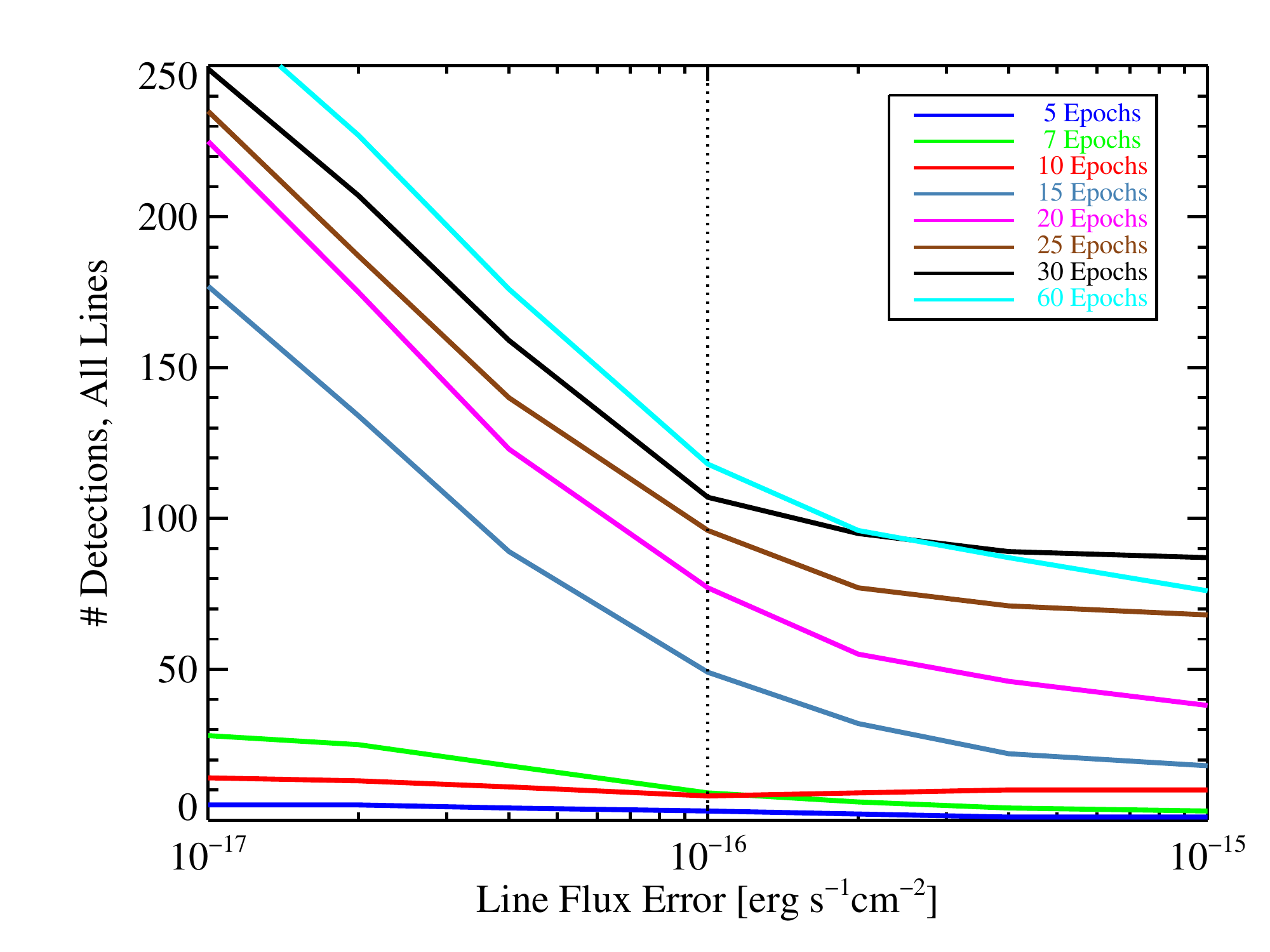}
    \includegraphics[width=0.48\textwidth]{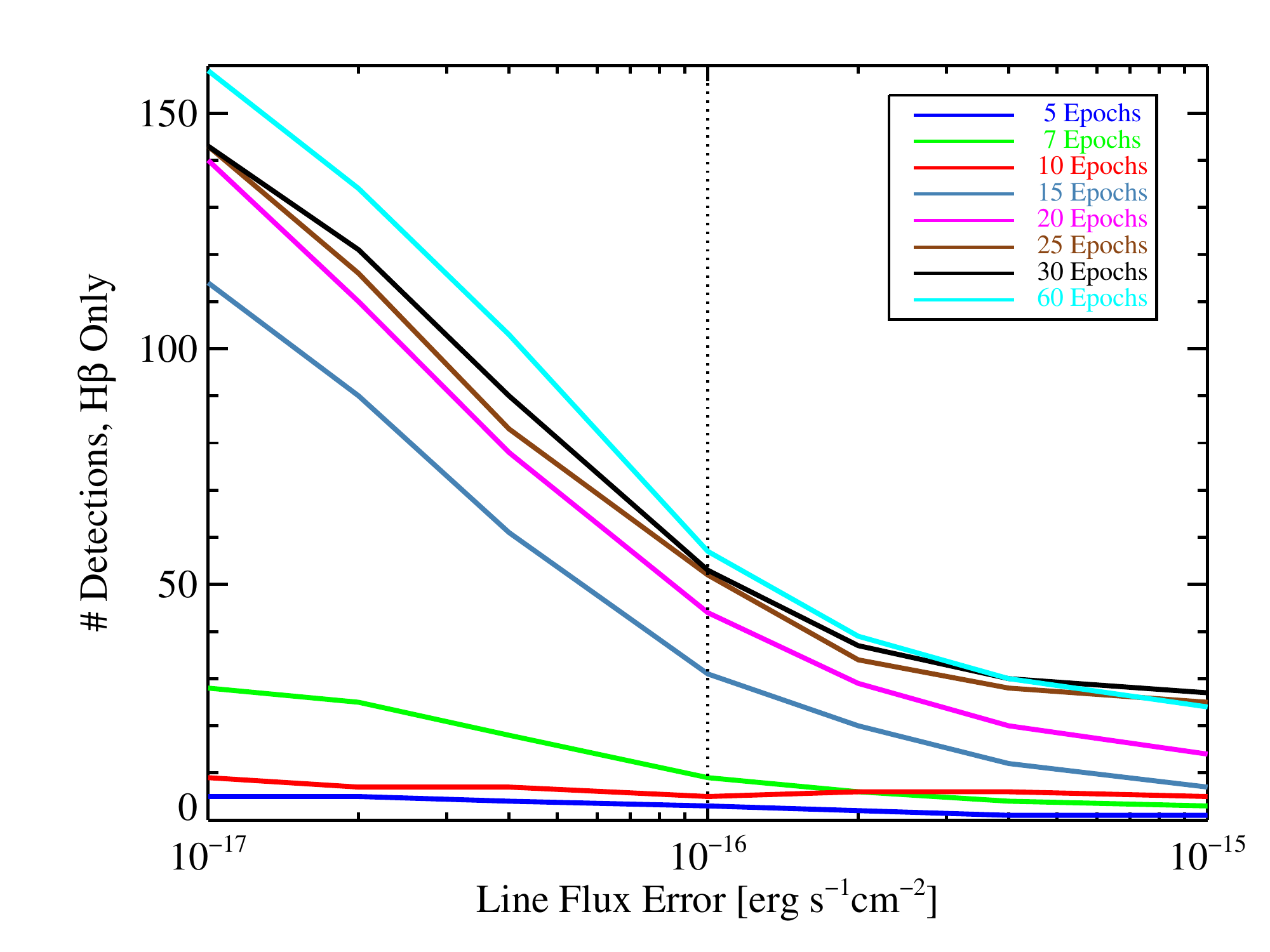}
    \caption{The maximum detection efficiency as a function of flux error, at fixed $N_{\rm ep}$, for a 6-month program. This maximum detection efficiency is achieved at the largest cadence that satisfies $N_{\rm ep}\Delta t\le 180$ days. Increasing the flux error leads to decreasing detection efficiency and degradation of the quality of the lag measurements. These results suggest that we cannot achieve a successful RM program with 10 epochs or less. {\em Top:} for all detections.  {\em Bottom:} for \hbeta\ detections only. }
    \label{fig:grid_det_err}
\end{figure}

Fig.\ \ref{fig:grid_det} summarizes the total number of detections for a plate as functions of flux errors, cadence and number of epochs. Given that \hbeta\ is the most likely line to detect a lag, we show the results for \hbeta\ separately in the right panel of Fig.\ \ref{fig:grid_det}. The results are averaged over 25 realizations. In this summary we only consider the 6-month program, therefore the total number of epochs and cadence are constrained to be $N_{\rm ep}\Delta t\le180$ days. Given a fixed $N_{\rm ep}$, higher cadence leads to more detections (since $t_{\rm span}=N_{\rm ep}\Delta t$ is increasing), and given fixed cadence, larger $N_{\rm ep}$ leads to more detections (because $t_{\rm span}$ is increasing, and more data points are used in the CCF calculations). Smaller errors lead to more detections. Fig.\ \ref{fig:grid_det_err} shows the increase of the maximum detection efficiency as the flux error decreases at fixed $N_{\rm ep}$. Such a maximum detection efficiency is generally achieved when the cadence is maximized at $N_{\rm ep}\Delta t\le 180$. In general we cannot achieve a successful RM program with 10 epochs or less, as already hinted at in \S\ref{sec:detection}.

\begin{figure}
\centering
    \includegraphics[width=0.45\textwidth]{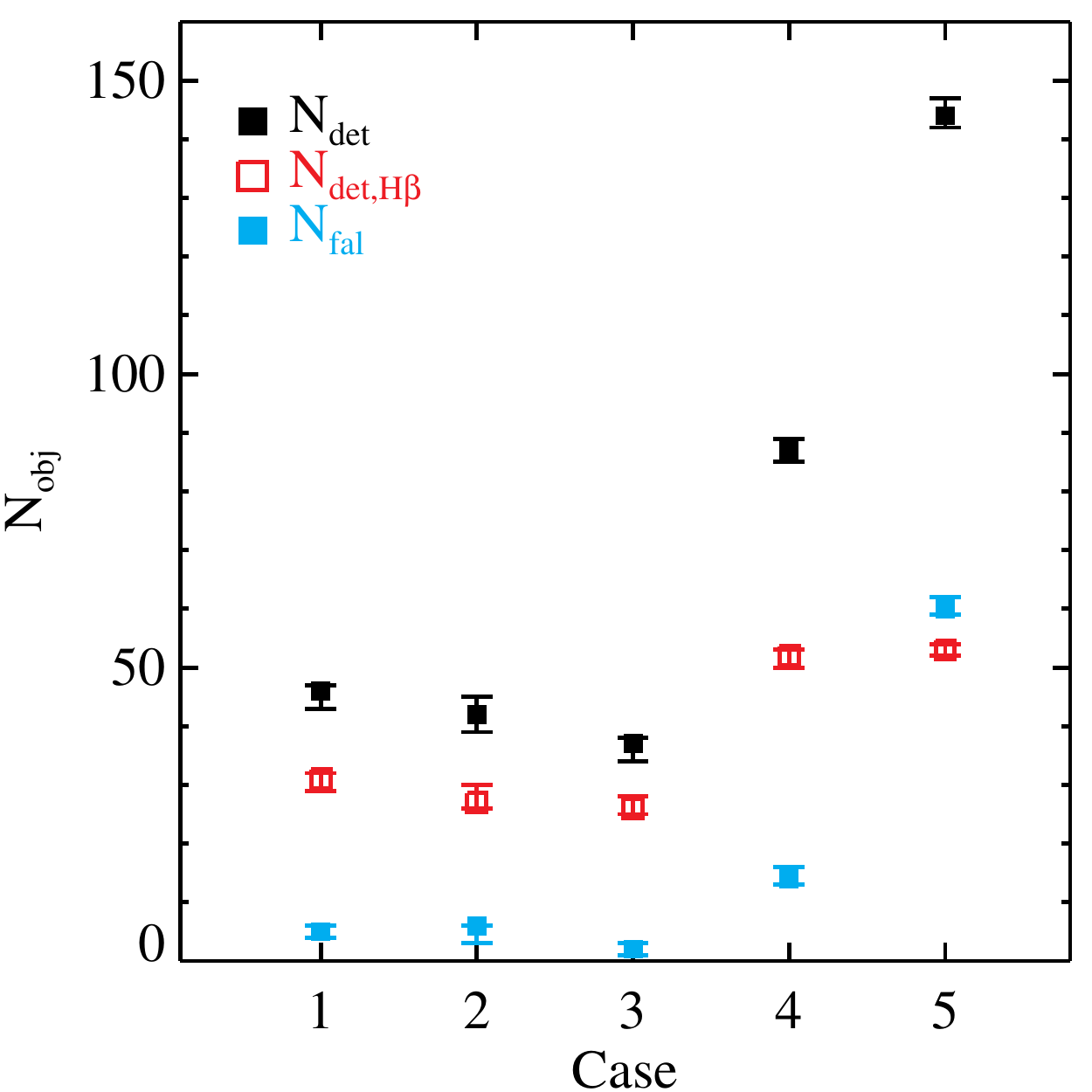}
     \caption{Tests in which we vary one aspect of the multi-object RM program. The black filled and open symbols are for all detections and \hbeta\ detections respectively, and the cyan points are for false detections. Ten trials are used to compute the mean and standard deviation for each case. Case 1: A reference 1-year program with a 12-day cadence and 15 epochs ({\em minimum requirement}). Case 2: same as Case 1, but with a larger transfer function width of $\sigma_{\rm tf}=0.2\tau$. Case 3: same as Case 1, but flux errors in continuum and line are perfectly correlated.
Case 4: same as Case 1, but with a more densely (e.g., daily) sampled continuum LC, covering the same period as the line LC. Case 5: same cadence as Case 1, but for two consecutive years (with a six-month gap), and 30 epochs in total. }
    \label{fig:test_conditions}
\end{figure}

\begin{figure}
\centering
    \includegraphics[width=0.48\textwidth]{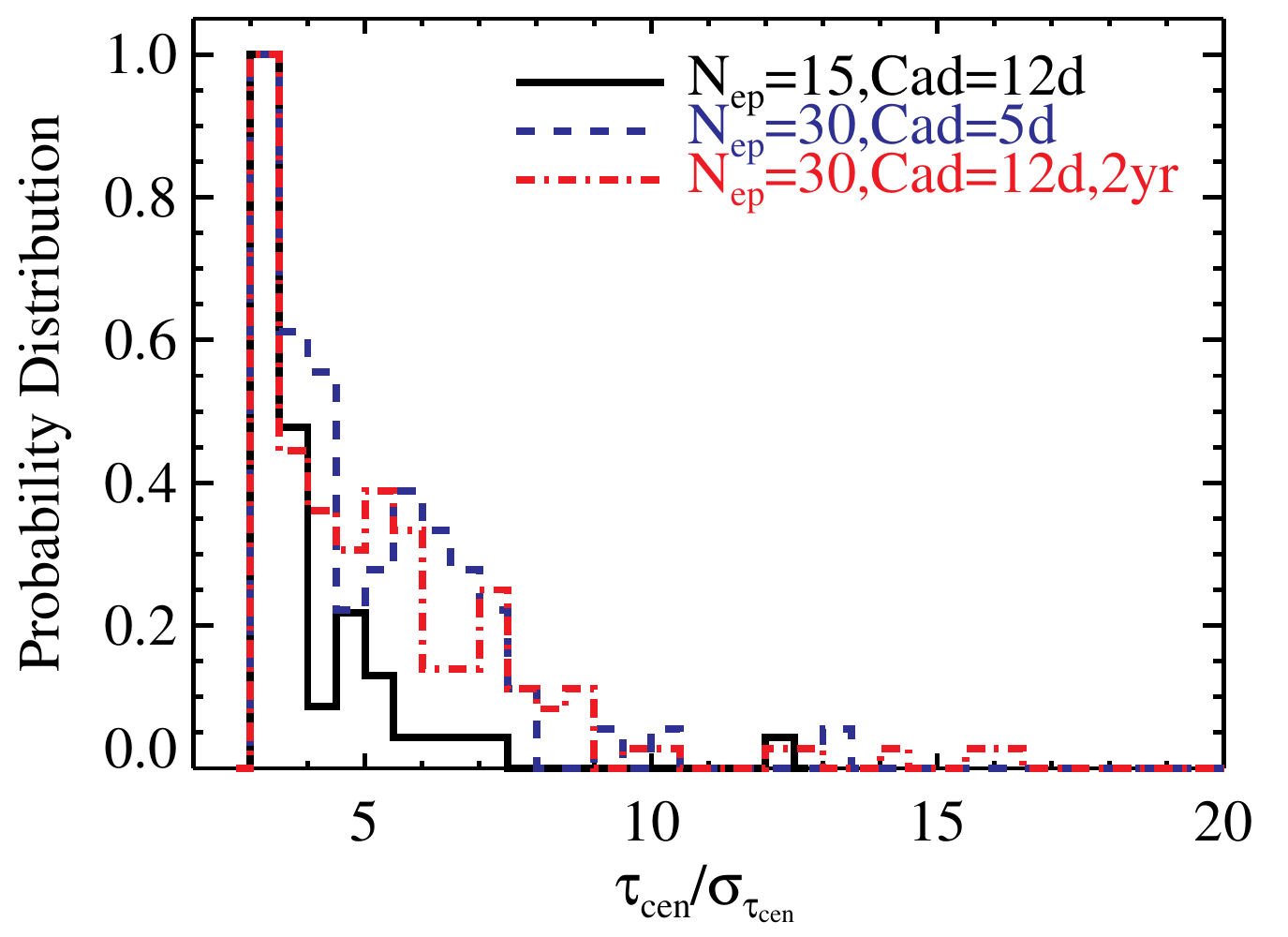}
     \caption{The distribution of lag measurement quality in terms of $\tau_{\rm cen}/\sigma_{\tau_{\rm cen}}$, for three different observing plans. The black histogram shows the default case, with 15 epochs and a 12-day cadence. With 30 epochs (red and cyan histograms), we can improve the measurement quality. The plan with 30 epochs in two consecutive years can deliver similar measurement quality as the plan with 30 epochs in one year (with a denser 5-day cadence), yet it can provide more detections with longer lags while at the same time detect a similar number of \hbeta\ lags (see Fig.\ \ref{fig:test_conditions}). }
    \label{fig:det_quality}
\end{figure}

Next, we consider the change due to varying one aspect of the program. The reference settings are:
12 days cadence, 15 epochs, fiducial (non-correlated) flux errors, simultaneous continuum and line measurements. This is the {\em minimum-requirement program} discussed previously, but the basic conclusions below also apply to other programs with more epochs. Fig.\ \ref{fig:test_conditions} presents the results for the following cases:

\begin{itemize}

\item[$\bullet$] Case 1: the reference case, with a 12-day cadence and 15 epochs.

\item[$\bullet$] Case 2: a larger transfer function width of $\sigma_{\rm tf}=0.2\tau$. 

\item[$\bullet$] Case 3: flux errors in continuum and line are perfectly correlated.

\item[$\bullet$] Case 4: a more densely (e.g., daily) sampled continuum LC, covering the same period as the line LC (i.e., not early continuum LC).

\item[$\bullet$] Case 5: same 12-day cadence, but for two consecutive years (with a six-month gap), i.e., with 30 epochs in total. 

\end{itemize}

Based on these tests, we can conclude the following: (1) the results are insensitive to a different transfer function or 
correlated errors in continuum and line fluxes; (2) a more densely-sampled continuum light curve significantly improves the results, and so it is highly desirable to obtain additional photometric LCs (for the continuum) for all targeted objects; (3) if there are 30 or more epochs, and if the program length is flexible, we should consider observing the same field in consecutive years. The gain in \hbeta\ detections is as large if we concentrate the 30 epochs within 6 months (see Fig.\ \ref{fig:grid_det} and Fig.\ \ref{fig:test_conditions}), but by extending the time baseline we can detect more objects with longer lags. In addition, the measurement quality is improved compared with the minimum-requirement 15-epoch case. Fig.\ \ref{fig:det_quality} shows the distributions of $\tau_{\rm mea}/\sigma_{\tau_{\rm mea}}$, i.e., the quality of the detections, for the simulated plate of quasars and for three cases (with fiducial flux errors): (1) a 12-day cadence with 15 epochs; (2) a 5-day cadence with 30 epochs; (3) a 12-day cadence and 15 epochs for two consecutive years (30 epochs in total). Clearly with more epochs, we can achieve higher quality lag measurements. Given 30 epochs, the 2-year plan tends to recover most of the short-lag detections in the 1-year plan with similar measurement quality, but also detects a larger number of longer lags, so it is preferred over the 1-year plan at a denser sampling rate. 


For Case 4, it will also be very useful to obtain early continuum light curves, such that more data points of the line LCs can be utilized in computing the CCF, and the gain in detections would be even greater than in Case 4.

We have also tested the impact of randomness in the LC sampling: small randomness (e.g., 2--3 days) in the sampling, which is unavoidable, does not materially affect the science return. The spectroscopic epochs do not have to be strictly evenly sampled, as usually expected under realistic weather conditions.  

\subsection{Multi-year Extension of the SDSS-RM program}\label{sec:exten}

\begin{figure*}
\centering
   \includegraphics[width=\textwidth]{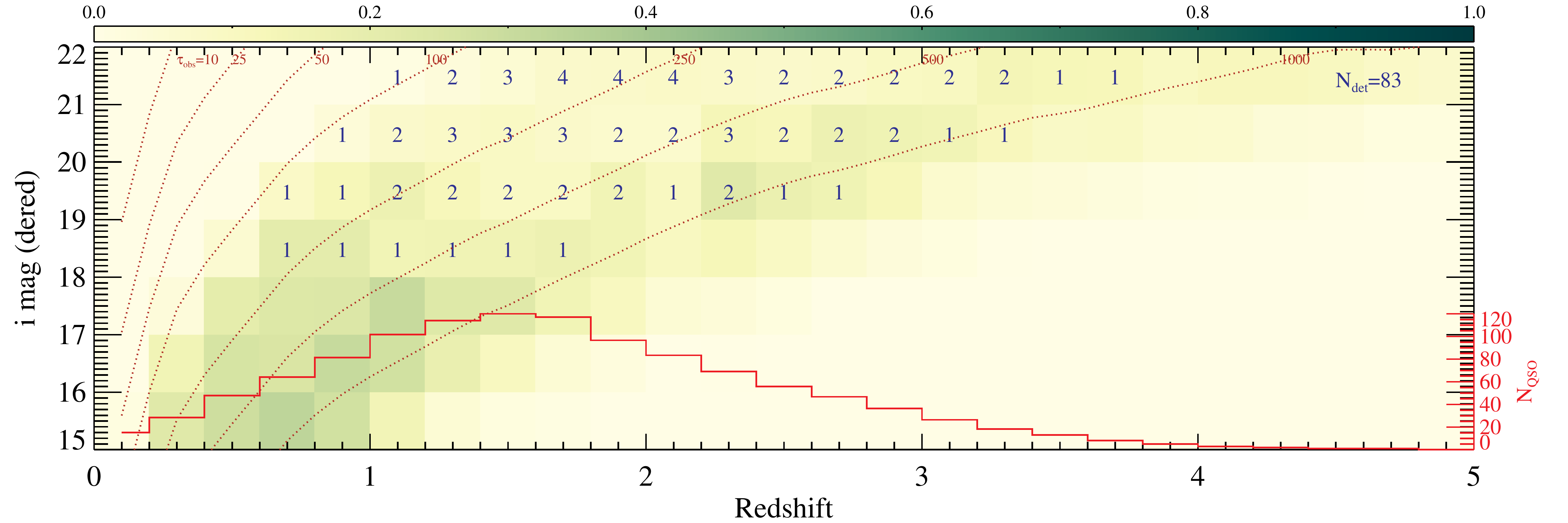}
   \includegraphics[width=\textwidth]{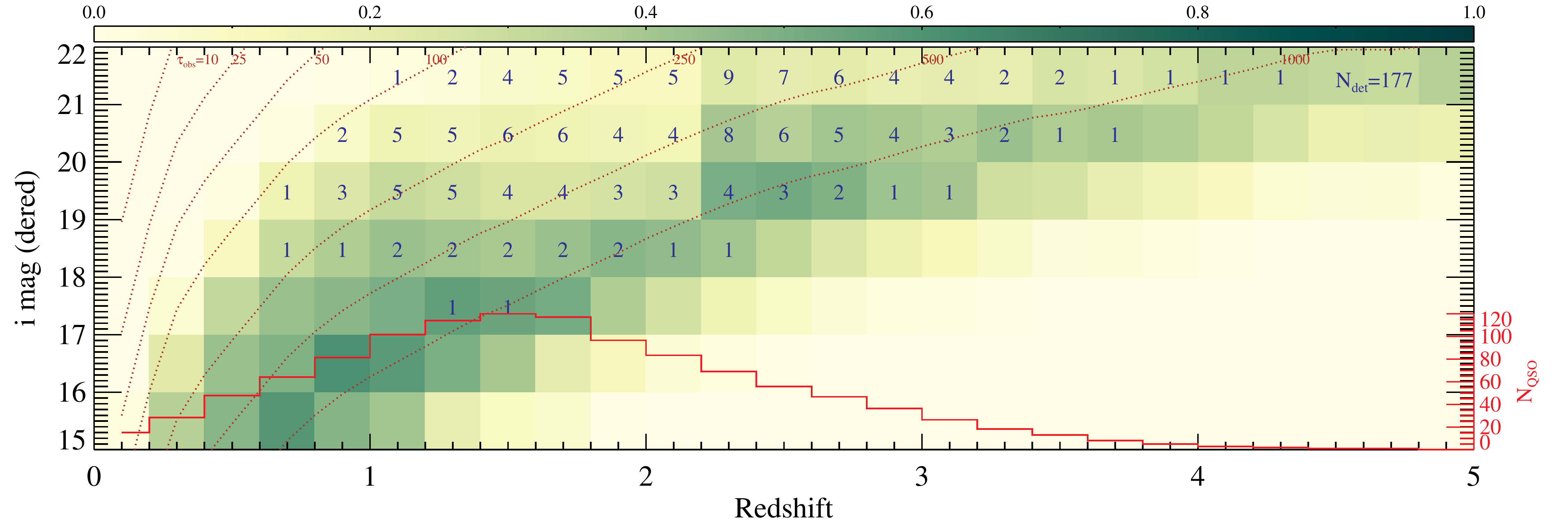}
   \includegraphics[width=\textwidth]{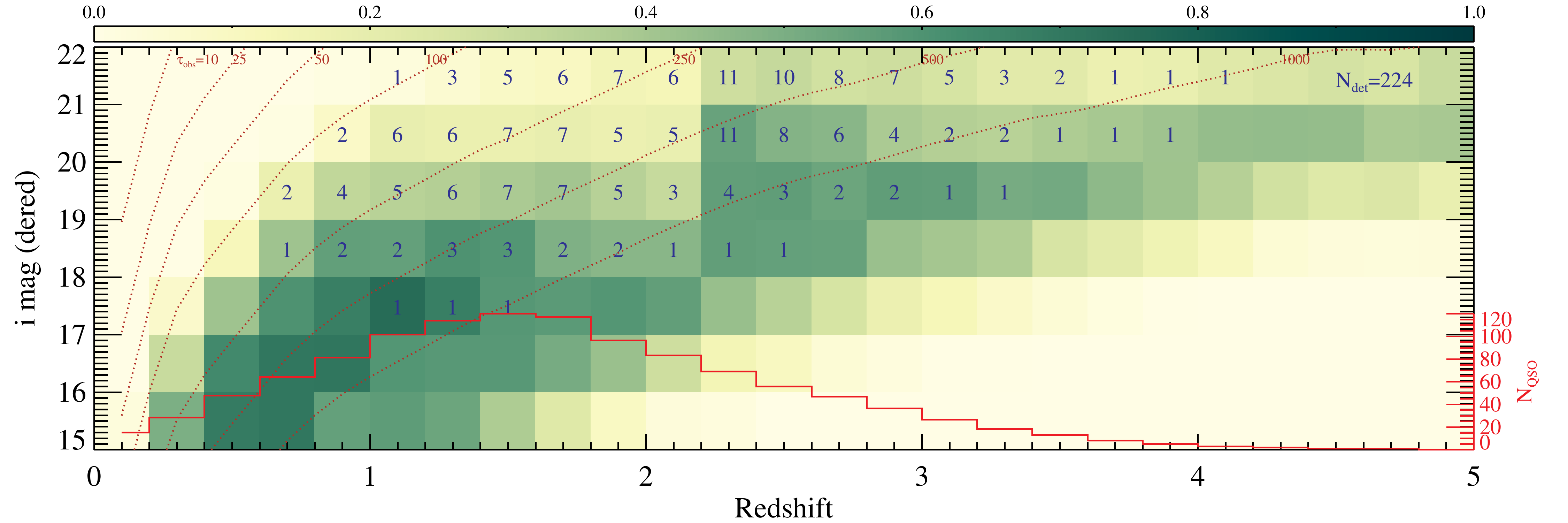}
   \caption{Detections of long lags ($>6$ months) for a multi-year program. Notation is the same as Fig.\ \ref{fig:de_examp_nqso}. In all simulations, we have 3-yr PanSTARRS early photometry (Year 1-3, 4-day cadence every 6 months, 135 photometric epochs in total), and the spectroscopic epochs in successive years. {\em Top}: 30 spectroscopic epochs in Year 4, with a $\sim 5$-day cadence for 6 months. {\em Middle:} with an additional 6 spectroscopic epochs spaced monthly each year in Year 5 and Year 6 (12 total additional spectroscopic epochs). {\em Bottom:} with an additional 6 spectroscopic epochs spaced monthly each year in Year 5-9 (30 total additional spectroscopic epochs). When the number of spectroscopic epochs is increased, the quality of the lag measurements is also increased. }
   \label{fig:de_examp_nqso_eboss}
\end{figure*}

The baseline SDSS-RM program has $\sim 30$ spectroscopic epochs on a single SDSS plate, obtained over the course of 6 months in 2014A semester (\S\ref{sec:implem}). Such a program is predicted to detect $\sim 100$ lags up to $z\approx 2$ (e.g., see Fig.\ \ref{fig:qso_plate} and Fig.\ \ref{fig:de_examp_nqso_more}). 

The actual SDSS-RM field coincides with one of the Panoramic Survey
Telescope and Rapid Response System 1 survey (Pan-STARRS1, or
PS1) Medium Deep fields that has been imaged in multi-bands in 2011--2013 with a cadence of several days (see \S\ref{sec:ps1}). These early photometric data provide an opportunity to detect long lags ($>6$ months) by combining them with the spectroscopy from the SDSS-RM program. The $\sim 80$ predicted detections of these long lags are shown as magenta triangles in Fig.\ \ref{fig:qso_plate}. 

However, quasars with these long lags typically vary on longer timescales than the 6-month baseline in the SDSS-RM program \citep[e.g.,][]{Kaspi_etal_2007}. Therefore to maximize the potential of the PS1 early photometry, an extended spectroscopy time baseline is highly desirable. The next stage of the SDSS survey (SDSS-IV) started in July 2014, and will continue through the next 4-6 years. One major component of the SDSS-IV is the eBOSS survey, which will obtain spectra for millions of galaxies and quasars with the BOSS spectrograph. This provides a natural opportunity to continue the SDSS-RM program in the eBOSS era, with reduced sampling rate for detecting long lags. 

To demonstrate the improvement in the detection of long lags with continued spectroscopic monitoring, we simulate an extended SDSS-RM program with various scenarios, starting from 2015A. The results are shown in Fig.\ \ref{fig:de_examp_nqso_eboss}. The three panels, from top to bottom, display the distribution of detected long lags in the redshift-magnitude plane, for three scenarios: (1) PS1 photometry $+$ the 6-month baseline SDSS-RM program (top); (2) PS1 photometry $+$ the 6-month baseline SDSS-RM program $+$ 2 year extension in eBOSS with 6 monthly-sampled spectroscopic epochs per year (middle); (3) PS1 photometry $+$ the 6-month baseline SDSS-RM program $+$ 5 year extension in eBOSS with 6 monthly-sampled spectroscopic epochs per year (bottom). With the best case (3), which only requires 30 additional plates in 2015-2019, we are able to detect hundreds of long lags, and almost fully cover the magnitude-luminosity plane of the flux-limited quasar sample. This would provide an unprecedented sample of quasars with RM measurements, and groundbreaking insights on quasar physics, the growth of supermassive black holes, as well as the coevolution of BHs and host galaxies. 

If we reduce the spectroscopic epochs in the eBOSS era to a lower demand, e.g., 1-3 epochs per year, there is still some, but reduced, gain in the detections of long lags. However, the quality of these lag measurements will not be as good as those with more epochs, and extra effort will be required to reduce aliasing in the detections.  

\section{The SDSS-RM Program: Implementation and observations}\label{sec:implem}

The SDSS-III collaboration issued a call for proposals during the summer of 2013 to utilize dark/grey time ($\sim250-300$ plate-hours of open-shutter time) in the last season of the BOSS survey. This opportunity arose due to the better-than-expected weather conditions in the past several seasons that led to the early completion of the BOSS survey. 

Our SDSS-RM program was approved as one of the ancillary programs that were scheduled during the Jan--July 2014 dark/grey time with the BOSS spectrograph. A total of 60 hours (or equivalently, 30 BOSS epochs) were awarded to this program, with an operational effort by the SDSS-III survey team to ensure the required cadence is met (weather permitted). The actual time allocation was adjusted slightly to compensate for weather losses. The BOSS pipeline-processed spectroscopic data from this program will be part of the final data release (DR12) of SDSS-III in December 2014.

\subsection{The RM field and the target sample}

Our RM field is the PS1 Medium Deep Field MD07 (RA$=213.704$, DEC$=+53.083$), which is a 7 deg$^2$ field within the CFHT-LS W3 field\footnote{http://www.cfht.hawaii.edu/Science/CFHTLS/}. This particular field was chosen because of the following advantages: (1) it has maximum visibility ($>6$ months) during 2014A, and its high declination allows the field to be observed at low airmass for a large range of Local Sidereal Time (LST); (2) a spectroscopically confirmed quasar sample matched to our program already exists (see below); (3) it has 3 years of dense ($\sim 4$-day cadence) PS1 photometric light curves in $griz$ (2011--2013), which provides the possibility to detect long lags ($\tau>6$ months; \S\ref{sec:exten}); (4) it has additional multi-wavelength data coverage for portions of the field from various ground-based and space-based facilities, such as the CFHT-LS D3 field and the Extended Groth Strip (EGS) deep HST field. 

This field is fully covered in the SDSS-I/II and SDSS-III BOSS footprints. Spectroscopy from SDSS-I/II/III provides $\sim 1000$ confirmed quasars in this field. Most of these quasars were targeted by the SDSS-III BOSS survey for spectroscopy (see Ross et~al.\ 2012 and Ahn et al.\ 2015 for details regarding the final quasar target selection in BOSS). In addition, variability selection based on PS1 light curves in MD07 was used to select quasar candidates and follow-up spectroscopy was obtained with MMT/Hectospec (Green et al., in preparation), which provided tens of additional quasars. Finally, there are a handful of quasars in the EGS field discovered by the DEEP2 survey \citep{Newman_etal_2013}, which were missed in the SDSS$+$MMT sample. These discovery spectra can be included in the time series analysis, but we note that most of the prior BOSS quasar spectra suffer from flux calibration problems that will need additional post-processing (see \S\ref{sec:boss_spec}), and the DEEP2 and MMT spectra are generally not well flux-calibrated to better than $10\%$.

\begin{figure}
\centering
    \includegraphics[width=0.45\textwidth]{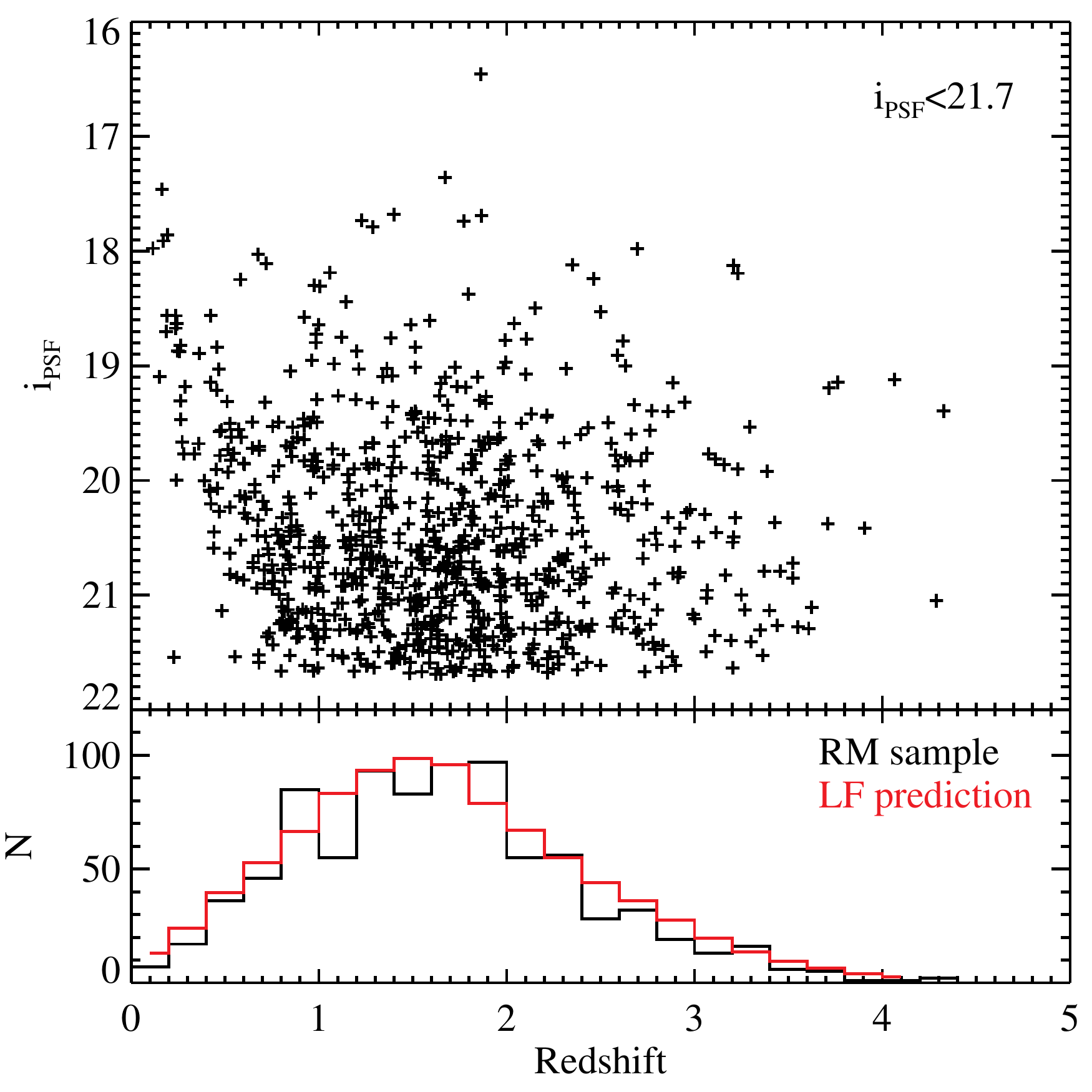}
     \caption{Distribution of the RM quasar targets in the magnitude-redshift plane. Bottom panel displays the redshift distribution, with the prediction from the luminosity function in \citet{Hopkins_etal_2007} shown in red, for $i_{\rm PSF}<21.7$ quasars. There is good agreement in the redshift distributions of our sample and the LF prediction. }
    \label{fig:iz_dist}
\end{figure}

Our final parent sample includes $\sim 1200$ quasars in the redshift range $0<z<5$, which were all visually inspected to confirm their quasar nature with at least one visible broad line. Fig.\ \ref{fig:iz_dist} shows the redshift and $i$-band magnitude distribution of these quasars. Although this sample selection is heterogeneous, the distribution of quasars in the $L-z$ plane roughly matches the expected distribution from the LF estimates in \citet{Hopkins_etal_2007}, which suggests that our sample is fairly complete down to $i\approx 22$. 

We selected a flux-limited subsample to form the final RM sample to be observed with the BOSS spectrograph. We imposed a flux limit of $i=21.7$, and removed objects that are in fiber collisions\footnote{Due to physical limitations of the BOSS spectrograph, two targets with an angular separation less than $62$\arcsec\ cannot be simultaneously assigned a fiber on a single plate. } with another target. When removing fiber collided objects, priorities were given to objects roughly according to the following considerations of quasar properties: a) $z<1$ over $z>1$; b) brighter $i$-mag over fainter ones; c) stronger broad lines over weaker lines; d) point sources over extended sources. These criteria were used to maximize the overall yield of lag detections. Criteria a) favors shorter lags, better resolved in the limited 6-month program, b) and c) are to maximize S/N in continuum and line fluxes, and d) is to mitigate host galaxy contamination. Since the number of targets exceeds that of available fibers, two rounds of tiling were performed (SOURCETYPE='RM\_TILE1' and 'RM\_TILE2' in Table \ref{table:sample}) to prioritize targets which are more likely to detect a time lag in the baseline program roughly based on the above criteria.  

The final RM sample contains 849 quasars, each of which was assigned a fiber on the RM plates. The remaining fibers were allocated to calibration sources, including 70 spectrophotometric standard stars, 80 sky fibers, and 1 LRG target (enforced by the BOSS tiling algorithm, but actually not tiled on any existing object in SDSS imaging). The spectrophotometric standards were chosen to be F stars, with $16<r_{\rm fib2}<19$, where $r_{\rm fib2}$ is the $r$-band magnitude within the 2\arcsec\ diameter BOSS fiber aperture. The total number of spectrophotometric standards is $\sim 3.5$ times the nominal number for BOSS plates, designed to improve the spectrophotometry by densely mapping the entire plate (see \S\ref{sec:specphoto}). 

\begin{figure}
\centering
    \includegraphics[width=0.45\textwidth]{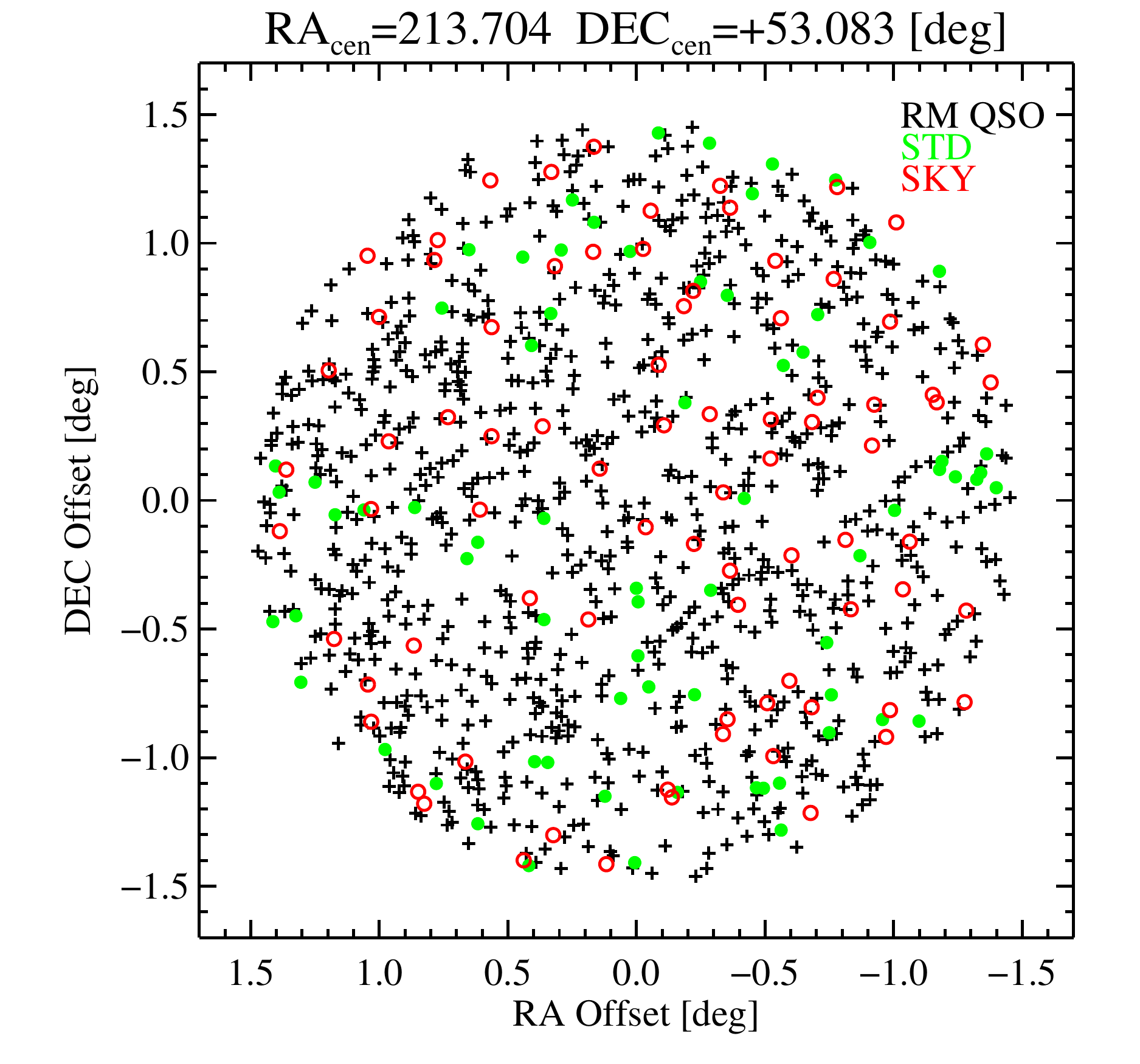}
     \caption{Distribution of tiled objects on the sky. }
    \label{fig:spat_dist}
\end{figure}

Table \ref{table:sample} summarizes the basic properties of our RM sample (and information of the calibration targets). Fig.\ \ref{fig:spat_dist} shows the positions of all targets on the plate. Our RM sample displays a diverse range of quasar properties, including classical broad-line quasars with blue continua, heavily reddened quasars with strong broad lines, and quasars with abnormal line flux ratios. A complete description of the sample properties (such as spectral diversity, variability and multi-wavelength properties) will be presented  in Shen et~al. (2015, in preparation). 

\begin{table*}
\caption{The tiled sample}\label{table:sample}
\centering
\scalebox{1.2}{
\begin{tabular}{lcc}
\hline\hline
Column & Format & Description \\
\hline
RMID &      LONG                 & Index of objects in this catalog [0-999]\\
RA       &        DOUBLE          &  Object RA [J2000; DR10 astrometry] \\
DEC     &       DOUBLE           &  Object DEC [J2000; DR10 astrometry]\\ 
Z          &    DOUBLE         & Redshift \\
SOURCETYPE     &  STRING        &   'RM\_TILE1', 'RM\_TILE2', 'LRG', 'STD', 'SKY' \\
PSFMAG   &       FLOAT ARRAY[5]   &     SDSS $ugriz$ PSF magnitude ($asinh$); undereddened \\
OBJCTYPE    &   LONG               &          0=unknown; 3=extended source; 6=point source \\
SAMPLE    &     STRING            &   'boss'=BOSS; 'dr7'=SDSS DR7; 'mmt'=MMT; 'egs'=EGS \\ 
   PLATE       &    LONG  ARRAY[32]    &  Plate number of 32 epochs \\
   FIBERID    &     LONG  ARRAY[32]    &  FiberID of 32 epochs \\
   MJD           &  LONG    ARRAY[32]     &  MJD of 32 epochs \\
   MEDSN    &    DOUBLE ARRAY[32]    &  Median S/N/pixel of each epoch (over full wavelength range) \\
\hline\\
\end{tabular}
}
\begin{tablenotes}
      \small
      \item NOTE. --- The catalog of the 1000 tiled targets in FITS format (RM target, F-stars and sky fibers). The first 849 rows are for the RM targets with SOURCETYPE='RM\_TILE1' or 'RM\_TILE2'. The PLATE, FIBERID and MJD columns record the spectroscopic identifications for all epochs of each target. Redshifts are measured by running the SDSS redshift pipeline \citep{Bolton_etal_2012} on the coadded spectra from all epochs and visually verified. Object positions are from the corrected SDSS astrometry discussed in DR10 \citep{Ahn_etal_2014}.
\end{tablenotes}
\end{table*}

\subsection{BOSS spectroscopy}\label{sec:boss_spec}

To obtain the full six months of coverage, we required three plates (with plate numbers 7338, 7339 and 7340) to be drilled with identical targets at hour angles designed to maximize LST coverage. Considering the declination of the target field and following the standard BOSS requirements for limiting atmospheric differential refraction across a plate \citep[][]{Dawson_etal_2013}, two plates designed with an hour angle HA$=\pm40^\circ$ produce visibility windows of one hour and 40 minutes. A third plate designed for observations at transit offers observing windows of $\sim 3$\,hrs in duration. These three plates allow observations in the hour angle range of $-50^\circ$ to $+50^\circ$, providing LST coverage from 163$^\circ<{\rm LST}< 263^\circ$. At least one plate would be visible on any night from December 24, 2013 until July 4, 2014, which extends slightly beyond the end of SDSS-III. These plates were observed repeatedly during the period of this program. However, the fiberID of each target will usually be different for each epoch (even with the same plate number), as fibers will be shuffled in each plate plugging. Table \ref{table:sample} provides information to track the fiber numbers of individual targets for each epoch. 

All targets were designed during tiling with the keywords $\lambda_{\rm eff}=5400$\ \AA\ and $z_{\rm offset}=0$ to ensure that the quasar targets are treated in the same spectrophotometric system as the calibration F-stars \citep[see][for technical details regarding the tiling process]{Dawson_etal_2013}. This approach is different from the BOSS main quasar survey, where the quasar targets are offset in the reference wavelength ($\lambda_{\rm eff}=4000$\ \AA) and distance from the focal plane to enhance the S/N in the blue region where the Ly$\alpha$ forest lies \citep[][]{Dawson_etal_2013}. If we had used $\lambda_{\rm eff}=4000\,$\AA\ and nonzero $z_{\rm offset}$, the flux calibration of quasars would have significantly larger systematic errors than the nominal $\sim 6\%$ spectrophotometric accuracy for BOSS \citep[][]{Paris_12}. 

The spectroscopic observations were acquired during 7 dark/grey runs from Jan 1, 2014 to July 3, 2014. A typical epoch consists of a minimum of eight 15-minute sub-exposures, and requires that the accumulated S/N$^2$ per plate exceeds twice the nominal threshold used in the BOSS main survey, i.e., S/N$^2_g>20$ (the average extinction-corrected S/N$^2$ per pixel in $g$ band evaluated at $g_{\rm psf}=21.2$ [$g_{\rm fib2}=22$]). For comparison, the typical depth of the BOSS DR9 quasar sample, which was used in the simulations, is S/N$^2_g\sim 15$. This required depth is to ensure good spectral quality at the limiting magnitude of our RM sample, and is usually achieved with the exception of only a few epochs. If the desired depth is not achieved during one night, the same plugging is observed again (if possible) during the next night and all exposures taken in the two consecutive nights are coadded to form a single epoch. The separation between epochs in a given run is roughly 3 -- 4 days, but varies from run to run under different weather conditions. Typically five epochs were taken for each run when weather permitted, totaling 32 epochs. Table \ref{table:boss_log} summarizes the spectroscopic observations. 

\begin{table*}
\caption{Log of BOSS RM Observations}\label{table:boss_log}
\centering
\scalebox{1.2}{
\begin{tabular}{lccccccc}
\hline\hline
Epoch\# & start MJD & end MJD & mean MJD & spPlate file & plate S/N$_g^2$ &  $N_{\rm coadd}$ & Seeing (\arcsec) \\
\hline
Dec/Jan & 2013/2014 & & & & & & \\
1   &  56659 &  56660  &  56660.209  & 7338-56660 & 18.0   & 7  & 1.7 \\
2   &  56664 &  56664  &  56664.513  & 7338-56664 & 9.9   & 5   & 2.2  \\
3*  &  56669 &  56669  &  56669.500  & 7338-56669 & 3.4     & 6  & 3.0 \\ 
Jan/Feb & & & & & & \\
4   &  56683 &  56683  &  56683.480  & 7339-56683 & 22.4   & 9  & 1.7 \\
5   &  56686 &  56686  &  56686.473  & 7339-56686 & 21.9   & 8  & 1.7 \\
6   &  56696 &  56697  & 56696.778   & 7339-56697 & 23.2   & 10 & 1.7 \\
Feb/Mar & & & & & & \\
7*  & 56713  & 56713   & 56713.426  & 7339-56713  & 0.5   & 2   & 1.4 \\
8   &  56715  &  56715   & 56715.388 & 7339-56715 & 26.1  & 8  & 1.4 \\
9   &  56717   & 56717   & 56717.334 & 7338-56717  & 25.3  & 9  & 1.6 \\
10  &  56720   & 56720 &   56720.446 &  7339-56720 & 22.6  & 8  & 1.7  \\  
11  &  56722  & 56722  & 56722.387 & 7339-56722  & 32.3   & 9  & 1.3 \\
12   & 56726   & 56726  &  56726.457 &  7340-56726 & 28.0   & 9  & 1.6 \\ 
Mar/Apr & & & & & & \\
13  &  56739  & 56739  & 56739.410 & 7339-56739  & 19.0 & 8  & 1.4 \\
14 &   56745  & 56745  & 56745.281 & 7338-56745 &  25.0  & 9 & 1.6 \\ 
15  &  56747  &  56747 & 56747.416 & 7339-56747  & 19.3  & 7  & 1.8 \\
16  &  56749  & 56749  &  56749.373 &  7339-56749  & 23.1  & 8 & 1.7 \\
17  &  56751  &  56751 &   56751.340  & 7339-56751  & 25.5  & 9  & 1.7 \\
18  &  56755  & 56755  &  56755.341  & 7339-56755  & 16.8  & 7  & 1.6 \\
Apr/May & & & & & & \\
19 &   56768  & 56768   & 56768.227 & 7339-56768 & 33.2 & 8  & 1.4 \\
20  &  56772   & 56772  &  56772.234 & 7339-56772 &  31.8 & 8 & 1.3  \\
21 &   56780  & 56780   & 56780.234  & 7339-56780  & 32.0  & 8 & 1.6 \\
22 &   56782  & 56782   &  56782.247 & 7339-56782 & 27.3   & 10 & 1.6 \\
23 &   56783 &   56783 &   56783.248 & 7339-56783  & 22.2   & 10  & 1.6 \\
May/Jun & & & & & & \\
24 & 56795  & 56795  &  56795.175 & 7339-56795  & 17.6   & 7 & 1.4 \\
25 &  56799 & 56799  &  56799.212 & 7339-56799  & 34.2   &  8 & 1.4 \\
26 &  56804 &  56804 &   56804.187 & 7339-56804  & 17.5  & 8 & 2.0 \\
27 &   56808  & 56808  &  56808.259  & 7339-56808  & 17.7  & 9 & 2.1 \\
28 &   56813  & 56813   &  56813.227 & 7339-56813  & 19.8  & 9 & 1.6 \\
Jun/Jul & & & & & & \\
29$^\dagger$ & 56825 & 56825 & 56825.186 & 7340-56825 & 25.0 & 8 & 1.7\\
30 & 56829 & 56829 & 56829.211 & 7340-56829 & 18.5  & 11 & 2.2 \\
31 & 56833 & 56833 & 56833.210 & 7340-56833 & 21.9 & 9 & 2.0 \\
32 & 56837 & 56837 & 56837.189 & 7340-56837 & 28.5 & 8 & 1.8\\
\hline\\
\end{tabular}
}
\begin{tablenotes}
      \small
      \item NOTE. --- $^*$ Epochs that have a plate S/N$_g^2<5$ due to poor observing conditions. Most of the targets on these plates
      should be discarded in the cross-correlation analysis due to the poor spectral quality. $^\dagger$ Epoch 56825 suffered from a software error with no simultaneous calibration exposures taken for science exposures. Afternoon calibration exposures were used in reduction. Except for Epochs 1 and 6, 
      all epochs were observed on the same night. The ``$N_{\rm coadd}$'' column indicates the number of coadded 15-min sub-exposures in each epoch. The last column is the median seeing during the observation estimated from the guide camera images. 
\end{tablenotes}
\end{table*}

\subsection{Spectroscopic data processing and calibration}\label{sec:specphoto}

The spectra taken by the BOSS spectrograph were processed with the latest BOSS spectroscopic pipeline {\em idlspec2d} v5\_7\_1, which performs flat-fielding, 1d extraction, wavelength calibration, sky subtraction and flux calibration using flat fields, arcs, F-star and sky fiber spectra. The details of the pipeline are described in Schlegel et al. (2015, in preparation). The individual sub-exposures are coadded for each epoch (plugging) with a logarithmic wavelength binning of $10^{-4}$ in $\log_{10}$. The spectra are stored in vacuum wavelength, calibrated to the heliocentric reference, with an accuracy of better than 5 $\kms$. The wavelength coverage of BOSS spectroscopy is 3650--10,400 \AA, with a spectral resolution of $R\sim 2000$. About $1\%$ of the total fibers were bad fibers which completely failed to produce a usable spectrum. 

The flat-fielded, wavelength-calibrated, and sky-subtracted 1d sub-exposures are stored in the {\tt spFrame*} files\footnote{See http://data.sdss3.org/datamodel/ for details regarding the format of BOSS data products.}, separately for each of the two spectrographs (1 and 2) and the two arms (blue and red), which are used as input files for the flux calibration procedure. The BOSS flux calibration uses the spectrophotometric standard F-stars on the plate to construct an average fluxing vector to scale the individual spectra. The construction of the average fluxing vector is achieved by fitting model stellar spectra to the spectrophotometric stars and tying the synthetic $r$-band flux to the PSF calibration flux from SDSS imaging. After each sub-exposure is scaled by the fluxing vector, they are scaled to the best exposure determined based on the S/N, then combined to form a coadded spectrum with the blue and red parts of the spectrum merged. The final step is to fit a positional-dependent distortion image to scale the coadded spectra to account for variations across the focal plane. The final spectra are output to a {\tt spPlate*} file that stores all the spectra on the plate. 

The BOSS pipeline is similar to that used for SDSS-I/II. However, due to the smaller fiber size (2\arcsec) of the BOSS spectrograph as opposed to SDSS-I/II (3\arcsec), and the combined effect of pointing errors and atmospheric differential refraction (ADR), the spectrophotometry accuracy degrades from 4\% in SDSS-I/II \citep[][]{DR6} to $\sim 6\%$ in SDSS-III \citep[][]{Dawson_etal_2013}. In our case, the situation is slightly worse because the RM plates are observed for longer time (hence having stronger ADR effects) than standard BOSS plates to accumulate more signal. 

\begin{figure}
\centering
    \includegraphics[width=0.45\textwidth]{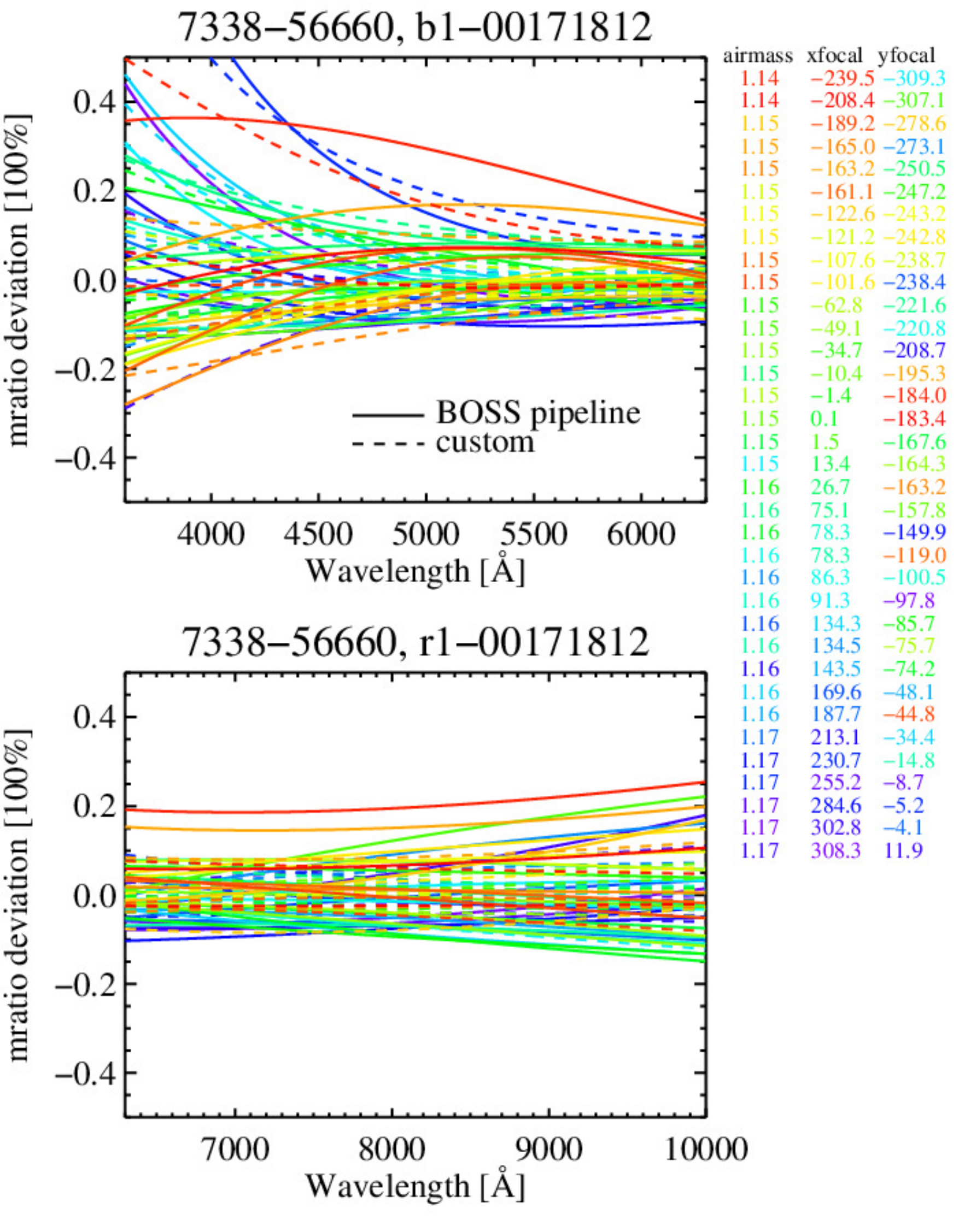}
     \caption{Improving BOSS pipeline flux calibration using standard stars. {\em Top:} the blue arm of BOSS spectrograph 1. {\em Bottom:} the red arm of BOSS spectrograph 1. Solid lines show the fractional deviations of the fluxing vectors of individual standard stars relative to the average fluxing vector, from one of the sub-exposures in an epoch. The average fluxing vector was adopted in the BOSS pipeline flux calibration for all objects. The scatter of these lines indicates the uncertainty of spectrophotometry expected in the pipeline calibration. The curves are color-coded from red to purple, progressing from one side of the focal plane to the other according to their fiber numbers. Although there are stochastic variations in these fluxing vectors, there is a general trend of the fluxing vectors as a function of focal position (xfocal and yfocal). Motivated by this, we fit a low-order polynomial function to the individual fluxing vectors as a function of the xy focal position at each wavelength, to account for the smooth spatial variation in the fluxing vectors. With this approach, the deviations of the model fluxing vectors (evaluated with the polynomial function at the locations of each star) from the true fluxing vectors for these standard stars are shown as dashed lines. As a result, the scatter of these deviations is reduced. For this exposure the improvement is most obvious in the red arm; for other exposures there are also noticeable improvements in the blue arm (especially for exposures at high airmass). The model fluxing vectors are used to flux calibrate all targets on the plate by evaluating the polynomial fit at the locations of each target. }
    \label{fig:mratio}
\end{figure}

To facilitate RM lag detections, we improve the BOSS pipeline spectrophotometry with a custom flux calibration, using the uncalibrated {\tt spFrame*} files as input. Our RM plate uses 3.5 times more spectrophotometric stars than the BOSS survey, which better samples the positional-dependent ADR across the focal plane. Thus instead of using an average fluxing vector as in the BOSS pipeline, we adopt a low-order 2D polynomial function to describe the spatial variation of the fluxing vector across the focal plane of the plate, and flux calibrate each object using the model fluxing vector evaluated at its focal position. To demonstrate why this custom flux calibration works, we show in Fig.\ \ref{fig:mratio} the dispersion in the flux vectors (ratio of model stellar spectra to observed flux for the F-stars) in one sub-exposure of a specific epoch. Although there are stochastic variations in these fluxing vectors due to pointing errors and other random processes, there is a general trend of the fluxing vector as a function of focal position. Our polynomial fit accounts for this systematic offset, and reduces the dispersion in the fluxing vectors of individual stars. In this way we improve the spectrophotometry for individual objects on the plate. 

\begin{figure}
\centering
    \includegraphics[width=0.45\textwidth]{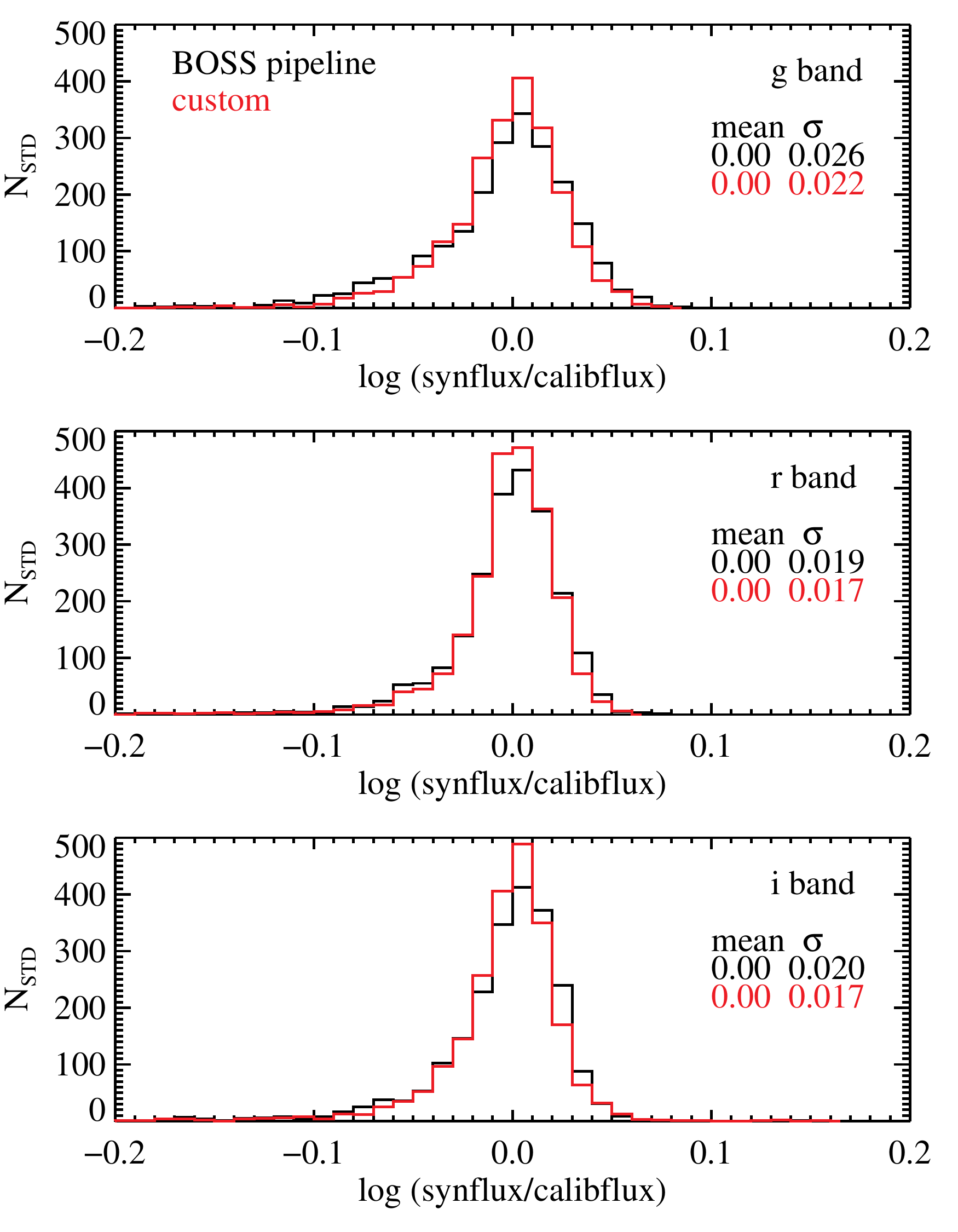}
     \caption{Comparisons between synthetic fluxes from spectroscopy and calibration fluxes in the $gri$ bands for the standard stars in all epochs. We fit a Gaussian to the distribution and report the mean and dispersion from the fit in each panel. The custom flux calibration (red) shows improvement over the pipeline calibration (black), as indicated by the smaller Gaussian dispersion. }
    \label{fig:syn_calib_std}
\end{figure}

Fig.\ \ref{fig:syn_calib_std} compares the spectrophotometry between the BOSS pipeline results and our custom calibration, for all the epochs in our program. We compare the synthetic flux in the $gri$ bands using the calibrated spectra of F-stars and the calibration PSF flux from SDSS imaging. The new calibration slightly improves the broad-band spectrophotometry rms errors from $\sim 6\%$ to $\sim 5\%$, and significantly reduces the fraction of outliers. For instance, the fraction of objects with a broad-band spectrophotometric error greater than 0.1 dex in any of the $gri$ bands declines from $4.6\%$ in the pipeline calibration to $2.9\%$ in our new calibration. 

\begin{figure}
\centering
    \includegraphics[width=0.45\textwidth]{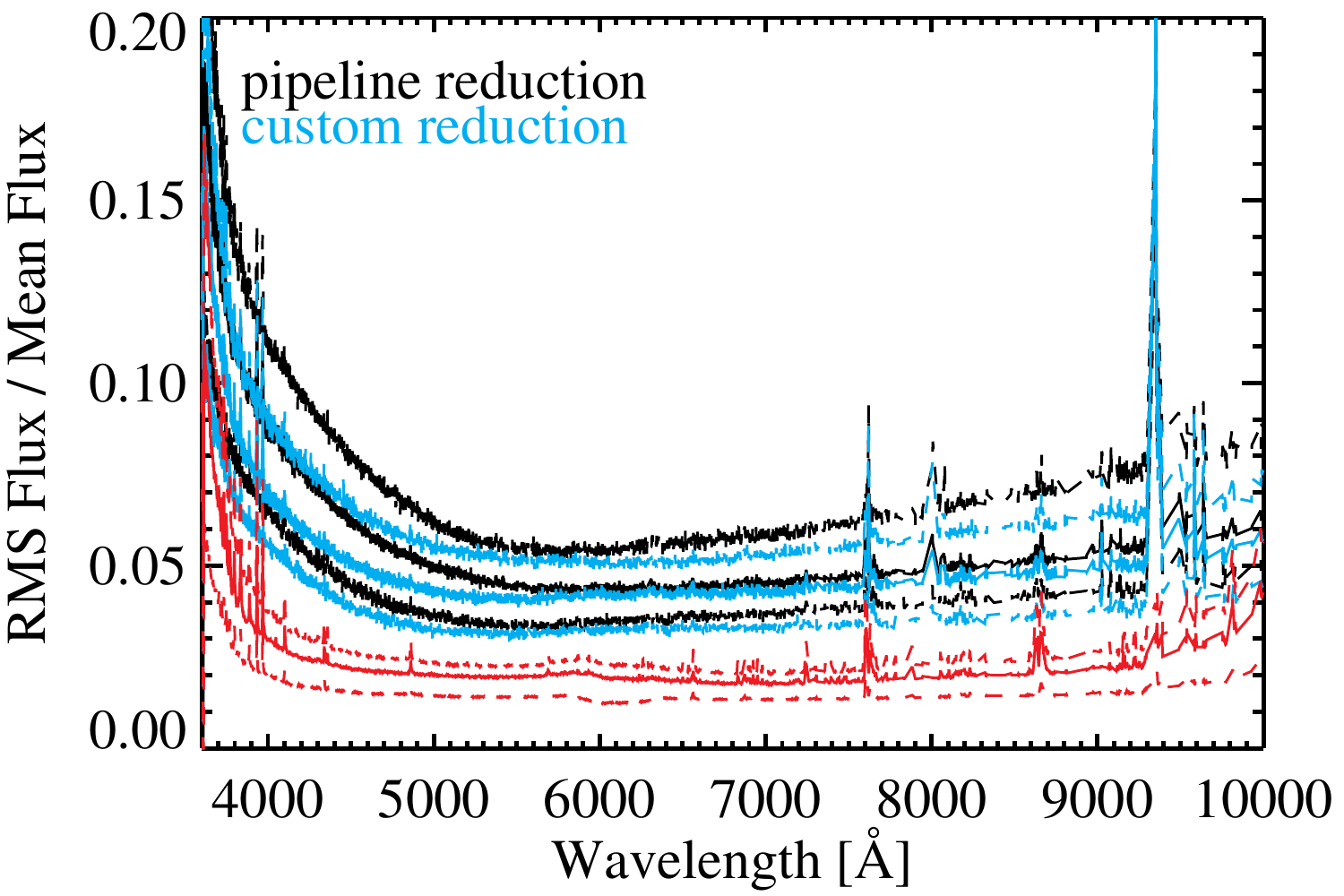}
     \caption{RMS spectra of standard F-stars. The black lines are from the BOSS pipeline reduction, while the cyan lines are from the new reduction with custom flux calibration and improved sky subtraction. The red lines are median statistical errors relative to the mean spectrum. Pixels identified with strong sky residuals (e.g., Table \ref{table:skymask}) have been interpolated over to reduce noisy spikes. Each set of three lines shows, from bottom, the 16\%, 50\% and 84\% percentiles of the distribution from all 70 stars. }
    \label{fig:rmsspec_std}
\end{figure}

Fig.\ \ref{fig:rmsspec_std} displays the rms spectra of the 70 standard stars from all but the two lowest S/N epochs (see Table \ref{table:boss_log}). The level of the rms variation reflects the systematic errors in flux calibration, since photon noise is sub-dominant for these standard stars. In general the spectrophotometric precision is consistent with our broad-band estimates -- at the few percent level, but becoming worse at blue wavelengths. Again, the new calibration shows improvement over the pipeline calibration, especially at the bluest and reddest wavelengths where the ADR effect is the largest.  

Using the standard stars in all epochs, we have noticed that very infrequently ($0.56\%$ of the time, or $0.82\%$ with the pipeline calibration), the spectrophotometry is catastrophically incorrect with a deviation from the calibration flux in at least one of the $gri$ bands greater than 0.3 dex. In essentially all these cases the spectral flux falls below the calibration flux. These rare cases occur randomly with no apparent correlation with the focal position, fiber or specific object. The exact cause of these flux anomalies is unknown, but we suspect that they were caused by the dropping of the fiber during science exposures. Although the occurrence rate of such flux anomalies is insignificant, we caution that such flux outliers for RM quasars should be removed from the time series analysis upon manual inspection.

\begin{figure*}
\centering
    \includegraphics[width=0.45\textwidth]{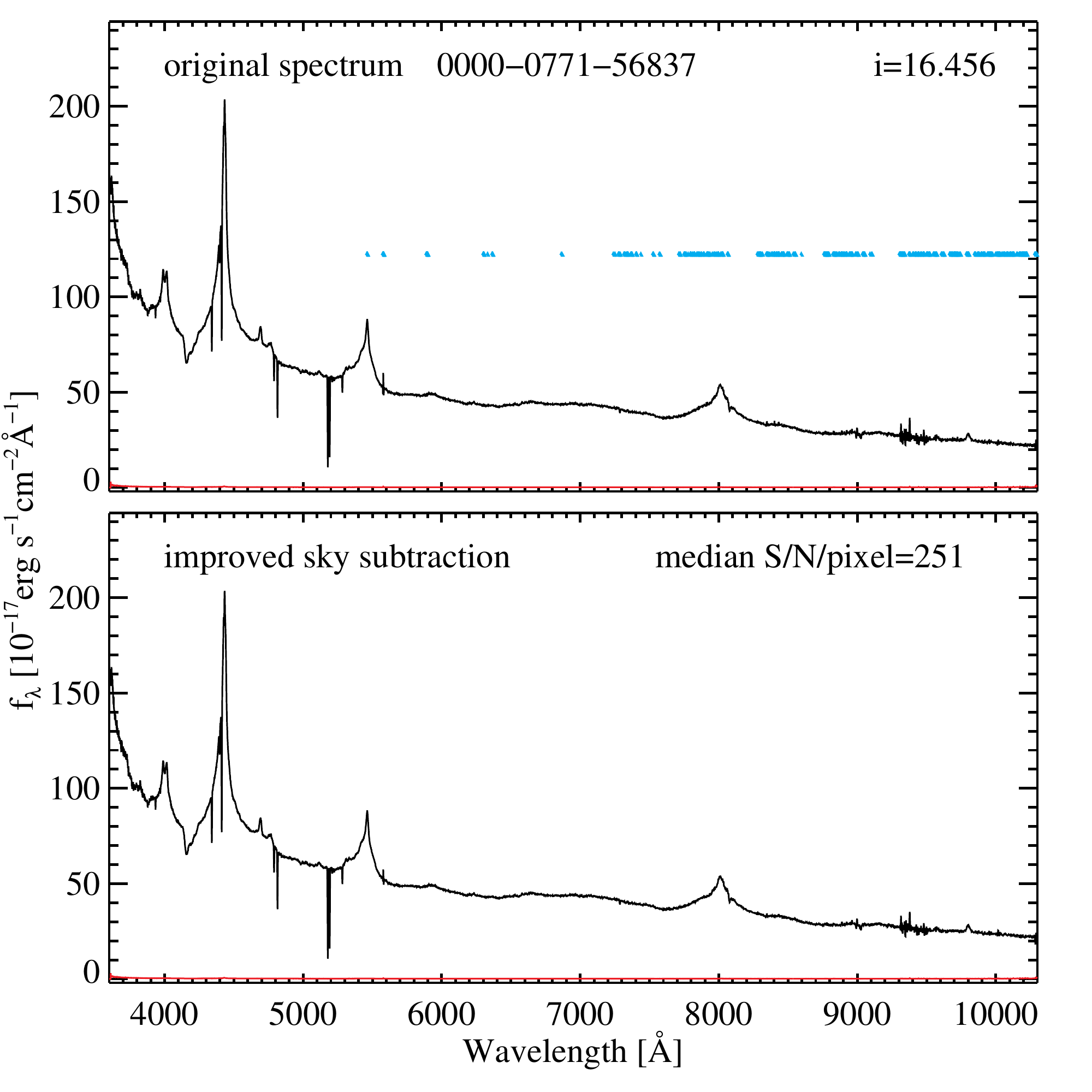}
    \includegraphics[width=0.45\textwidth]{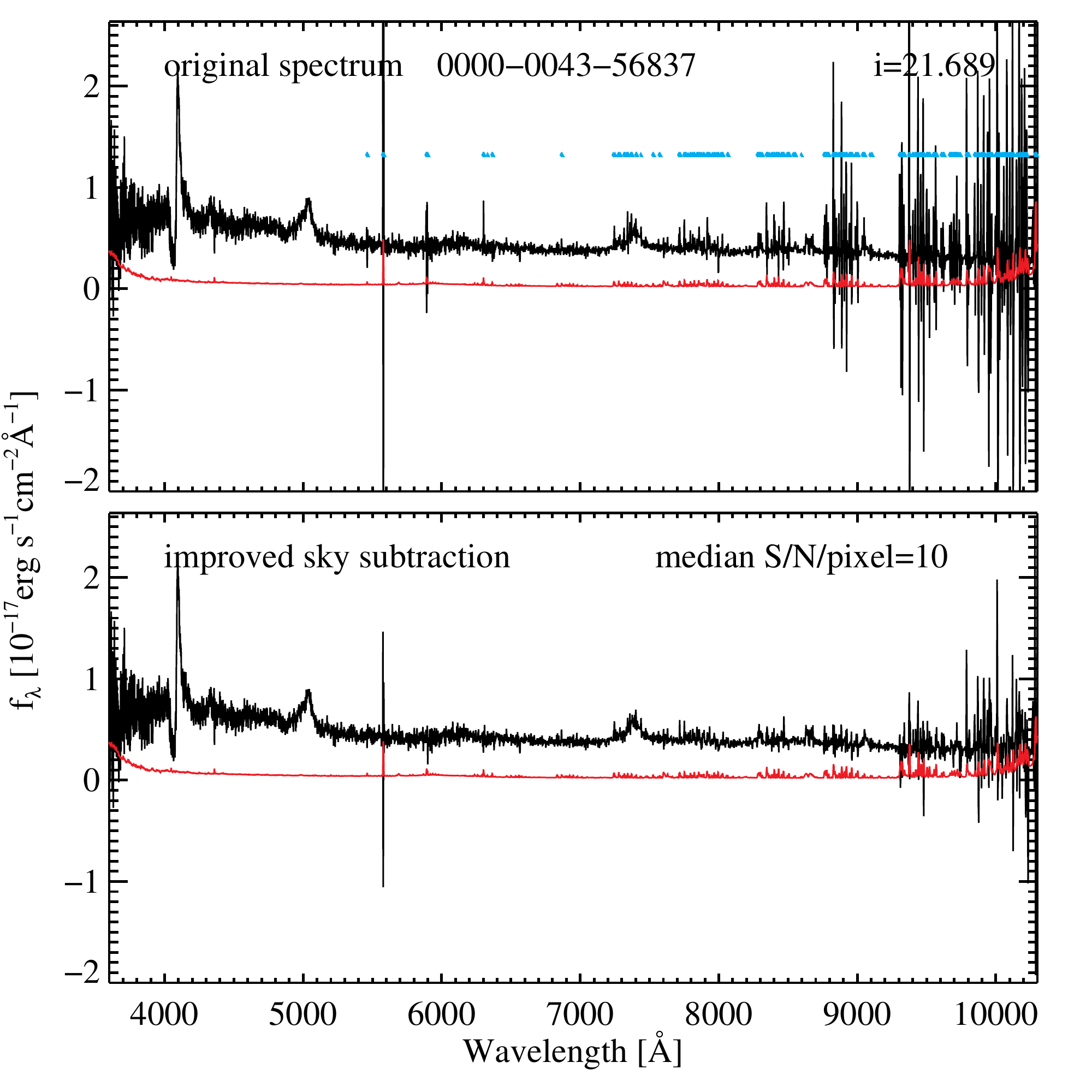}
     \caption{Effects of the improved sky subtraction using the \citet{Wild_Hewett_2005} approach. The black lines are
 the coadded spectra using all individual epochs with (bottom) and without (top) the improved sky subtraction, for two examples in our RM sample, one bright quasar (left) and one faint quasar (right). The improvement is generally better for fainter objects. The red lines are the flux density errors. The cyan symbols in the top panels indicate the locations of sky pixels (Table \ref{table:skymask}). The median S/N per pixel is shown in the bottom panels. }
    \label{fig:skysub}
\end{figure*}

Finally, the BOSS spectra extend to $\sim 1$\ \micron\ where atmospheric OH emission introduces significant residuals in sky subtraction in the reddest part of the spectrum. To remove systematic OH line subtraction residuals, we followed the approach in \citet{Wild_Hewett_2005}. The reader is referred to that paper for technical details. In short, we use all the sky spectra from this RM program ($80$ per epoch) to identify pixels at wavelengths longer than 5000\,\AA\ where significant sky subtraction residuals are present \citep[e.g.,][]{Wild_Hewett_2005}. These sky pixels are listed in Table \ref{table:skymask}. We then perform a principal component analysis on the set of sky spectra at these sky pixels. The resulting principal components (eigenspectra) are used to reconstruct the sky residuals in a given object spectrum, which are subtracted from the original spectrum to produce a new spectrum with reduced sky residuals at these sky pixels. The effect of the improved sky subtraction is demonstrated in Fig.\ \ref{fig:skysub}, where we compare the coadded spectra of two examples from our RM sample between the original and sky-subtraction improved version. Similar improvement is achieved for individual epochs. 

\begin{table}
\caption{Mask of sky pixels}\label{table:skymask}
\centering
\scalebox{1.2}{
\begin{tabular}{lc}
\hline\hline
$\lambda$ [\AA] & $\log_{10}\lambda$ [\AA] \\
\hline
5458.835 & 3.7371 \\
5460.092 & 3.7372 \\
5462.607 & 3.7374 \\
5465.123 & 3.7376 \\
5573.140 & 3.7461 \\
5574.424 & 3.7462 \\
5575.707 & 3.7463 \\
5576.991 & 3.7464 \\
\hline\\
\end{tabular}
}
\begin{tablenotes}
      \small
      \item NOTE. --- Central wavelengths of sky pixels within [5000,10300]\,\AA, generated using the sky spectra from the SDSS-RM program and following the approach in \citet{Wild_Hewett_2005}. A portion is shown here, and the full table is available in the online version. 
\end{tablenotes}
\end{table}

\begin{figure}
\centering
    \includegraphics[width=0.45\textwidth]{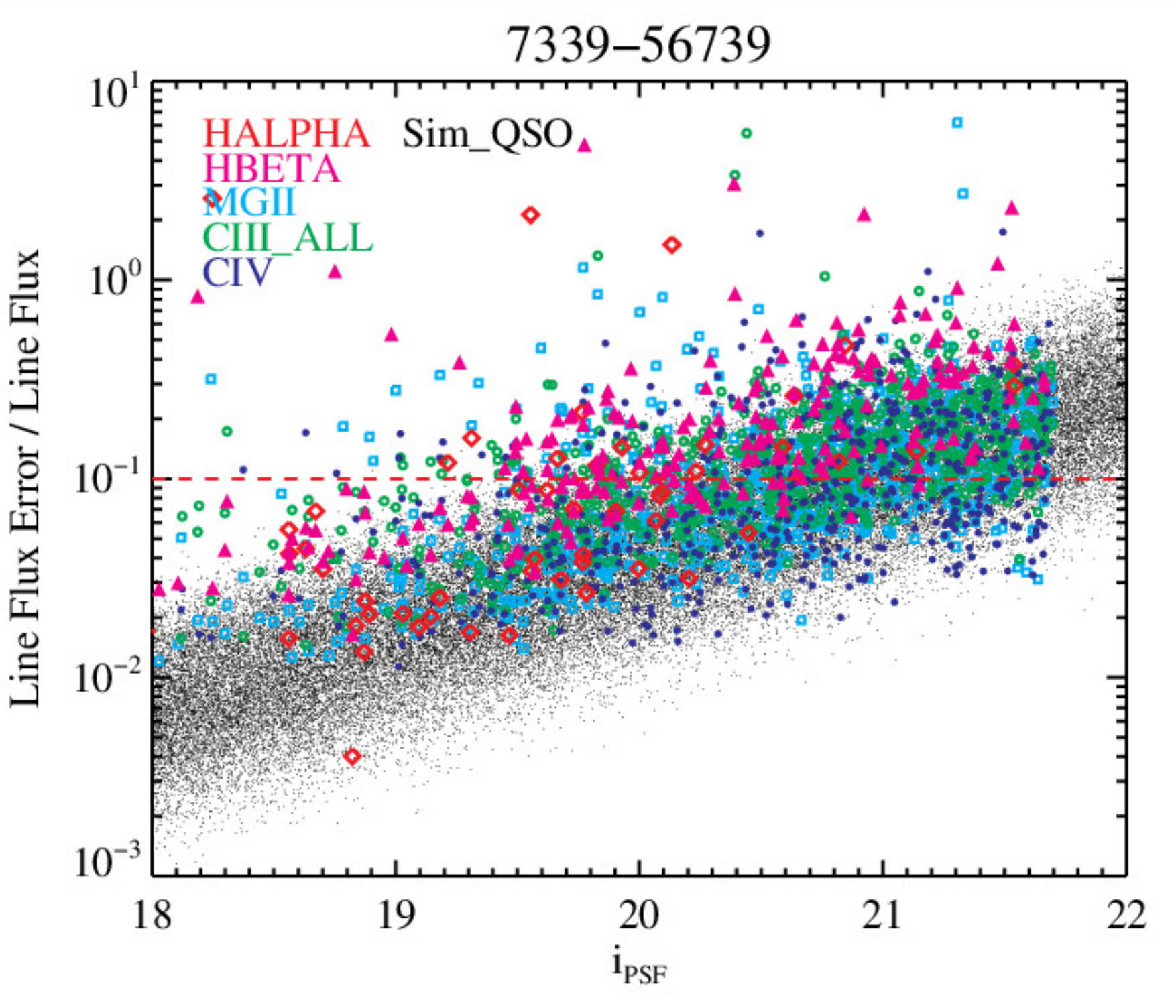}
     \caption{Fractional line flux measurement errors from the spectral fitting procedure, for different lines, on a representative epoch (MJD=56739). The black dots are the simulated quasars used in our forecast (see \S\ref{sec:design}). The measurement errors of the RM quasars are roughly consistent with expectation. }
    \label{fig:line_err_fit}
\end{figure}

Fig.\ \ref{fig:line_err_fit} presents the line flux measurement errors from one typical epoch (MJD$=56739$) for several broad lines based on the line fitting procedure described in Shen et~al.\ (2015, in preparation), and compares the results to those used in the simulated sample. There is general agreement between the actual measurement errors and those assumed in the simulations, which indicates that the observations reached the designed depth. The actual flux measurement errors from line fitting are slightly larger, especially for the Balmer lines, because of the simplification of error estimation in the simulations, as described in \S\ref{sec:det_lag}.

The re-calibrated spectra with the custom flux calibration form the basis of our spectral analysis. Additional improvement of the spectrophotometry using strong narrow emission lines and/or overlapping real-time photometry will be performed to improve the flux measurements for RM time series analyses. 

In the RM data products, we will provide the recalibrated spectra ({\tt spPlate*} files) with and without the Wild \& Hewett sky subtraction. In addition to the individual epochs, we will also provide the final coadded spectra using all epochs, generated using {\em idlspec2d}. The coadded spectra are organized in the same way as individual epochs, with an assigned plate number of $0000$ and an MJD corresponding to the last coadded epoch; fibers 1-1000 in the coadded plate correspond to objects in the same order as in the master sample catalog (Table \ref{table:sample}). The coadded data provide high S/N spectra for the RM quasars, which allows a range of applications including quasar absorption line studies and detailed spectral analysis. The original {\tt spFrame*} files and BOSS pipeline reductions can be accessed as usual from the DR12 server; improved spectroscopic data products will be distributed via the SDSS-RM site.  

\subsection{Further refinement of spectrophotometry}\label{sec:prepspec}


The SDSS-RM spectra taken through fibers suffer from time- and wavelength-dependent light losses. The initial spectrophotometric calibration described in \S\ref{sec:specphoto}, based on spectra of comparison stars distributed across the field of view, is interpolated to the position of each fiber. To check the accuracy of this initial calibration, and of the associated error spectra, we fit the resulting quasar spectra with a simple model that allows for continuum and broad line variations and small time-dependent systematic errors. This is an updated version of the PrepSpec procedure which has been used in earlier RM studies \citep[e.g.,][]{Bentz_etal_2010b}. We model the variable spectrum by
\begin{equation}
S(t,\lambda) = \bar{S}(\lambda)
	+ N(\lambda)
	+ C(t,\lambda)
	+ B(t)\, B(\lambda)
\ .
\end{equation}
Here the non-variable components are decomposed into a ``mean'' spectrum $\bar{S}(\lambda)$, and a narrow-line spectrum $N(\lambda)$ which is non-zero inside specified velocity intervals around specified narrow emission lines, e.g. \OIII\ $\lambda\lambda4959$, 5007. We represent continuum variations $C(t,\lambda)$ by low-order polynomials in wavelength with time-dependent coefficients. We model broad-line variations as a line profile $B(\lambda)$ scaled by a lightcurve $B(t)$, with $B(\lambda)$ smoothed by a spline fit and vanishing outside a specified velocity range, and $B(t)$ normalized to a mean of 0 and standard deviation 1. Several variable broad lines with distinct lightcurves are included when required. We fit the model by iterated optimal scaling \citep[e.g.,][]{Horne_1986} of the lightcurve and spectral patterns, using $1/\sigma^2$ weights to account for the error bar $\sigma$ on each datum. The fit is well defined with fewer parameters than data points. A greyscale image of residuals after fitting the above model can reveal residuals around the narrow lines due to ``jitter'' in the wavelength shift, spectral blurring, and photometric scaling.\footnote{One implicit assumption here is that the narrow line flux remains constant over the course of the program, and that the narrow-line region is compact enough such that seeing variations and aperture effects will not change the narrow line flux within the fiber.} We fit these residual systematic errors by including 3 additional patterns in the model:
\begin{equation}
\mu(t,\lambda)
	= p(t)
	\left(
	S - \Delta\lambda(t)\, \fracd{\partial S}{\partial\lambda}
	 + \Delta b(t)\, \fracd{\partial^2 S}{\partial\lambda^2}
	\right)
\ .
\end{equation}
Here $p(t)$ is the photometric scale factor, $\Delta\lambda(t)$ models sub-pixel wavelength shifts, and $\Delta b(t)$ models changes in spectral blurring. We restrict these fits to a velocity range around the narrow emission lines.
The coefficients can be low-order polynomials in $\lambda$ if required.

For objects with sufficiently strong narrow emission lines the results of the above procedure indicate that residual photometric errors are at the 5\% level, consistent with the estimates based on standard stars in \S\ref{sec:specphoto}, and that the flux error estimates are generally reliable, with reduced $\chi^2$ close to 1. We intend to apply these corrections to the final spectra where warranted, and to look for correlations with other properties to see if the calibrations can be extended to targets that do not have sufficiently strong narrow emission lines (e.g. higher redshift targets).


\subsection{Imaging}\label{sec:imaging}

Supporting photometric monitoring of our RM sample has the following benefits: (1) it improves the precision of the continuum light curves, as opposed to those measured from spectroscopy; (2) it avoids correlated errors in the continuum and line flux measurements due to flux calibration errors in spectroscopy; (3) it is observationally less expensive than spectroscopy and hence can provide more complete coverage of the time baseline of the program, which in turn will increase the probability of lag detections. 

The main portion of the RM imaging of our program is obtained with the CFHT/MegaCam and the Steward Observatory Bok 2.3m/90Prime -- both are degree-size wide-field imagers suitable for monitoring the entire RM field. Given that our RM sample covers a wide redshift range, the photometric monitoring is performed in the SDSS $g$ band and $i$ band to sample the quasar continuum at low and high redshifts, respectively. The $g$ band observations have higher priority than $i$ band observations since these are for the low-$z$ quasars, for which a lag detection is more likely given the limited time baseline of our spectroscopy. CFHT/MegaCam is a dark time instrument, and covers the same period as the BOSS spectroscopy, while Bok/90Prime is primarily covering bright time, between the spectroscopic observations. We also had 6 nights of KPNO-4m (Mayall)/MOSAIC-1.1 observation to perform multi-band ($Ugriz$) imaging of the RM field to understand better the systematics in the BOSS flux calibration.

\begin{figure*}
\centering
    \includegraphics[width=1.0\textwidth]{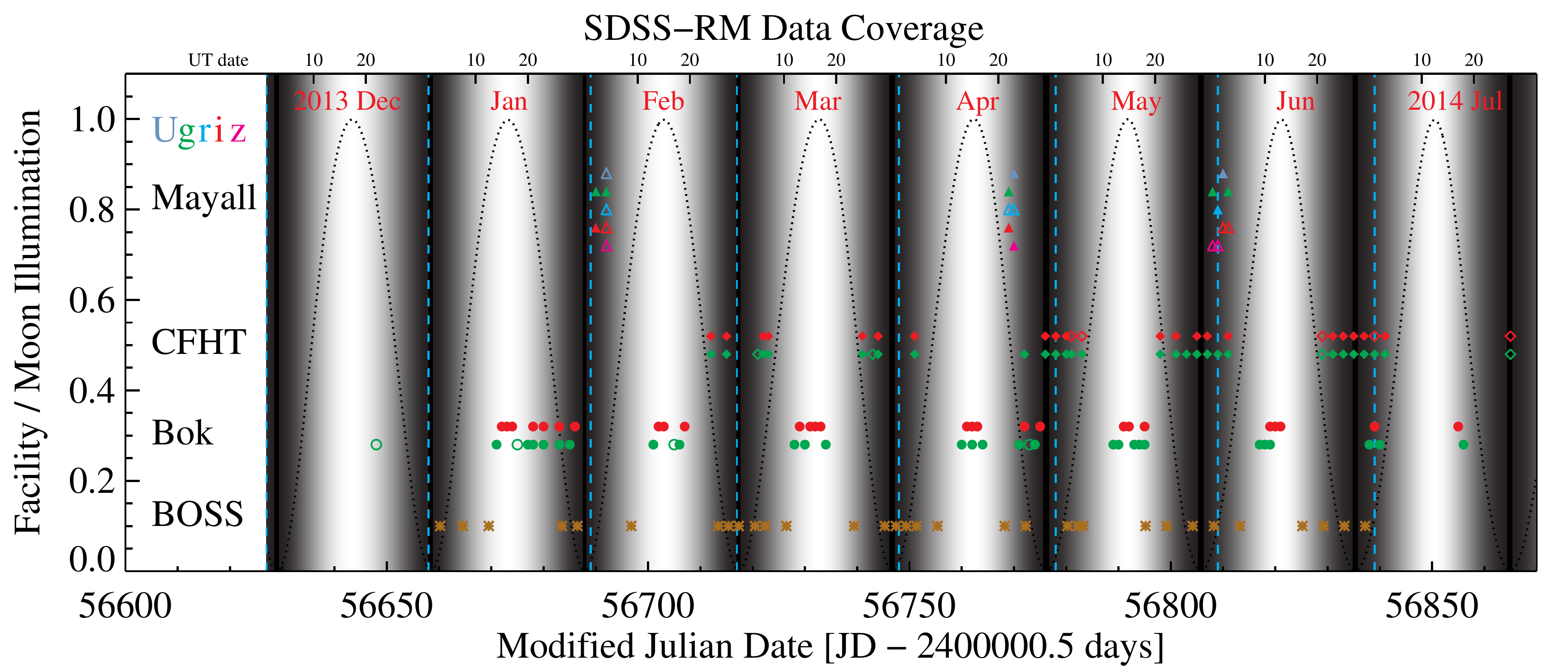}
     \caption{Coverage of the BOSS spectroscopy and imaging observations from Bok, CFHT and Mayall for the 6-month SDSS-RM baseline program. Epochs from different facilities are vertically offset for clarity. Filled and open symbols indicate full and partial coverage of the RM field, respectively. Different colors indicate different filters. The dotted lines are the moon illumination fraction as a function of time in each month, which is also reflected in the grayscale shades (darker means darker skies).}
    \label{fig:coverage}
\end{figure*}

Below we provide brief descriptions of the observing strategy and outcome for each of the three facilities. Tables \ref{table:img_log_bok}-\ref{table:img_log_mayall} summarizes the basic information of these photometric observations, and Fig.\ \ref{fig:coverage} displays the coverage of the imaging and spectroscopic observations. As we are currently processing these imaging data, a detailed description on the data processing and homogenization, and the production of photometric catalogs and light curves will be presented in a forthcoming paper (Kinemuchi et~al.\ 2015, in preparation). 

\subsubsection{Bok observations}


The Bok/90Prime instrument is a prime focus wide-field imager with a $\sim 1\times 1$ deg FoV and a plate scale of 0\farcs45/pixel on the 2.3m Bok telescope. The 90Prime consists of four individual 4Kx4K CCDs, each divided into four amplifiers for readout. A total of $\sim 60$ nights were allocated to the RM program during Jan--July 2014. Most of these nights were scheduled during bright blocks, with a few nights during dark time.  


The observations were performed in classical mode by observers from the SDSS-RM team, and summarized in Table \ref{table:img_log_bok}. The observing strategy is similar to the CFHT observations, with slight differences in dither positions and exposure times. The exposure sequence consists of $2\times150$ sec exposures at each of the two dither positions of a single pointing, with 9 pointings to cover the entire RM field. During the middle of the semester when the RM field is visible for most of the night, two or more passes of a full sequence were executed to increase the S/N. During each Bok run, the $g$ and $i$ observations were typically executed on alternate nights to deliver a 2-night cadence, with the exception of nights within one night of full Moon, during which only $i$-band observations were conducted. With these exposure times, a typical photometric accuracy of $\sim 4\%$ was achieved at the limiting magnitude of our sample.   

Weather conditions were exceptionally good for the Bok runs. By the end of the program, there were 31 epochs in $g$ and 27 epochs in $i$ from the Bok observations. 

\subsubsection{CFHT observations}

The CFHT/MegaCam \citep{Aune_etal_2003} is a wide-field optical imager with a $1\times1$ deg FoV and a pixel size of 0\farcs187. A total of 81.3 hrs were jointly awarded to our RM program by three CFHT time allocation committees (Canada, France and China) during the Jan--Jul 2014 period. Our program was scheduled in 6 dark MegaCam runs during this period.

The observations were carried out in service mode by the CFHT Queued Service Observations team. A total of 9 pointings were used to cover the entire RM field. The exposures at each pointing consist of two dither positions to cover CCD gaps, and two exposures were taken at each dither position: one fixed 30 (40) sec short exposure in $g$ ($i$) for bright RM targets and one dynamic long exposure (on average 55 sec in $g$ and 77 sec in $i$) to achieve a desired S/N of 25 at $i=22$ and $g=22.25$. 

The $g$ and $i$ observations were executed in Observation Groups (OGs) in each run. Completion of the $g$ observation group triggered, via a relational execution link (REEL), an $i$ observation group to be executed within 12 hr (24 hr in Dec/Jan and Jun/Jul). Most of the time the $g$ and $i$ OGs were executed on the same night. If no $i$ epoch was obtained within 24 hr of a $g$ epoch, the $i$ epoch was skipped in favor of a new $g$ epoch two nights later. Under poor weather conditions, the requested cadence increased to nightly near the end of a MegaCam run to yield 4 epochs in that run. 


The CFHT observations were significantly affected by bad weather early in 2014A. By the end of the program, we had observed $64\%$ of the allocated time, totaling 26 full (and 4 partial) epochs in $g$ and 20 full (and 5 partial) epochs in $i$. Most of these data were acquired under excellent seeing conditions ($\lesssim$0\farcs7; see Table \ref{table:img_log_cfht}).  

\subsubsection{Mayall observations}

The Mayall/Mosaic-1.1 imager is a wide-field imaging camera with a FoV of 36\arcmin\ by 36\arcmin\ and a pixel size of 0\farcs26. A total of 6 dark nights were awarded to our program, including 4 second-half nights on Jan 31--Feb 3, 2 full nights on Apr 21--22, and 4 first-half nights on May 30--Jun 2. In contrast to the CFHT and Bok observations, Mosaic-1.1 imaging was conducted in Johnson $U$ and SDSS $griz$ bands. We aim to quantify better the systematics in the BOSS spectrophotometry using these real-time multi-band imaging that roughly overlap with the SDSS-RM spectroscopy. 

The observations were performed by observers from the SDSS-RM team, and are summarized in Table \ref{table:img_log_mayall}. The full RM field is divided into four quadrants, in each of which a sequence of 8 pointings in a spiral pattern was executed, with two dithered exposures in each pointing. The typical single-frame exposure times were: $80$\,s ($U$), $40$\,s ($g$), $60$\,s ($r$), $80$\,s ($i$) and $100$\,s ($z$). With some weather losses, we obtained full or partial coverage of the RM field in the five bands during the three dark runs. 

\subsubsection{Pan-STARRS 1 (PS1) early photometry}\label{sec:ps1}


In addition to imaging data from this RM program, we will utilize early multi-epoch optical photometry from the
Medium Deep Field (MDF) survey of PS1.  Kaiser et~al.\ (2010) describe the overall PS1 system, including
hardware and software. Optical design of the 1.8m diameter $f/4.4$ primary
and 0.9m secondary mirrors, with a 3.3 degree diameter field of view, are
described in Hodapp et al. (2004), while the PS1 imager with its array
of sixty 4800$\times$4800, 10 micron (0\farcs258) pixel detectors is
described in Tonry \& Onaka (2009). The PS1 observations are obtained
through a set of five broadband filters we refer to as simply $grizy$
(Stubbs et al. 2010).  While similar to the SDSS filters, there are
some important differences, particularly in the $g$ band (Tonry et
al. 2012), and the reddest $y$ filter has no SDSS counterpart.

In addition to covering the entire sky at DEC$>-30^\circ$ in 5 bands
(3$\pi$ survey), PS1 conducted a Medium Deep Survey (MDS) consisting
of 10 single fields imaged at a higher cadence of about 4 days in each
filter, to a 5$\sigma$ depth near $\sim$23 mag in the $griz$ filters, and
$\sim$22\,mag in the $y$-filter (with observations taken near full
moon).  Accessible fields are observed with a staggered 3-day cadence
in each of the $griz$ bands during dark or grey time ($g_{\rm PS1}$ and $r_{\rm PS1}$
on the first day, $i_{\rm PS1}$ on the second day, $z_{\rm PS1}$ on the third day, and
then repeat with $g_{\rm PS1}$ and $r_{\rm PS1}$), and in the $y_{\rm PS1}$ band during bright time.
We use the ``\"ubercalibration'' described by Schlafly et al. (2012),
calibrated to 0.01 mag or better. PS1 observed MD07 (the SDSS-RM field) in 2011--2013
($\sim 6$ months each year), in multiple filters; the 5$\sigma$ depth
in a single exposure is $g_{\rm PS1}=22.8$, $r_{\rm PS1}=22.7$,
$i_{\rm PS1}=22.0$, and $z_{\rm PS1}=22.0$.

\begin{table}
\caption{Log of Bok observations}\label{table:img_log_bok}
\centering
\scalebox{1.0}{
\begin{tabular}{lcccc}
\hline\hline
\colhead{MJD} & \colhead{UT} & \colhead{Filter(s)} & \colhead{Seeing} & \colhead{Comments}  
\\
\hline
56648 &  20131222 &   g &    3.5"       & partial coverage\\
56671 &  20140114 &   g &        1.6" &     \\               
56672 &  20140115 &   i &        1.7" &     \\               
56673 &  20140116 &   i &        1.6" &     \\               
56674 &  20140117 &   i &        1.3" &     \\               
56675 &  20140118 &   g &        1.2" &    partial coverage \\
56677 &  20140120 &   g &        1.5" &     \\               
56678 &  20140121 &   g,i  &     1.4" &     \\               
56680 &  20140123 &   g,i   &    1.8" &     \\               
56683 &  20140126 &   g,i   &    1.3" &     \\               
56685 &  20140128 &   g &        1.5" &     \\               
56686 &  20140129 &   i &        1.4" &     \\               
56701 &  20140213 &   g &        1.5" &     \\               
56702 &  20140214 &   i &         1.2" &     \\             
56703 &  20140215 &   i &      1.3" &     \\                
56705 &  20140217 &   g &      1.4" &    partial coverage\\ 
56706 &  20140218 &   g &   1.8" &     \\                    
56707 &  20140219 &   i &   1.7" &     \\                    
56728 &  20140312 &   g &   1.4" &     \\                    
56729 &  20140313 &   i &   1.3" &     \\                   
56730 &  20140314 &   g &  2.0" &     \\                    
56731 &  20140315 &   i &   1.4" &     \\                   
56732 &  20140316 &   i &   2.5" &     \\                   
56733 &  20140317 &   i &   1.4" &     \\                   
56734 &  20140318 &   g &   1.4" &     \\                    
56760 &  20140413 &   g &   2.1" &     \\                    
56761 &  20140414 &   i &   1.5" &     \\                    
56762 &  20140415 &   g,i & 1.8" &     \\                    
56763 &  20140416 &   i &   1.1" &     \\                    
56764 &  20140417 &   g &   1.8" &     \\                    
56771 &  20140424 &   g &   1.2" &     \\                    
56772 &  20140425 &   i &   1.2" &     \\                    
56773 &  20140426 &   g &   2.3" &    partial coverage \\    
56774 &  20140427 &   g &   1.8" &     \\                    
56775 &  20140428 &   i &   1.7" &     \\                    
56789 &  20140512 &   g &   2.2" &     \\                    
56790 &  20140513 &   g &   1.9" &     \\                    
56791 &  20140514 &   i &   2.0" &     \\                    
56792 &  20140515 &   i &   1.3" &     \\                    
56793 &  20140516 &   g &   1.5" &     \\                    
56794 &  20140517 &   g &   1.3" &     \\                    
56795 &  20140518 &   g,i & 1.5" &     \\                    
56817 &  20140609 &   g &   1.4" &     \\                    
56818 &  20140610 &   g &   1.4" &     \\                    
56819 &  20140611 &   g,i & 1.8" &     \\                    
56820 &  20140612 &   i &   1.2" &     \\                    
56821 &  20140613 &   i &   1.6" &     \\                    
56838 &  20140630 &  g &  1.6" & \\                         
56839 &  20140701 &  i  &  1.4" &  \\                       
56840 &  20140702 &  g  &  1.7" &  \\                       
56855 &  20140717 &  i  &  1.4"  &  \\                      
56856 &  20140718  & g  &  2.1" & \\                        
\hline\\
\end{tabular}
}
\begin{tablenotes}
      \small
      \item NOTE. --- ``partial coverage'' indicates the RM field is partially covered during an epoch. 
\end{tablenotes}
\end{table}

\begin{table}
\caption{Log of CFHT observations}\label{table:img_log_cfht}
\centering
\scalebox{1.0}{
\begin{tabular}{lcccc}
\hline\hline
\colhead{MJD} & \colhead{UT} & \colhead{Filter(s)} & \colhead{Seeing} & \colhead{Comments}  
\\
\hline
56712 &  20140224  &  g,i &  0.7" &  \\                             
56715 &  20140227  &  g,i  & 1.0" &        \\                       
56721 &  20140305  &  g &     1.2" & partial coverage \\            
56722 &  20140306  &  g,i  &  0.7" &  \\                            
56723 &  20140307  &  g,i  &  0.6" &  \\                            
56741 &  20140325  &  g,i  &  1.0" &  \\                            
56743 &  20140327  &  g &   0.8" & partial coverage \\              
56744 &  20140328  &  g,i  &  0.7" &  \\                            
56751 &  20140404  &  g,i  &  0.7" &  \\                            
56772 &  20140425  &  g &     0.9" &  \\                            
56776 &  20140429  &  g,i &   0.6" &  \\                            
56778 &  20140501  &  g,i  &  0.7" &  \\                            
56780 &  20140503  &  g,i  &  0.4" &  \\                            
56781 &  20140504  &  g,i  &  0.5" & partial coverage in i  \\      
56783 &  20140506  &  g,i  &  0.5" & partial coverage in i \\       
56798 &  20140521  &  g,i  &  0.6" &  \\                            
56801 &  20140524  &  g,i  &  0.8" & \\                             
56803 &  20140526  &  g &     0.9" & \\                             
56805 &  20140528  &  g,i  &  0.7" & \\                             
56807 &  20140530  &  g,i  &  0.8" & \\                             
56809 &  20140601  &  g &     0.8" & \\                             
56811 &  20140603  &  g,i  &  0.5" & \\                             
56829 &  20140621  &  g,i  &  0.6" & partial coverage (CFHT-LS D3)\\ 
56831 &  20140623  &  g,i  &  0.5" & \\                             
56833 &  20140625  &  g,i  &  0.6" & \\                             
56835 &  20140627  &  g,i  &  0.6" & \\                             
56837 &  20140629  &  g,i  &  0.6" & \\                             
56839 &  20140701  &  g,i  &  0.8" & partial coverage in i \\       
56841 &  20140703  &  g,i  &  0.6" & \\                             
56865 &  20140727  &  g,i  &  0.8" & partial coverage (CFHT-LS D3)\\
\hline\\
\end{tabular}
}
\begin{tablenotes}
      \small
      \item NOTE. --- ``partial coverage'' indicates the RM field is partially covered during an epoch. Epochs MJD$=56829$ and $56865$ were pointing at the CFHT-LS D3 field to observe additional calibration sources, which partially covers the RM field. 
\end{tablenotes}
\end{table}

\begin{table}
\caption{Log of Mayall observations}\label{table:img_log_mayall}
\centering
\scalebox{1.0}{
\begin{tabular}{lcccc}
\hline\hline
\colhead{MJD} & \colhead{UT} & \colhead{Filter(s)} & \colhead{Seeing} & \colhead{Comments}  
\\
\hline                                                                      
56690 &  20140202 &   g,i &  2.0" &    \\                                 
56692 &  20140204  &  U,g,r,i,z &   1.0" & partial in U,r,i,z \\         
56769 &  20140422  &  g,r,i &   1.1" & partial in r \\                  
56770 &  20140423  &  U,r,z & 1.7" & partial in r \\                    
56808 &  20140531  &  g,z & 1.5" & partial in z \\                      
56809 &  20140601  &  r,z & 1.0" & partial in z \\                      
56810 &  20140602  &  U,i & 0.8" & partial in i \\                      
56811 &  20140603  &  g,i & 1.5" & partial in i \\    
\hline\\
\end{tabular}
}
\begin{tablenotes}
      \small
      \item NOTE. -- ``partial'' indicates the RM field is partially covered during an epoch.
\end{tablenotes}
\end{table}

\section{Discussion}\label{sec:disc}

The primary science goal of the SDSS-RM program is to measure time lags between the broad line and continuum variability. These lag measurements will enable an array of applications to understand the structure of quasar BLRs and the co-evolution of galaxies and BHs. Below we summarize the RM-related primary science cases of this program:

\begin{enumerate}

\item[$\bullet$] We will measure the typical sizes of the BLR with RM lags, and compare the results among different broad lines. We aim to detect RM lags for a significant number ($\sim 100$ under ideal situation) of $z<2$ quasars in our sample from the 6-month program, and many more by combining the PS1 early light curves and continued spectroscopic monitoring of the RM field. This sample of RM detections can be used to study the BLR structure and broad line stratification, and to test kinematic models of the BLR. Although our RM data are unlikely to have sufficient quality for velocity-resolved time delays for individual objects, we will try composite RM \citep[e.g.,][]{Fine_etal_2013} to constrain the average properties of velocity-resolved lags for our quasar sample. 

\item[$\bullet$] We will examine the empirical relation between BLR lags and continuum luminosity (the $R-L$ relation) found in early RM work. With our sample, we will investigate the scatter (and its dependence on intrinsic quasar properties) in the $R-L$ relation based on the \hbeta\ line using an unbiased RM target selection, and we will also investigate the $R-L$ relations based on other broad lines (including \MgII\ and \CIV), which are poorly constrained at present. These improved $R-L$ relations will enable the design of single-epoch BH mass estimators that can be used with confidence in a variety of applications over a large range of redshift and luminosity. 

\item[$\bullet$] The RM lags will provide direct BH mass estimates for a substantial sample of quasars at $z>0.3$. These will be the first sample of quasars in this redshift regime with more reliable BH mass estimates than those estimated indirectly from scaling relations. By measuring the host galaxy properties (either from spectral decomposition or high resolution imaging decomposition), we will be able to investigate the correlations between BH mass and host properties beyond the local Universe. This exercise avoids the usage of the less reliable and less accurate BH mass estimates based on single-epoch spectroscopy, and will assess the possibility of evolving scaling relations between BH mass and host properties in a more robust manner. 
\end{enumerate} 

With these potential significant advances in mind, we fully appreciate the caveats and challenges in our program. This multi-object RM program is the first of its kind, and works in a regime with little past experience. The quasars in our sample are substantially fainter (in flux) than the local RM AGNs; the spectral quality, cadence and time baseline are substantially worse than in traditional RM work. Although our simulations predict a large number of detections, some simple assumptions in our simulations (such as the \MgII\ and \CIV\ lag timescales) must be examined with real data. Nevertheless, the spectroscopic and photometric data sets from this program are among the best time-domain data sets for AGN and quasar variability studies on the timescales sampled. These data sets will enable a broad range of ancillary science:

\begin{enumerate}

\item[$\bullet$] Photometric and spectral variability of quasars. The RM project will create a large and uniform sample of quasars with superb photometric and spectroscopic monitoring data covering timescales of days to months. Such a sample will provide crucial information on quasar variability and its correlations with other quasar properties. A particular application is to search for short-time variability in broad absorption line troughs of quasars \citep[e.g.,][]{Capellupo_etal_2011}, which will shed light on the physical processes regulating the fast outflows launched near the BH.  

\item[$\bullet$] Deep spectroscopy of quasars in the coadded data. The equivalent total exposure is $\sim 10$\,hrs on a 6m telescope. These high S/N coadded spectra will be used to perform spectral decomposition of quasar and host light, measure accurate spectral properties and single-epoch BH mass estimates, and study quasar absorption lines. 

\item[$\bullet$] High-resolution and deep imaging of quasar hosts and immediate environments. The CFHT coadded data are sufficiently deep and have good image quality ($<$0\farcs7) to address this issue. Such coadded imaging will surpass the coadded SDSS images in Stripe 82 and provide better constraints on the quasar host properties using imaging decomposition techniques. 

\item[$\bullet$] This program will promote multi-wavelength synergy in the RM field, such as HST imaging to measure host properties, X-ray observations of RM quasars, and infrared monitoring to detect optical-IR lags to infer the size of the dust torus. 

\end{enumerate}

\section{Summary}\label{sec:sum}

We have presented a technical overview of the SDSS-RM project, the first multi-object RM program on the SDSS-III BOSS spectrograph. This program has obtained more than 30 spectroscopic epochs for a flux-limited sample of 849 quasars within a single 7 deg$^2$ field, and $\sim 60$ epochs of 2-band ($g$ and $i$) photometric observations from CFHT, Bok and KPNO-Mayall, over $\sim 6$ months in 2014A. We described the design (\S\ref{sec:design}) and implementation (\S\ref{sec:implem}) of this program, and outlined additional efforts in improving the spectral quality for RM analysis. In successive papers, we will present detailed sample characterization (Shen et~al.\ 2015), photometry data reduction and products (Kinemuchi et~al.\ 2015), as well as science results from this program. We will also try to extend the program length by continuing the spectroscopic and photometric monitoring of the same RM field in the SDSS-IV era (which started in July 2014). 

Data products from the SDSS-RM program (aside from those included in the standard SDSS-III DR12), including improved spectroscopic calibration, photometric images and catalogs, light curves, etc, will be made available successively at http://www.sdssrm.org/.

\acknowledgements  We thank Aaron Barth and Shai Kaspi for useful comments that helped us improve the program design, and David Schlegel, Stephen Bailey and Adam Bolton for help with the BOSS pipeline. Support for the work of YS was provided by NASA through Hubble Fellowship grant number HST-HF-51314.0, awarded by the Space Telescope Science Institute, which is operated by the Association of Universities for Research in Astronomy, Inc., for NASA, under contract NAS 5-26555. WNB acknowledges support from NSF grant AST-1108604. KDD acknowledges support by the NSF through award AST-1302093. LCH acknowledges support by the Chinese Academy of Science through grant No. XDB09030102 (Emergence of Cosmological Structures) from the Strategic Priority Research Program and by the National Natural Science Foundation of China through grant No. 11473002. BMP acknowledges support from the NSF through grant AST-1008882.

Funding for SDSS-III has been provided by the Alfred P. Sloan Foundation, the
Participating Institutions, the National Science Foundation, and the U.S.
Department of Energy Office of Science. The SDSS-III web site is
http://www.sdss3.org/.

SDSS-III is managed by the Astrophysical Research Consortium for the
Participating Institutions of the SDSS-III Collaboration including the
University of Arizona, the Brazilian Participation Group, Brookhaven National
Laboratory, University of Cambridge, Carnegie Mellon University, University
of Florida, the French Participation Group, the German Participation Group,
Harvard University, the Instituto de Astrofisica de Canarias, the Michigan
State/Notre Dame/JINA Participation Group, Johns Hopkins University, Lawrence
Berkeley National Laboratory, Max Planck Institute for Astrophysics, Max
Planck Institute for Extraterrestrial Physics, New Mexico State University,
New York University, Ohio State University, Pennsylvania State University,
University of Portsmouth, Princeton University, the Spanish Participation
Group, University of Tokyo, University of Utah, Vanderbilt University,
University of Virginia, University of Washington, and Yale University.

We thank the Bok and CFHT Canadian, Chinese and French TACs for their support. This research uses data obtained through the Telescope Access Program (TAP), which is funded by the National Astronomical Observatories, Chinese Academy of Sciences, and the Special Fund for Astronomy from the Ministry of Finance in China.


\end{document}